\documentclass[twocolumn,times,tighten]{aastex61}

\usepackage[all]{hypcap} %Figure refs breaks deluxetables; use %\capstartfalse %\capstarttrue to fix it

\usepackage{amsmath,amsfonts, amssymb}

\usepackage{mathrsfs}

\usepackage{dsfont}

\usepackage{url}
\usepackage{paralist}
\usepackage{float}
\restylefloat{deluxtable}

\usepackage{subfigure}

\shorttitle{Precision distances to dwarf galaxies and globular clusters from Pan-STARRS1 3$\pi$ RR Lyrae}
\shortauthors{Hernitschek et~al.}

\begin{document}


\title{Precision distances to dwarf galaxies and globular clusters from Pan-STARRS1 3$\pi$ RR Lyrae}


\author{Nina Hernitschek}
\affiliation{Division of Physics, Mathematics and Astronomy, Caltech, Pasadena, CA 91125}

\author{Judith G. Cohen}
\affiliation{Division of Physics, Mathematics and Astronomy, Caltech, Pasadena, CA 91125}

\author{Hans-Walter Rix}
\affiliation{Max-Planck-Institut f{\"u}r Astronomie, K{\"o}nigstuhl 17, 69117 Heidelberg, Germany}

\author{Eugene Magnier}
\affiliation{Institute for Astronomy, University of Hawai’i at Manoa, Honolulu, HI 96822, USA}

\author{Nigel Metcalfe}
\affiliation{Department of Physics, University of Durham, South Road, Durham DH1 3LE, UK}

\author{Richard Wainscoat}
\affiliation{Institute for Astronomy, University of Hawai’i at Manoa, Honolulu, HI 96822, USA}

\author{Christopher Waters}
\affiliation{Department of Astrophysical Sciences, 4 Ivy Lane, Princeton University, Princeton, NJ 08544, USA}
%
%\author{Nick Kaiser}
%\affiliation{Institute for Astronomy, University of Hawai’i at Manoa, Honolulu, HI 96822, USA}
%
%\author{John L. Tonry}
%\affiliation{Institute for Astronomy, University of Hawai’i at Manoa, Honolulu, HI 96822, USA}
%
\author{Rolf-Peter Kudritzki}
\affiliation{Institute for Astronomy, University of Hawai’i at Manoa, Honolulu, HI 96822, USA}
%
%\author{Klaus Hodapp}
%\affiliation{Institute for Astronomy, University of Hawai’i at Manoa, Honolulu, HI 96822, USA}
%
%\author{Ken Chambers}
%\affiliation{Institute for Astronomy, University of Hawai’i at Manoa, Honolulu, HI 96822, USA}
%
%\author{Heather Flewelling}
%\affiliation{Institute for Astronomy, University of Hawai’i at Manoa, Honolulu, HI 96822, USA}
%
\author{William Burgett}
\affiliation{GMTO Corporation (Pasadena), 465 N. Halstead Street, Suite 250, Pasadena, CA 91107, USA}

\correspondingauthor{Nina Hernitschek}
\email{ninah@astro.caltech.edu}



%  250 words maximum
\begin{abstract}
We present new spatial models and distance estimates for globular clusters (GC) and dwarf spheroidals (dSphs) orbiting our Galaxy based on RR Lyrae (RRab) stars in the Pan-STARRS1 (PS1) 3$\pi$ survey.
Using the PS1 sample of RRab stars from \cite{Sesar2017} in 16 globular clusters and 5 dwarf galaxies, we fit structural models in $(l,b,D)$ space;
for 13 globular clusters and 6 dwarf galaxies, we give only their mean heliocentric distance $D$. We verify the accuracy of the period-luminosity (PL) relations used in \cite{Sesar2017} to constrain the distance to those stars, and compare them to period-luminosity-metallicity (PLZ) relations using metallicities from \cite{Carretta2009}.
We compare our \cite{Sesar2017} distances to the parallax-based \textit{Gaia} DR2 distance estimates from \cite{BailerJones2018}, and find our distances to be consistent and considerably more precise.
\end{abstract}


%\keywords{  quasars: supermassive black holes ---  galaxies: photometry}

\section{Introduction}
\label{sec:Introduction}

In this paper, we exploit the opportunities provided by the Pan-STARRS1 (PS1) 3$\pi$ RR Lyrae catalog \citep{Sesar2017} to explore the Galactic halo.
Using that large data set of RRab stars covering 3/4 of the sky out to more than 130 kpc, in \cite{Hernitschek2018} we looked at the seemingly smooth part of the stellar halo. We now focus on the RRab stars in globular clusters (GC) and dwarf galaxies in order to determine their distances and constrain their structure. For most of these objects, RRab stars may be the most precise distance tracers. As shown in Section \ref{sec:GaiaDR2Distances}, all GC and dwarf galaxies we study here are beyond any parallax precision of \textit{Gaia} DR2, which is according to \cite{BailerJones2018} only valid within a heliocentric distance range of about 5 kpc.
The main advantages of such a study are that due to the large angular extent and depth of PS1 3$\pi$, the source of observational data as well as the methodology are the same for all GC and dSph we analyze.
The aim of this paper is to precisely constrain the position and extent of each of these overdensities from RRab stars, where we determine these properties for 29 GC and 11 dwarf spheroidal (dSph) galaxies, and if this is not possible, to at least to give their mean heliocentric distances. Furthermore, from the globular clusters we demonstrate that the period-luminosity relations used by \cite{Sesar2017} to determine the distances of the RRab stars are precise and their predicted magnitudes well within the claimed uncertainties.

RR Lyrae stars (RRL) as distance indicators have the advantage of being low-mass stars found in both early- and late-type stellar systems \citep{vandenBergh1999}. 
When periods, or periods and metallicities, for RRL are known, their distances can be estimated using period-luminosity (PL) or period-luminosity-metallicity (PLZ) relations.
Visual magnitude-metallicity relations are described for example in \cite{Chaboyer1996}, \cite{Bono2003}, \cite{Cacciari2003}.
In the infrared, the extinction is smaller and the amplitude of variation is smaller. In the near infrared (NIR), well-defined PL relations can be found \citep{Longmore1986, Bono2001, Bono2003, Catelan2004, Braga2015, Sesar2017}, extending to the mid-infrared with only small scatter \citep{Madore2013, Klein2014}.

These individual distance estimates can then be used to calculate distances to and the shape of globular clusters as well as dwarf galaxies.
Dwarf galaxies orbiting our Milky Way are not only interesting objects per se, but can also provide us with important information to determine the Milky Way's total mass and dynamics.

\cite{Dambis2014} used WISE as well as zero-points from HST parallaxes to derive PL relations from 360 RRL belonging to 15 globular clusters.

More recent work on the distances of globular clusters was carried out by e.g. \cite{Neeley2015} who investigated mid-IR period-luminosity-metallicity (PLZ) relations for the globular cluster NGC 6121 (M4), which is not in our sample as it lies outside of the PS1 3$\pi$ footprint. Their work contains PL relations for both the RRab and RRc samples as well as for the combined RRab+RRc sample. In their 2017 paper \citep{Neeley2017}, they also present theoretical mid-IR PLZ relations for RR Lyrae stars at Spitzer and WISE wavelengths and show
that the mid-IR PLZ relations can provide distance estimates to individual RR Lyrae stars with uncertainties better than 2\%. This is comparable to the result by \cite{Sesar2017} derived for the RRab stars with optical photometry used in this paper.
For the dSph Sculptor, which lies outside of the PS1 3$\pi$ footprint, \cite{Garofalo2018} derived mid-IR PLZ relations and calculated the distance to the dwarf galaxy 
as part of the Spitzer SHMASH project, which targeted in total four dSph (Ursa Minor dSph, Bootes dSph, Sculptor dSph and Carina dSph).

This paper is structured as follows. In Section \ref{sec:PS1}, we introduce the PS1 3$\pi$ survey and lay out the properties of the PS1 RRab stars. This is followed in Section \ref{sec:SampleOfGlobularClustersAndDwarfGalaxies} by a description of the spatial area covered by PS1 3$\pi$, and the list of GCs and dSph that are included in it.  In Section \ref{sec:DensityFitting}, we present the method of fitting a series of parameterized stellar density models and the probabilistic approach to constrain their model parameters. We describe  fitting tests on mock data in Section \ref{sec:TestsOnMockData}. In Section \ref{sec:ResultsfromFitting}, the position and extent are estimated for each GC and dSph. We present various analyses based on these fitted parameters. In Section \ref{sec:MeanDistancestoOtherGlobularClustersandDwarfGalaxies}, we calculate mean distances for those dwarf galaxies and GC which don't have enough sources for the fitting process. We also discuss the implications on period-luminosity relations in Sec. \ref{sec:PeriodLuminosityRelations}, and compare our distance estimates to those derived using the second \textit{Gaia} data release in Sec. \ref{sec:GaiaDR2Distances}. {A discussion and summary of our results is given in Section \ref{sec:Discussion}.}
The main part of the paper is followed by a large Appendix containing figures as well as tables.

\section{RR Lyrae Stars from the PS1 Survey}
\label{sec:PS1}

As in our previous papers, e.g. \cite{Hernitschek2017} and \cite{Hernitschek2018}, our analysis is based on a sample of highly likely RRab stars, as selected by \cite{Sesar2017} from the Pan-STARRS1 3$\pi$ survey. 
Here, we briefly describe the pertinent properties of the PS1 3$\pi$ survey and the RR Lyrae light curves obtained, and recapitulate briefly the process of selecting the likely RRab for the catalog of PS1 RRab stars, as laid out in \cite{Sesar2017}. We also briefly characterize the obtained candidate sample.

The Pan-STARRS 1 (PS1) survey \citep{Kaiser2010} collected multi-epoch, multi-color observations undertaking a number of surveys, among which the PS1 3$\pi$ survey \citep{Chambers2016} is currently the largest. It has observed the entire sky north of declination $-30^{\circ}$ in five filter bands ($g_{\rm P1},r_{\rm P1},i_{\rm P1},z_{\rm P1},y_{\rm P1}$) 
with a 5$\sigma$ single epoch depth of about 22.0, 22.0, 21.9, 21.0 and 19.8 magnitudes in $g_{\rm P1},r_{\rm P1},i_{\rm P1},z_{\rm P1}$, and $y_{\rm P1}$, respectively  \citep{Stubbs2010, Tonry2012}. 

Starting with a sample of more than  $1.1\times 10^9$ PS1 3$\pi$ sources, 
\cite{Hernitschek2016} and \cite{Sesar2017} subsequently selected a sample of 
44,403 likely RRab stars, of which $\sim$17,500
are at $R_{\mathrm{gc}} \geq 20$~kpc, by applying machine-learning techniques based on 
light-curve characteristics. RRab stars are the most common type of RR Lyrae, 
making up ${\sim}91\%$ of all observed RR Lyrae \citep{Smith2004}, and 
displaying the steep rises in brightness typical of RR Lyrae.

The identification of the RRab stars is highly effective, and the sample of RRab stars is pure ($90\%$), and complete ($\geq80\%$ at 80~kpc) at high galactic latitudes. 
Distances to these stars were calculated based on flux-averaged $i_{\mathrm{P1}}$ magnitudes, corrected for dust extinction using extinction 
coefficients of \cite{Schlafly2011} and the dust map of \cite{Schlafly2014}. The distance estimates are precise to $3\%$, based on newly derived PL relations for the optical/near-infrared PS1 bands \citep{Sesar2017}. Overall, this results in the widest (3/4 of the sky) and deepest (reaching $> 120$~kpc) sample of RR Lyrae stars to date, allowing us to observe them globally across the Milky Way. Out of these sources, 1093 exist beyond a Galactocentric distance of 80 kpc, with 238 beyond 100 kpc, which enables us to estimate distances to and extents of dwarf galaxies and globular clusters.
RRc stars, which were also selected by \cite{Sesar2017}, are not included in our study.

\section{Sample of Globular Clusters and Dwarf Galaxies}
\label{sec:SampleOfGlobularClustersAndDwarfGalaxies}

PS1 3$\pi$, as it covers the entire sky north of declination $-30^{\circ}$ out to more than 130 kpc, gives us access to many globular clusters (GC) and several dwarf galaxies. 

\begin{figure*}
\begin{center}  
\includegraphics[]{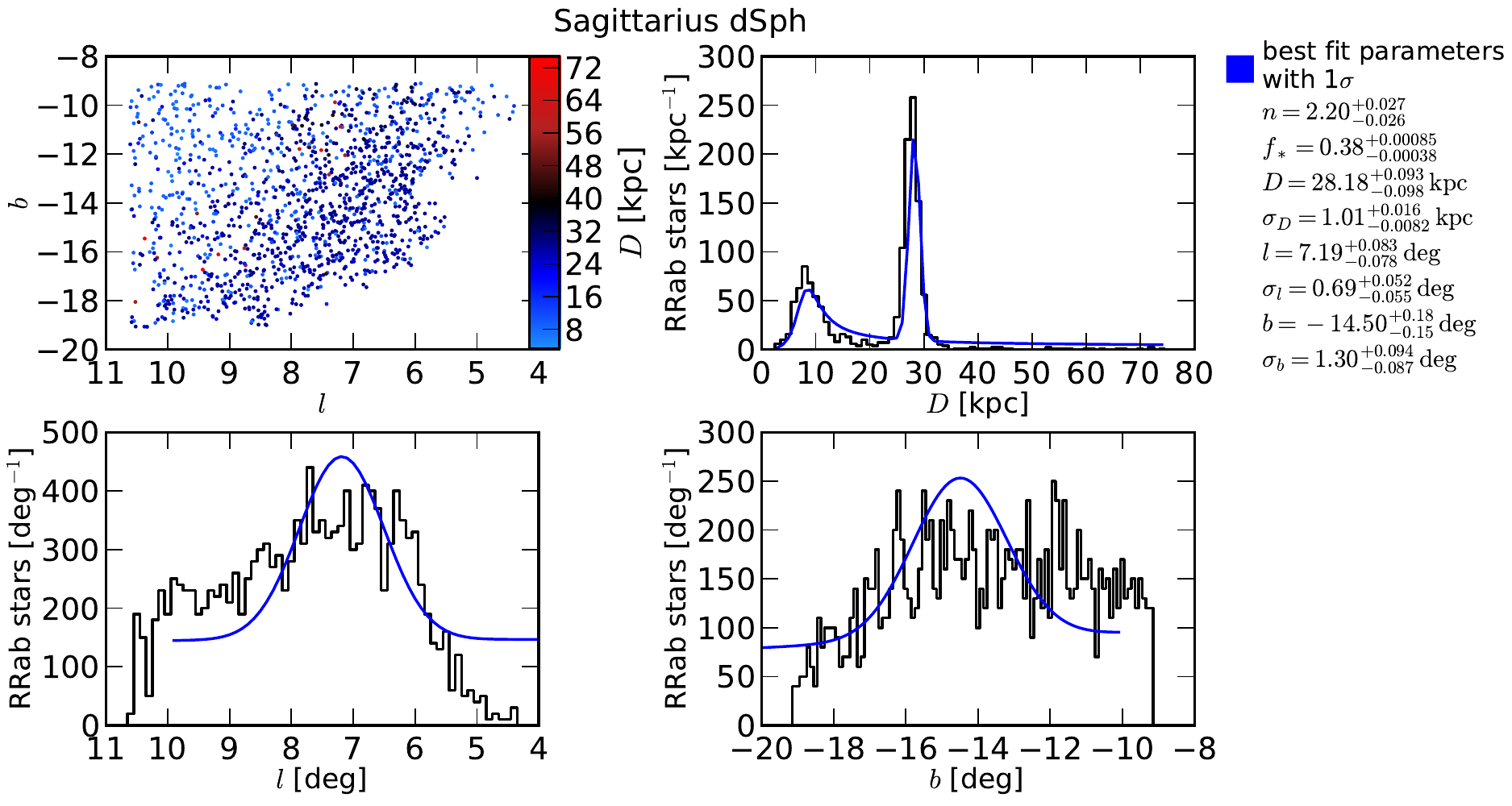}
\caption{{
The dwarf spheroidal Sagittarius dSph. The first panel shows a map of 1413 RRab stars near the dwarf galaxy; the stars are color-coded according to their heliocentric distance.
The other three panels show the histograms in $l$, $b$, $D$ for the stars from the first panel. Overplotted is the best-fit model from Sec. \ref{sec:DensityFitting} with the parameters given on the right. As the Sagittarius dSph lies near the edge of the PS1 3$\pi$ footprint, we cannot successfully fit its on-sky position $(l,b)$, but we still can fit its heliocentric distance $D$.}
\label{fig:Sagittarius_dSph_fits}}
\end{center}  
\end{figure*}

\begin{figure*}
\begin{center}  
\subfigure[]
    {
\includegraphics[]{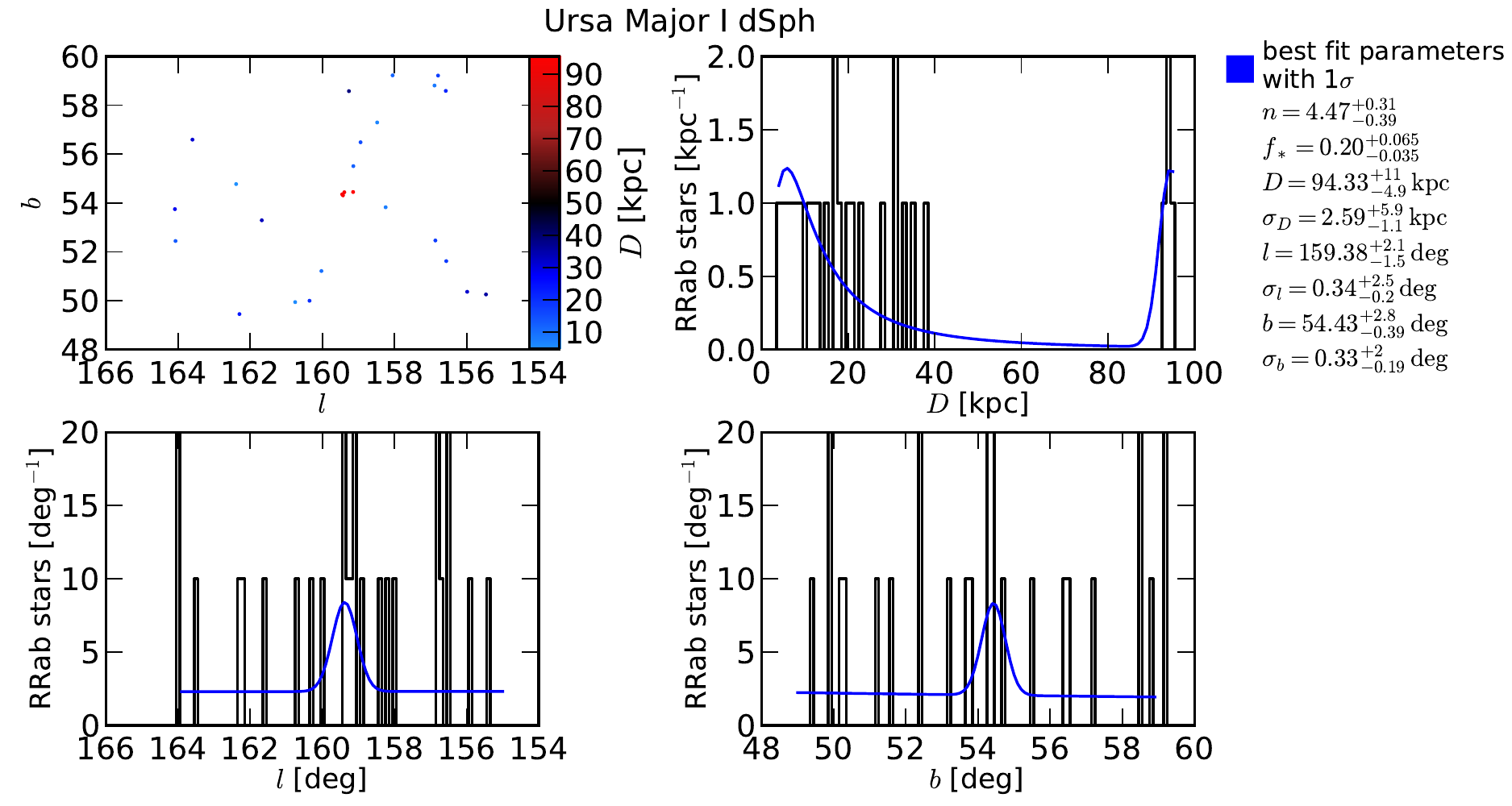}
}
\subfigure[]
    {
\includegraphics[]{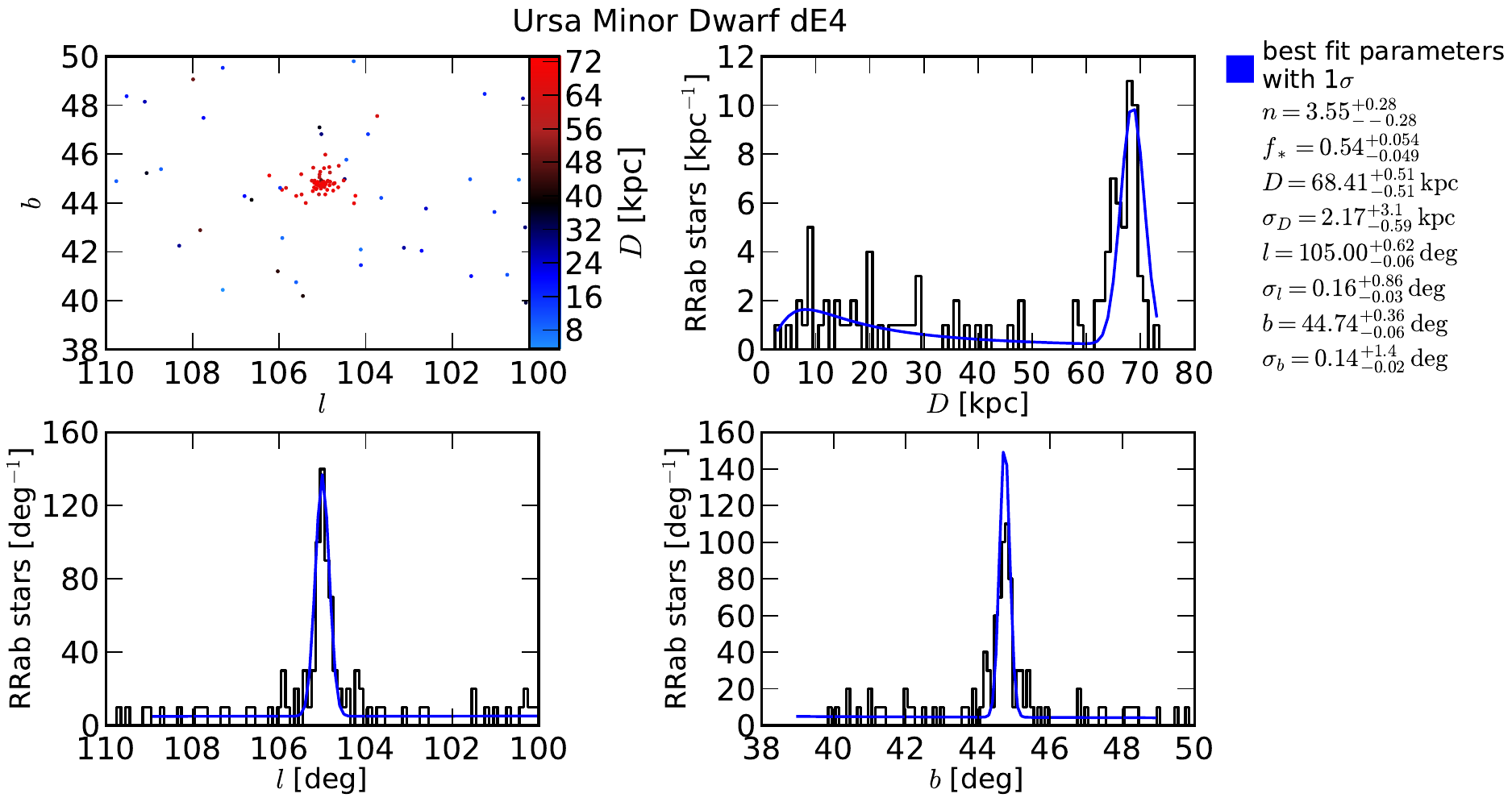}
}
\caption{{
The dwarf spheroidals Ursa Major I dSph (a) and Ursa Minor Dwarf dE4 (b). The first panel shows a map of RRab stars near the dwarf galaxy, 26 for Ursa Major I dSph, and 98 stars in the figure for Ursa Minor Dwarf dE4, respectively. The stars are color-coded according to their heliocentric distance.
The other three panels show the histograms in $l$, $b$, $D$ for the stars from the first panel. Overplotted is the best-fit model from Sec. \ref{sec:DensityFitting} with the parameters given on the right.\newline
In the case of Ursa Major I dSph, the best-fit model in the lower left panel doesn't seem to match the histogram quite well. However, this is only an effect due to the marginalization in the histogram, as it shows all sources independent of their distance, while the Gaussian is centered on the fitted $(l,b,D)$.
}
\label{fig:Ursa_Minor_Dwarf_dE4_fits_Ursa_Major_I_dSph_fits}}
\end{center}  
\end{figure*}

\begin{figure*}
\begin{center}  
\subfigure[]
    {
\includegraphics[]{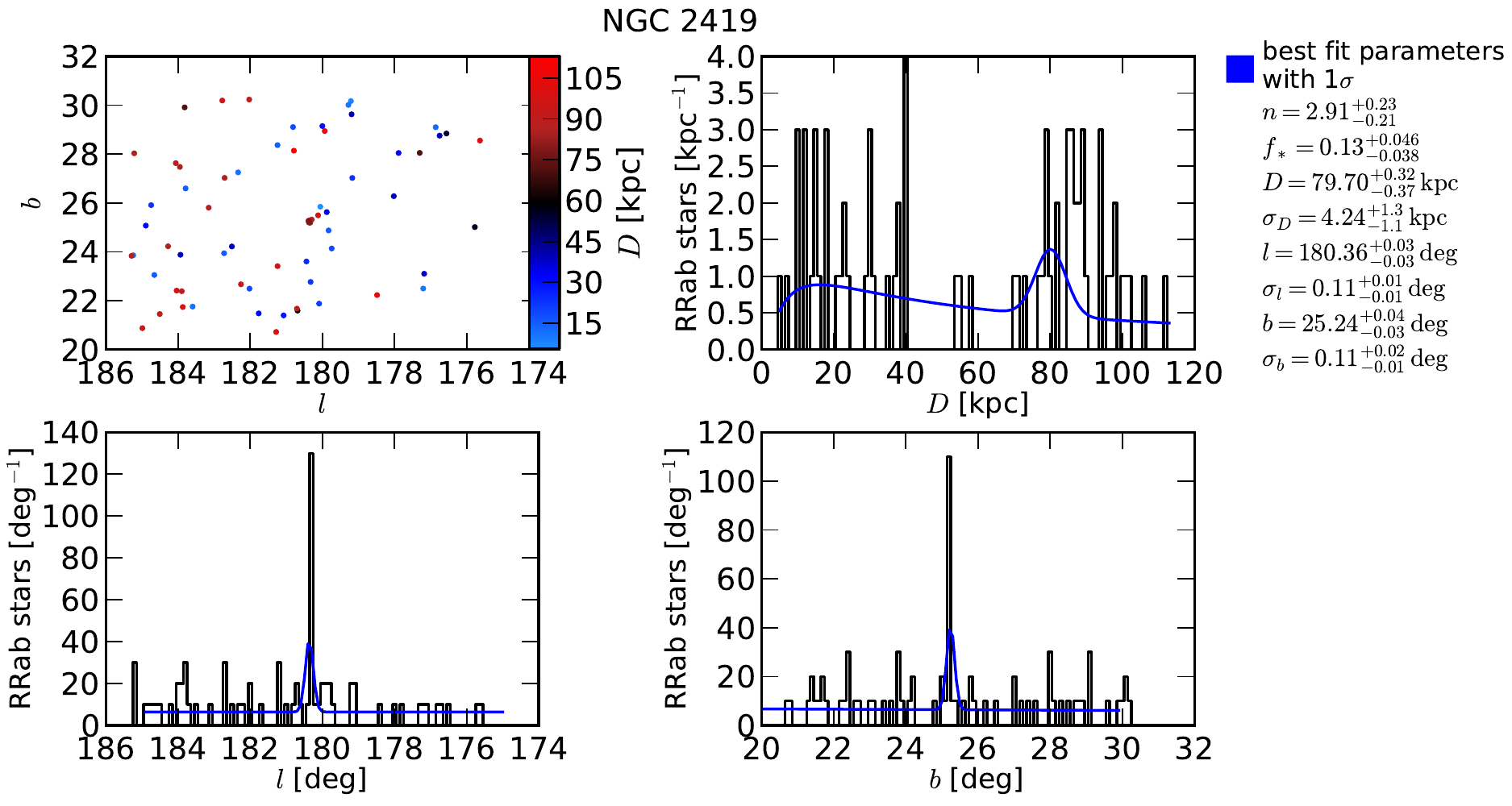}
}
\subfigure[]
    {
\includegraphics[]{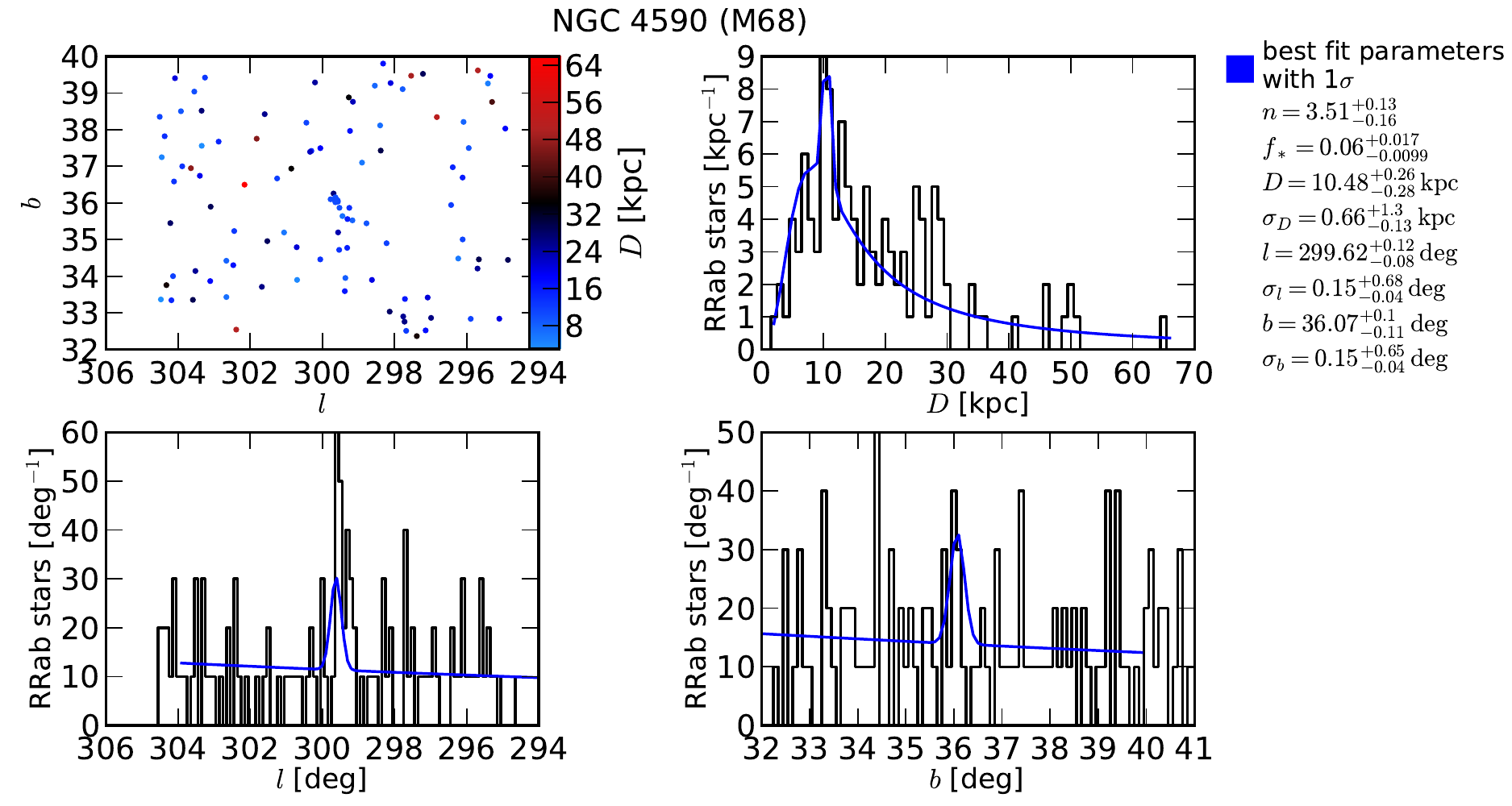}
}
\caption{{
The globular clusters NGC 2419 (a) and NGC 4590 (M68) (b). The first panel shows a map of RRab stars near the globular cluster, 74 stars in the figure for NGC 2419, and 119 for NGC 4590, respectively. 
The other three panels show the marginalized histograms in $l$, $b$, $D$ for the stars from the first panel. Overplotted is the best-fit model from Sec. \ref{sec:DensityFitting} with the parameters given on the right.\newline
For NGC 4590, the best-fit model in the lower left panel doesn't seem to match the histogram quite well. However, this is only an effect due to the marginalization in the histogram, as it shows all sources independent of their distance, while the Gaussian is centered on the fitted $(l,b,D)$.
}
\label{fig:NGC_2419_NGC_4590__M_68_fits}}
\end{center}  
\end{figure*}

\begin{figure*}
\begin{center}  
\subfigure[]
    {
\includegraphics[]{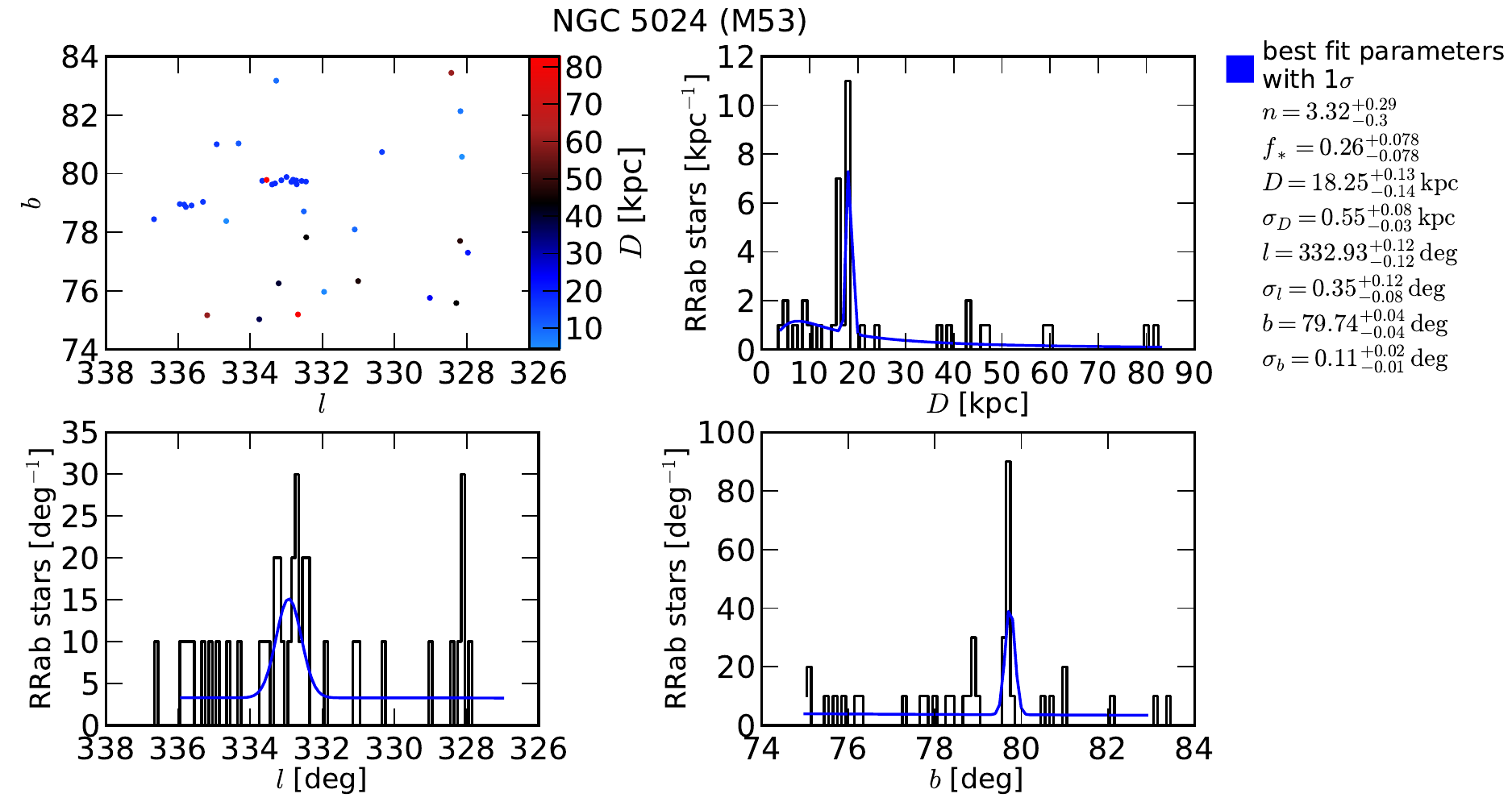}
}
\subfigure[]
    {
\includegraphics[]{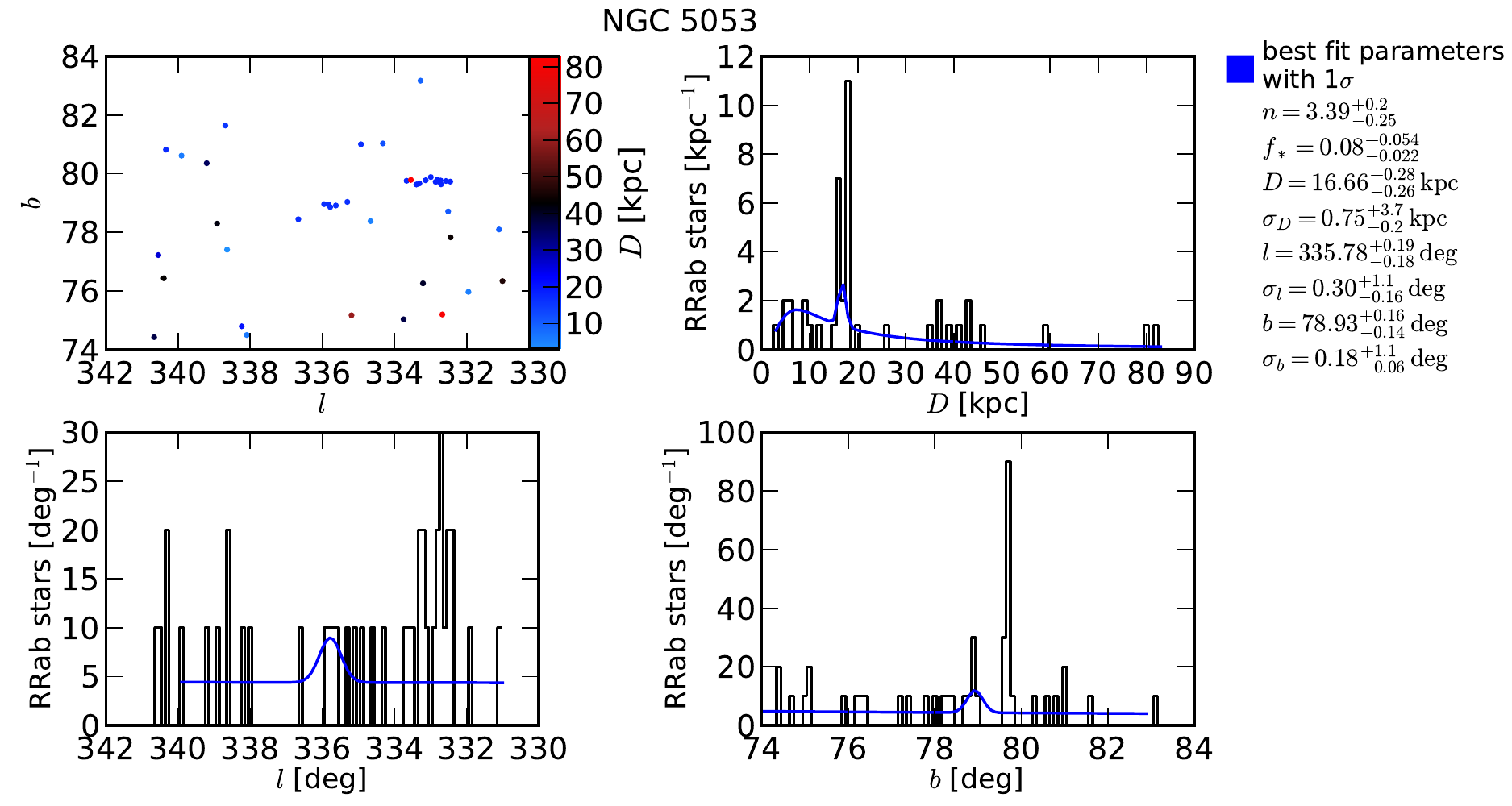}
}
\caption{{
The globular clusters NGC 5024 (M53) (a) and NGC 5053 (b). The first panel shows a map of RRab stars near the globular cluster, 40 stars in the figure for NGC 5024, and 75 stars for NGC 5272, respectively. The stars are color-coded according to their heliocentric distance. In the Cartesian projection, both NGC 5024 and NGC 5053 appear to be elongated in $l$ direction because of its high latitude. 
The other three panels show the histograms in $l$, $b$, $D$ for the stars from the first panel. Overplotted is the best-fit model from Sec. \ref{sec:DensityFitting} with the parameters given on the right.}
\label{fig:NGC_5024__M53_fits_NGC_5053_fits}}
\end{center}  
\end{figure*}

\begin{figure*}
\begin{center}  
\subfigure[]
    {
\includegraphics[]{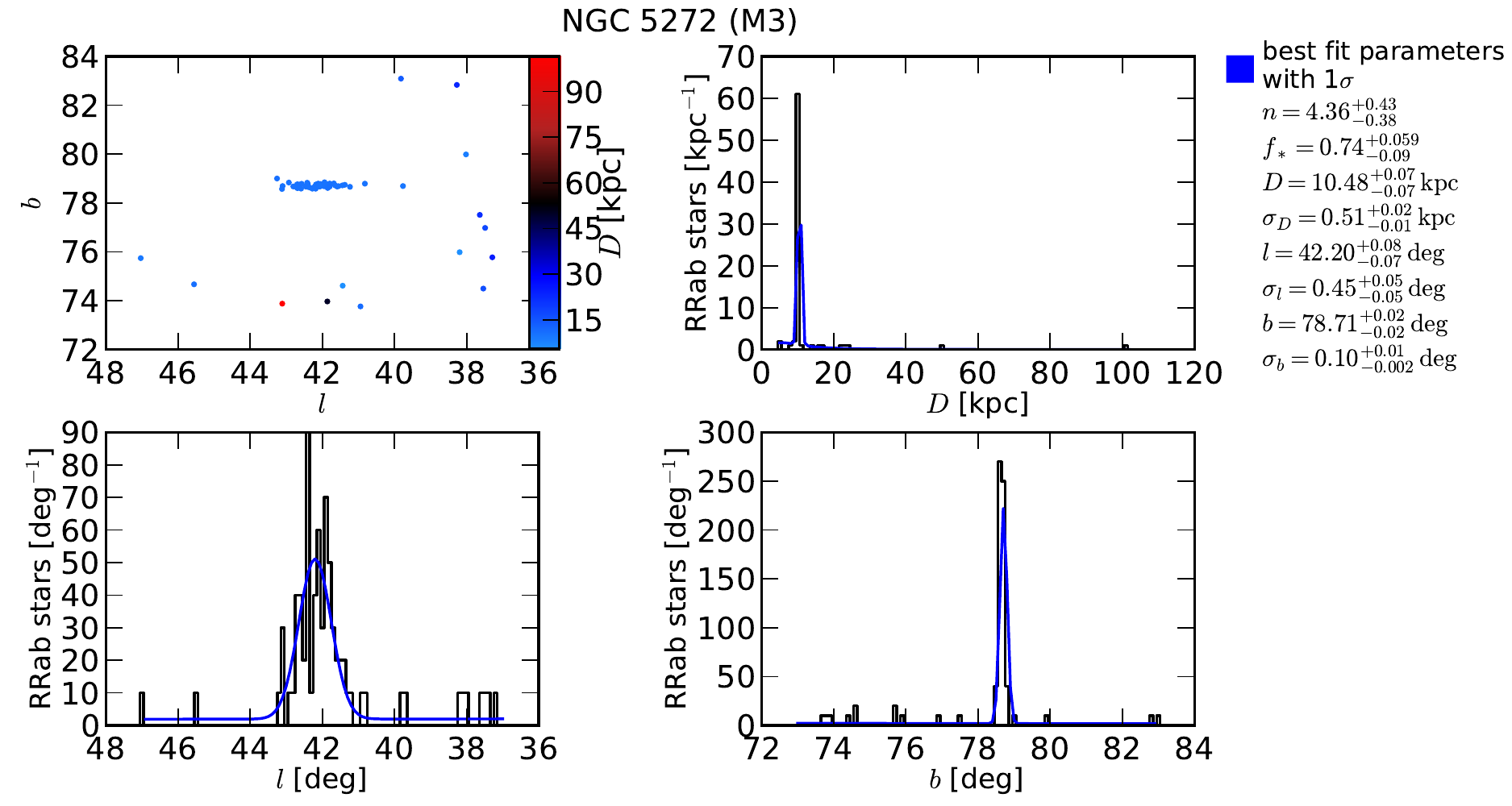}
}
\subfigure[]
    {
\includegraphics[]{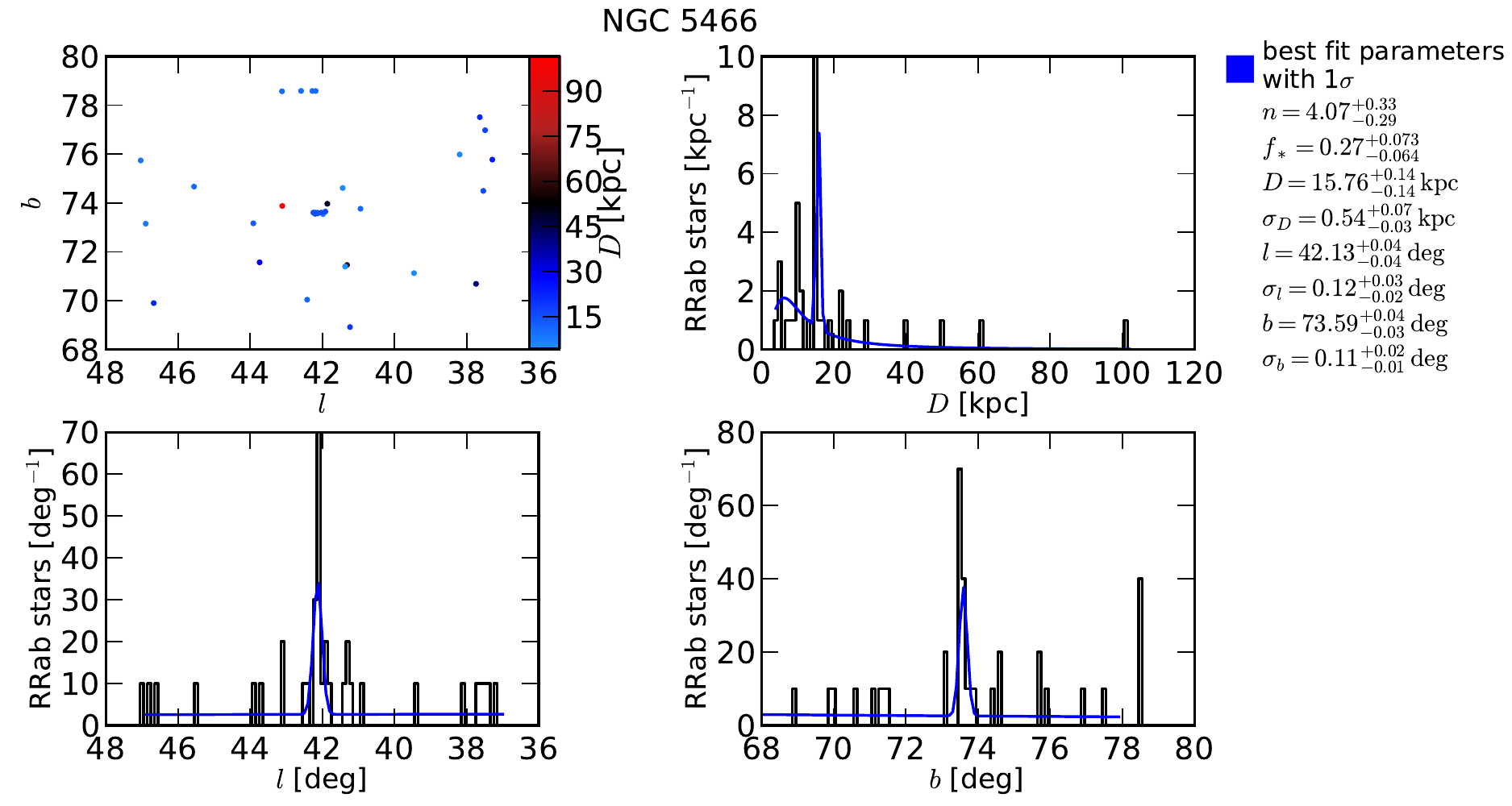}
}
\caption{{The globular clusters NGC 5272 (M3) (a) and NGC 5466 (M5) (b). The first panel shows a map of RRab stars near the globular cluster, 75 in the figure for NGC 5272, and 36 for NGC 5904, respectively. 
In the Cartesian projection, both NGC 5272 and NGC 5466 appear to be elongated in $l$ direction because of its high latitude. 
The other three panels show the histograms in $l$, $b$, $D$ for the stars from the first panel. Overplotted is the best-fit model from Sec. \ref{sec:DensityFitting} with the parameters given on the right.}
\label{fig:NGC_5272_M_3__fits_NGC_5466_fits}}
\end{center}  
\end{figure*}

\begin{figure*}
\begin{center}  
\subfigure[]
    {
\includegraphics[]{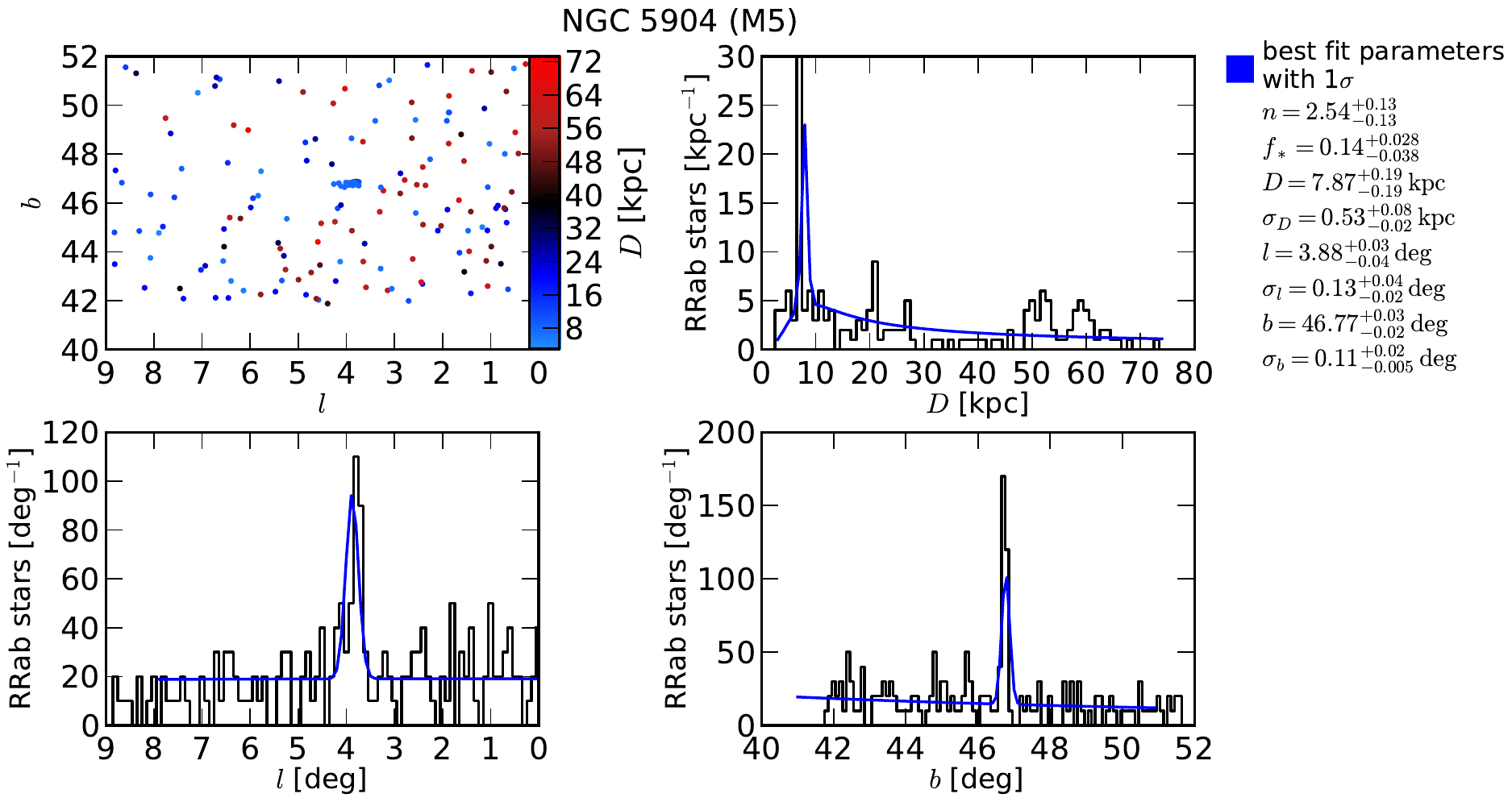}
}
\subfigure[]
    {
\includegraphics[]{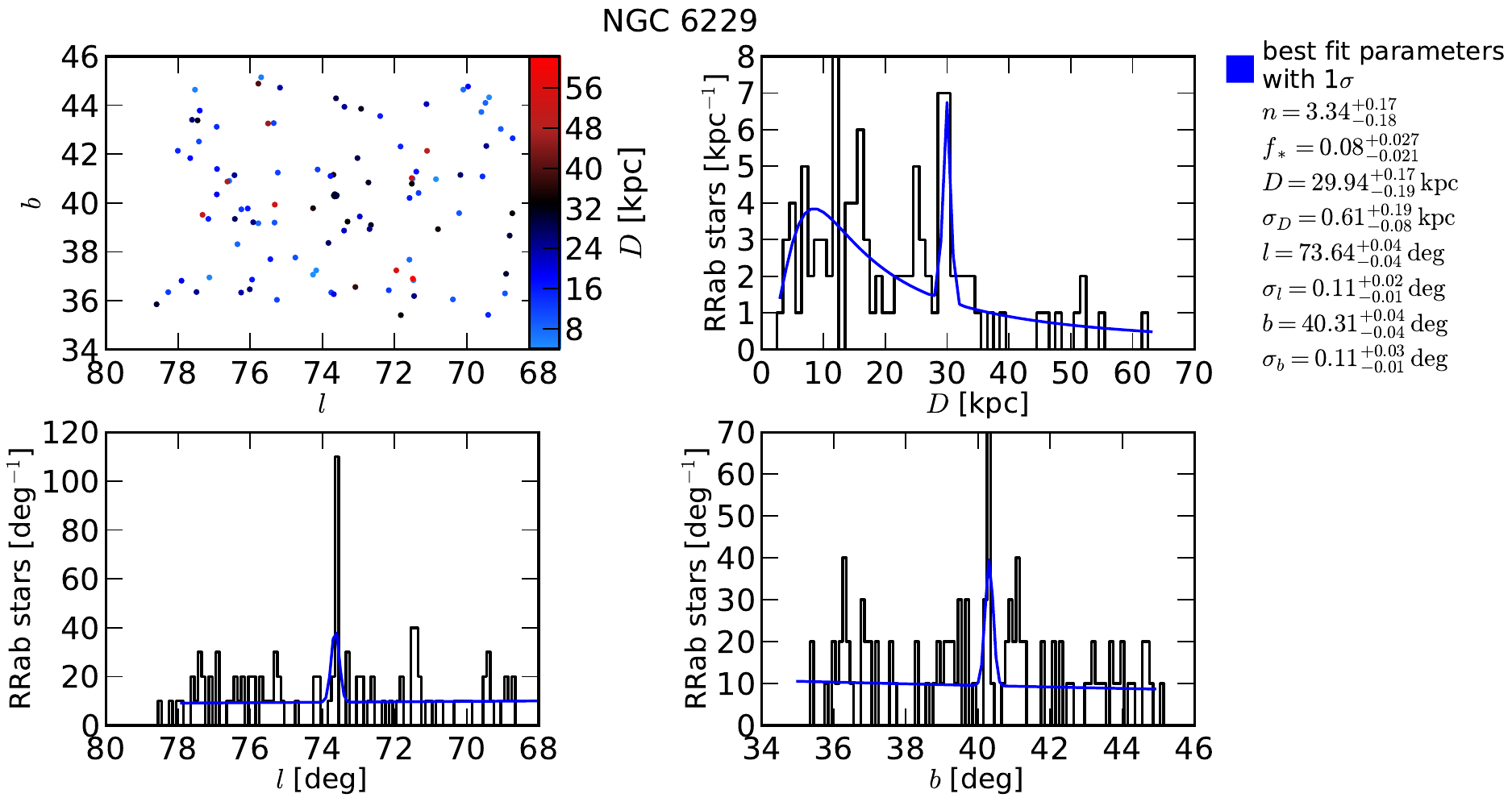}
}
\caption{{
The globular clusters NGC 5904 (M5) (a) and NGC 6229 (b). The first panel shows a map of RRab stars near the globular cluster, 177 in the figure for NGC 5904, and 104 for NGC 6229, respectively. The other three panels show the histograms in $l$, $b$, $D$ for the stars from the first panel. Overplotted is the best-fit model from Sec. \ref{sec:DensityFitting} with the parameters given on the right.\newline
For NGC 5904, the best-fit model in the lower left panel doesn't seem to match the histogram quite well. However, this is only an effect due to the marginalization in the histogram, as it shows all sources independent of their distance, while the Gaussian is centered on the fitted $(l,b,D)$.}
\label{fig:NGC_    _fits_NGC_6229_fits}}
\end{center}  
\end{figure*}

\begin{figure*}
\begin{center}  
\subfigure[]
    {
\includegraphics[]{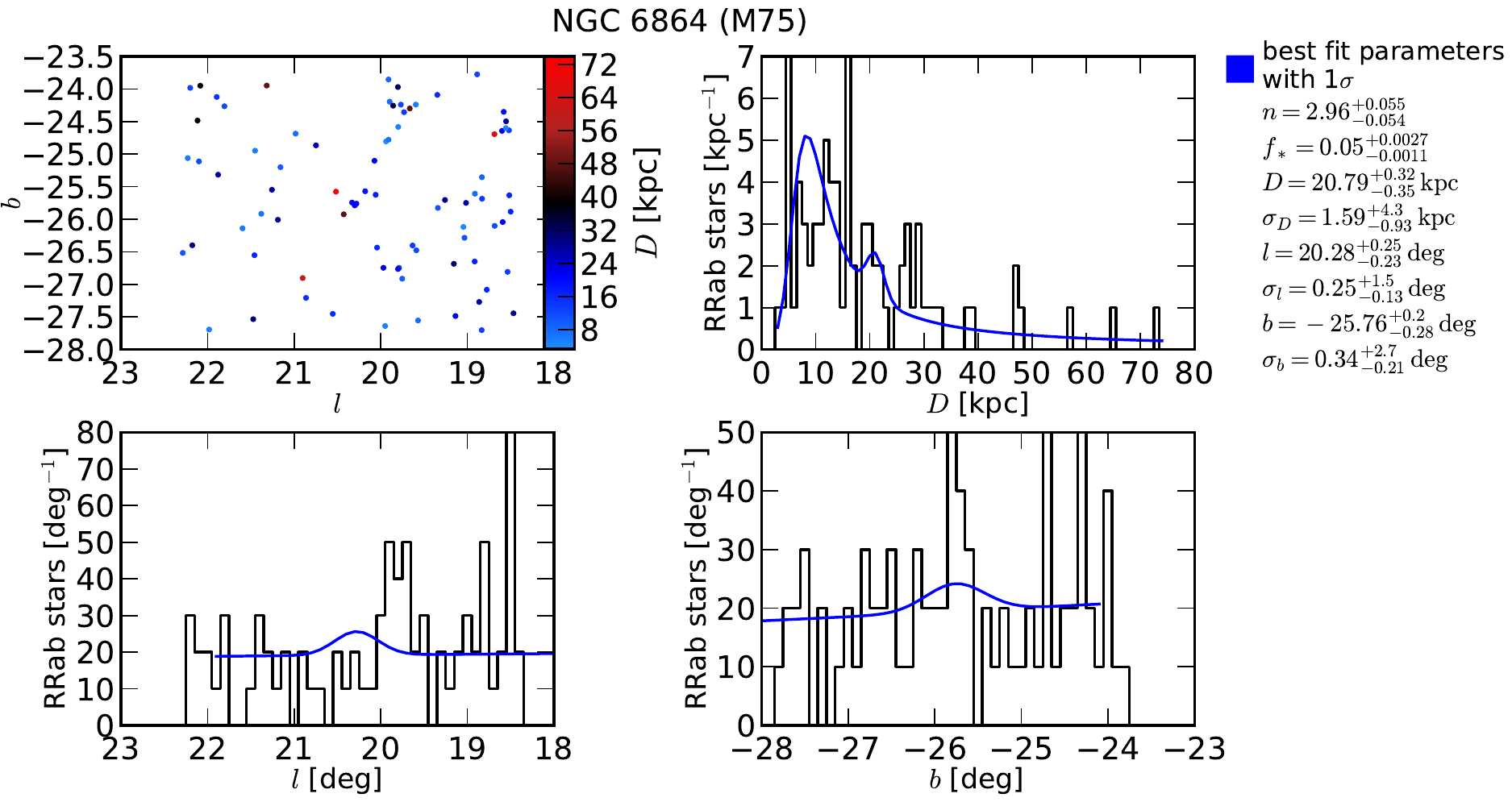}
}
\subfigure[]
    {
\includegraphics[]{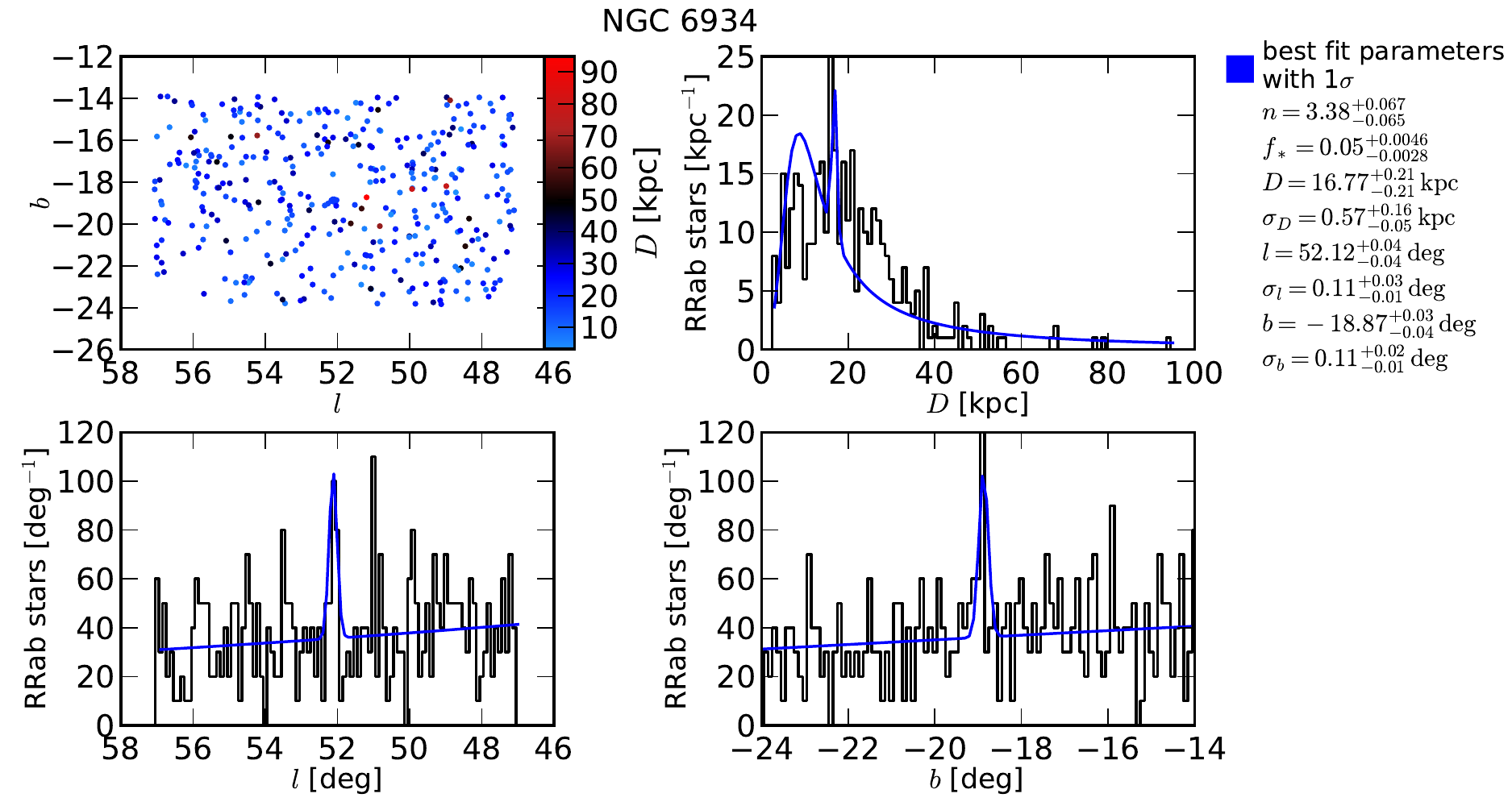}
}
\caption{{
The globular clusters NGC 6864 (M75) (a) and NGC 6934 (b). The first panel shows a map of RRab stars near the globular cluster, 81 in the figure for NGC 6864, and 378 for NGC 6934, respectively. The other three panels show the histograms in $l$, $b$, $D$ for the stars from the first panel. Overplotted is the best-fit model from Sec. \ref{sec:DensityFitting} with the parameters given on the right. For NGC 6864, the best-fit model in the lower left panel doesn't seem to match the histogram quite well. However, this is only an effect due to the marginalization in the histogram, as it shows all sources independent of their distance, while the Gaussian is centered on the fitted $(l,b,D)$.
}
\label{fig:NGC_6864__M_75_fits_NGC_6934_fits}}
\end{center}  
\end{figure*}

\begin{figure*}
\begin{center}  
\subfigure[]
    {
\includegraphics[]{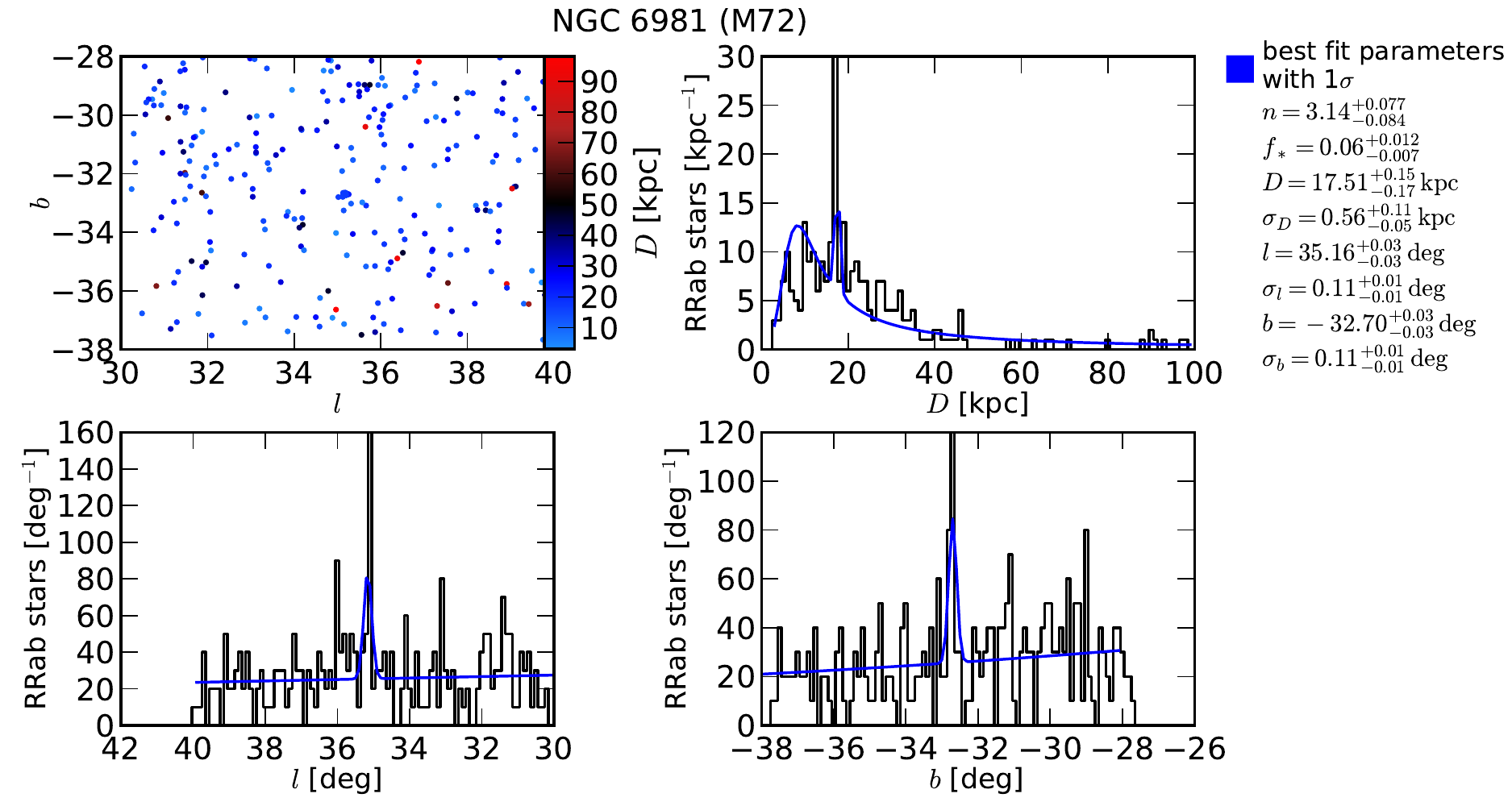}
}
\subfigure[]
    {
\includegraphics[]{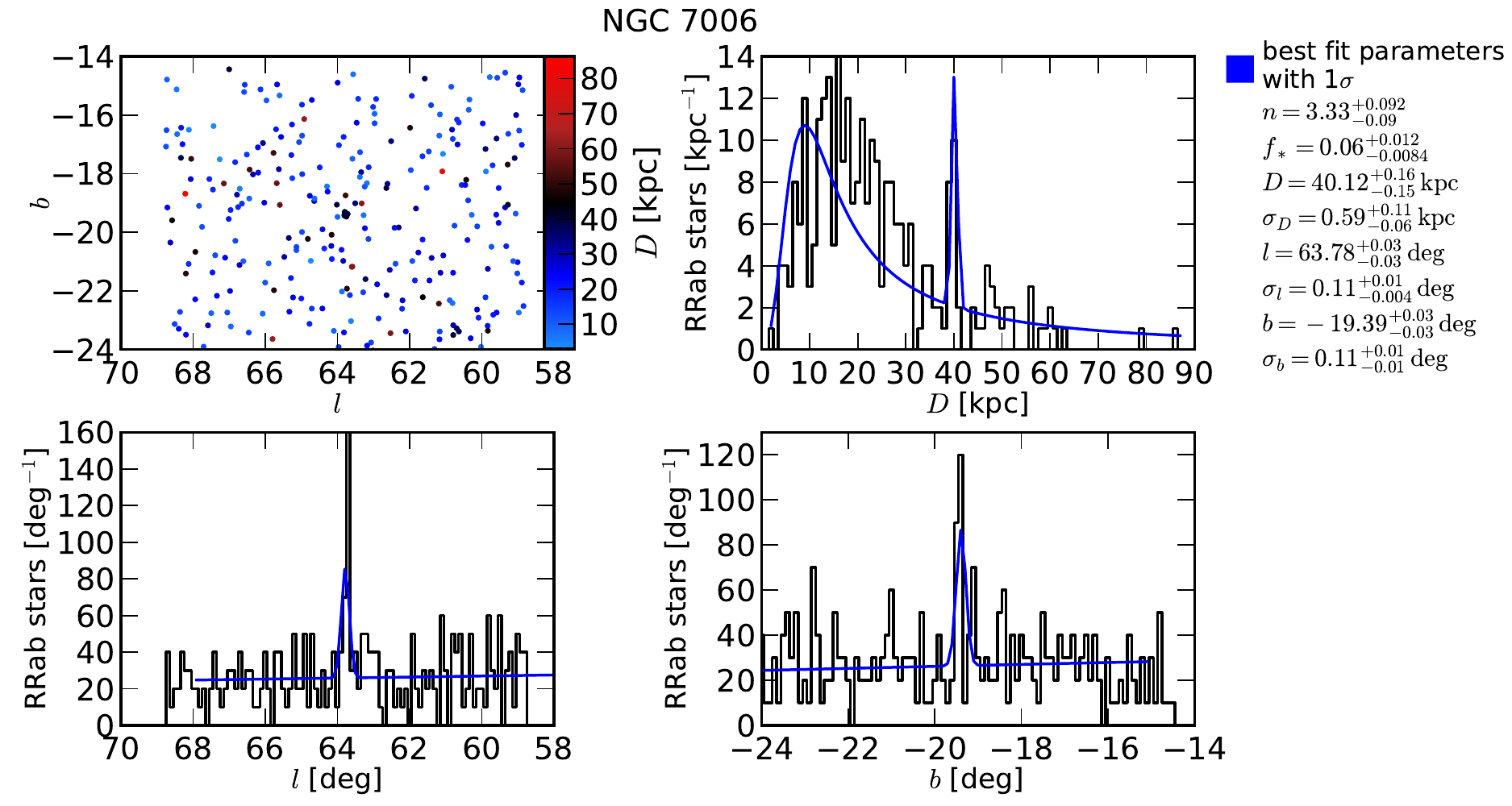}
}
\caption{{
The globular clusters NGC 6981 (M72) (a) and NGC 7006 (b). The first panel shows a map of RRab stars near the globular cluster, 271 in the figure for NGC 6981, and 278 for NGC 7006, respectively. The other three panels show the histograms in $l$, $b$, $D$ for the stars from the first panel. Overplotted is the best-fit model from Sec. \ref{sec:DensityFitting} with the parameters given on the right.}
\label{fig:NGC_6981__M_72_fits_NGC_7006_fits}}
\end{center}  
\end{figure*}

\begin{figure*}
\begin{center}  
\subfigure[]
    {
\includegraphics[]{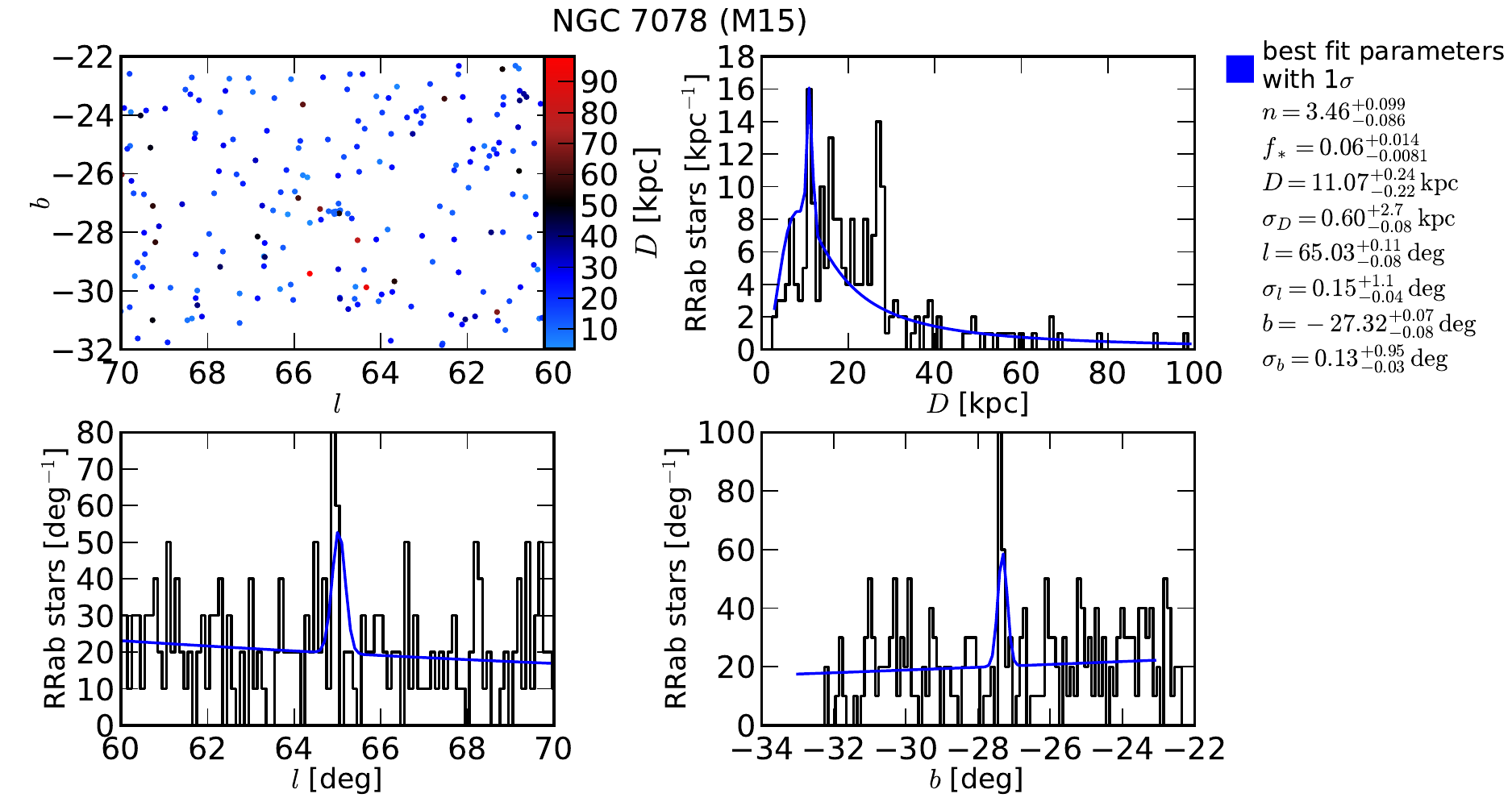}
}
\subfigure[]
    {
\includegraphics[]{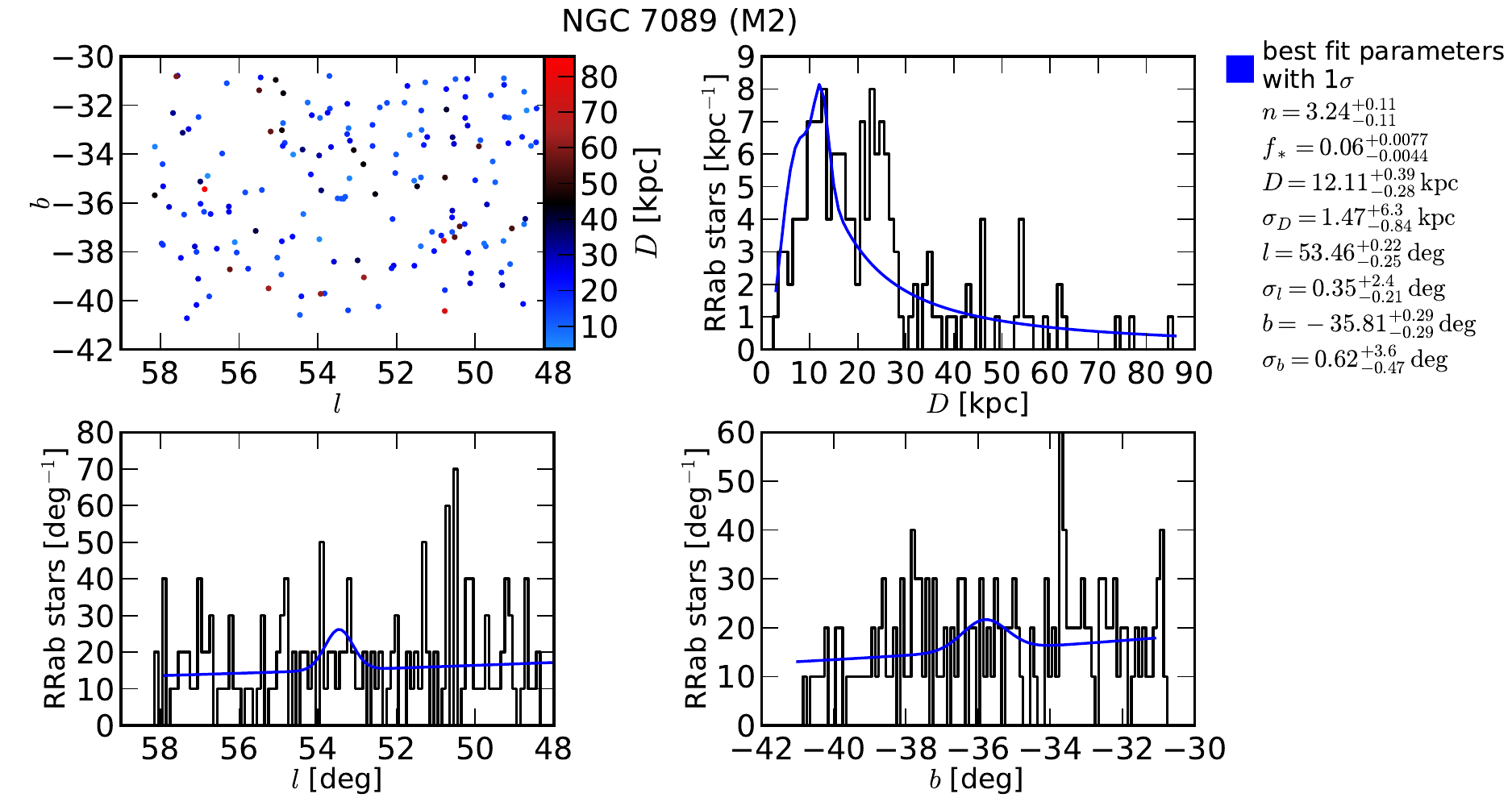}
}
\caption{{
The globular clusters NGC 7078 (M15) (a) and NGC 7089 (M2) (b). The first panel shows a map of RRab stars near the globular cluster, 211 in the figure for NGC 7078, and 163 for NGC 7089, respectively. The other three panels show the histograms in $l$, $b$, $D$ for the stars from the first panel. Overplotted is the best-fit model from Sec. \ref{sec:DensityFitting} with the parameters given on the right.\newline
For NGC 7089, the best-fit model in the lower left panel doesn't seem to match the histogram quite well. However, this is only an effect due to the marginalization in the histogram, as it shows all sources independent of their distance, while the Gaussian is centered on the fitted $(l,b,D)$.
}
\label{fig:NGC_7078__M_15_fits_NGC_7089__M_2_fits}}
\end{center}  
\end{figure*}

\begin{figure*}
\begin{center}  
\subfigure[]
    {
\includegraphics[]{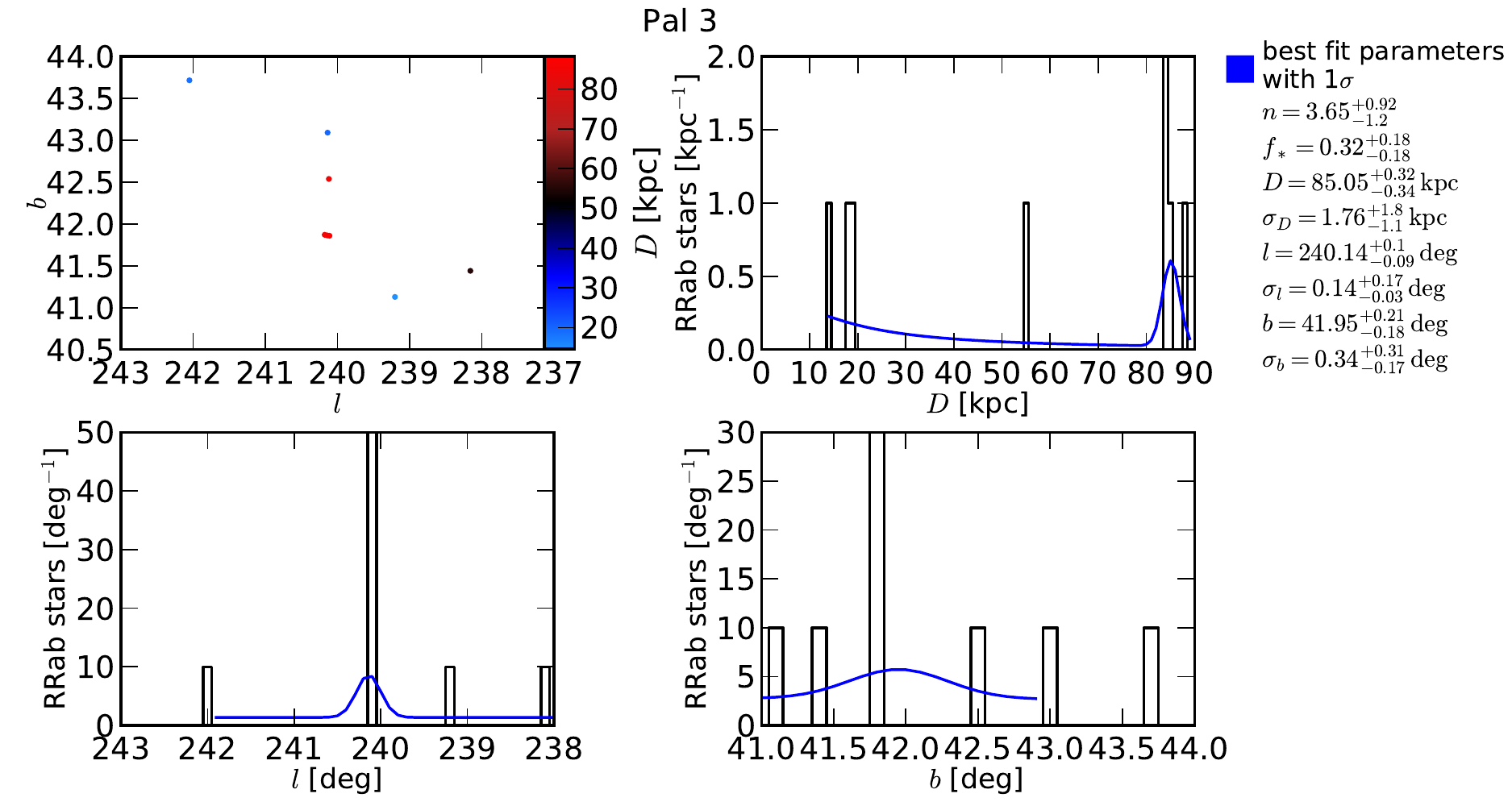}
}
\subfigure[]
    {
\includegraphics[]{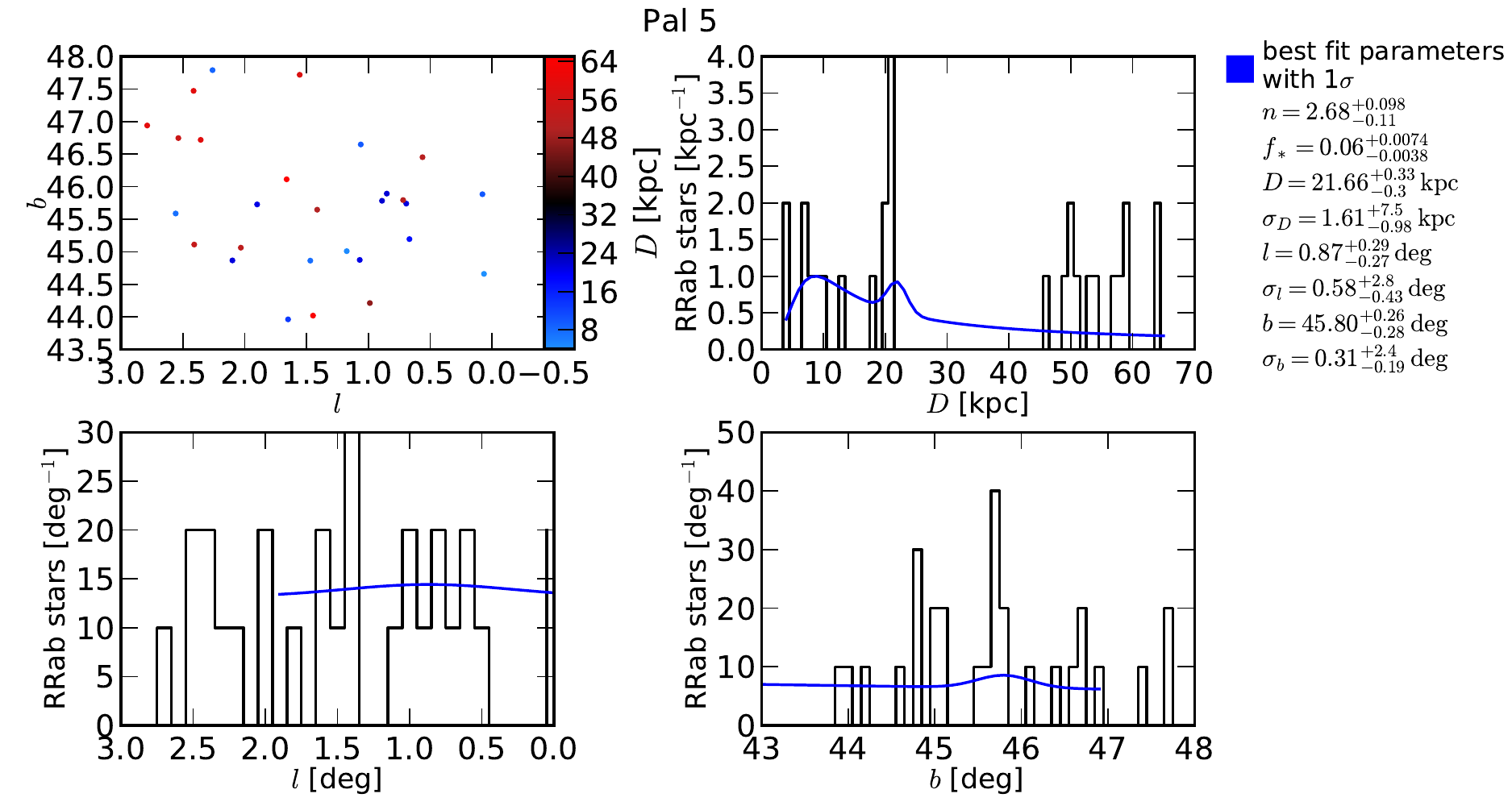}
}
\caption{{
The globular clusters Pal 3 (a) and Pal 5 (b). The first panel shows a map of RRab stars near the globular cluster, 8 in the figure for Pal 3, and 28 for Pal 5, respectively. The other three panels show the histograms in $l$, $b$, $D$ for the stars from the first panel. Overplotted is the best-fit model from Sec. \ref{sec:DensityFitting} with the parameters given on the right.}
\label{fig:Pal_3_fits_Pal_5_fits}}
\end{center}  
\end{figure*}

To select those overdensities for which fitting their density might be feasible, we started with a list of dwarf galaxies within 3 Mpc by \cite{McConnachie2012}, its update from 2014\footnote{\url{https://www.astrosci.ca/users/alan/Nearby_Dwarfs_Database.html}}
and a list of currently known globular clusters from the 2010 update of \cite{Harris1996_2010}. We excluded the ones outside the PS1 3$\pi$ footprint and made plots of regions on the sky around for each of the remaining ones.
If an overdensity is apparent in the PS1 RRab sample from visual inspection, we consider it for further analysis. We ended up with a list of 11 dwarf galaxies and 29 globular clusters within PS1 3$\pi$, as given in Tables \ref{tab:dwarfs_planned} and \ref{tab:gc_planned}.
Plots of a subset of these overdensities are found in the Appendix, Fig. \ref{fig:Sextans_dSph_fits_Draco_dSph_fits} to \ref{fig:Pal_3_fits_Pal_5_fits}.
For plotting those overdensities, we chose a Cartesian projection of $(l,b)$ to more easily compare to their marginalized distributions in $l$ and $b$ (histograms).
With this choice, high-latitude overdensities such as NGC 5024, NGC 5053 and NGC 5272 appear to be elongated in the $l$ direction.

\begin{figure*}
\begin{center}  
\includegraphics[trim=0cm 0.3cm 0cm 0.3cm, clip=true]{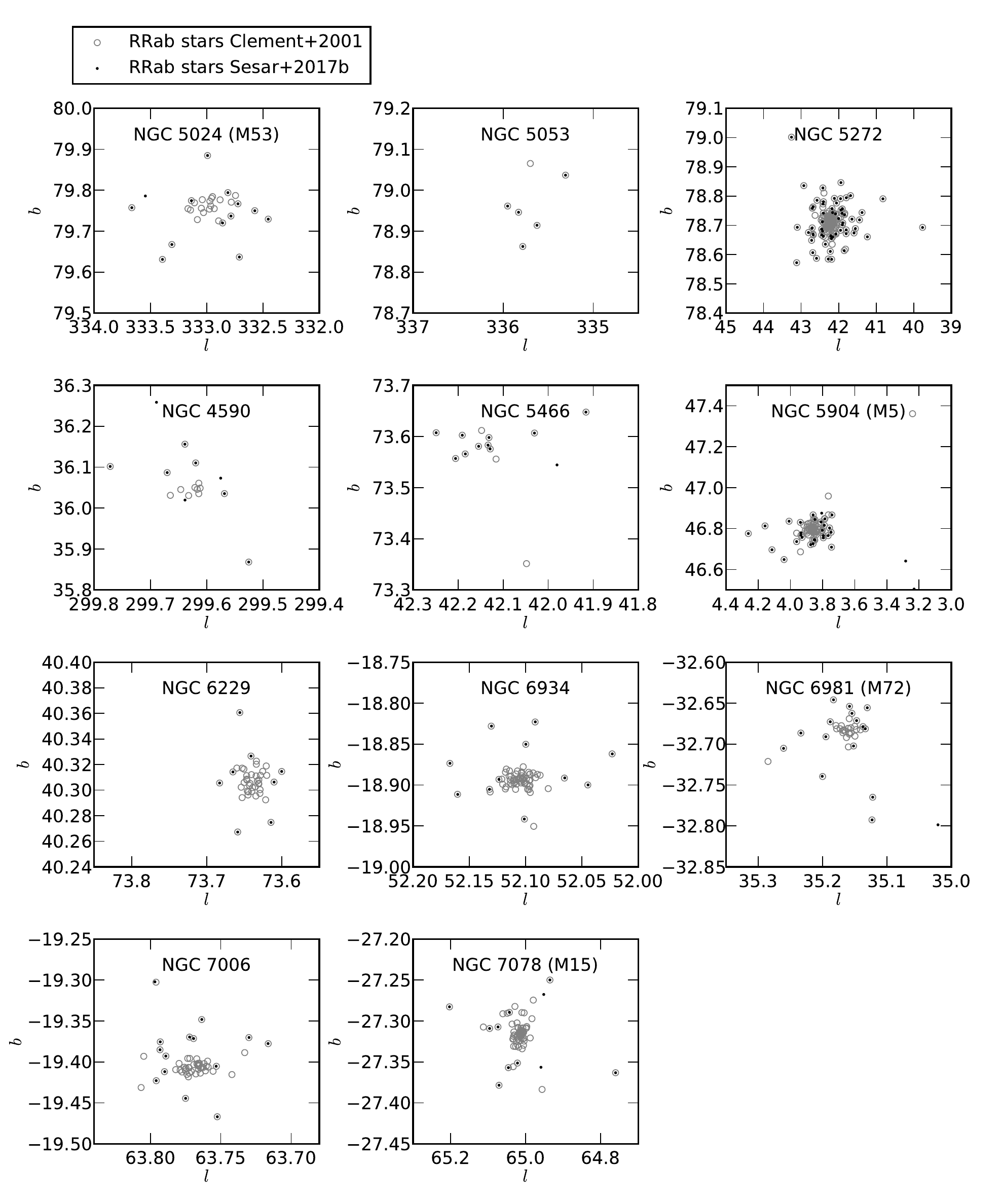}
\caption{{Comparison of the RRab stars available in our PS1 RRab catalog \citep{Sesar2017} as used for this work, to the RRab stars in the Catalogue of Variable Stars in Galactic Globular Clusters \citep{Clement2001}. For the 11 globular clusters available in both catalogs, we find that our catalog misses most RRab in the central regions of the globular clusters. These regions are too compact, and thus most stars did not pass the quality criteria for the PS1 RRab catalog.
}
\label{fig:gc_compare}}
\end{center}  
\end{figure*}

We expected that the central regions of the globular clusters might be not well represented due to crowding and blending effects that made many RRab stars fail to meet the criteria for inclusion in the catalogue of PS1 RRab stars \citep{Sesar2017}. To test that, we compared our sample with the Catalogue of Variable Stars in Globular Clusters from \cite{Clement2001}.
There are 11 globular clusters detectable with RRab stars both in the catalog of PS1 RRab stars and the Catalogue of Variable Stars in Globular Clusters.
As shown in Fig. \ref{fig:gc_compare}, we find that for each of the 11 globular clusters available in both catalogs, our catalog misses most RRab in the globular cluster's central regions. The dSph galaxies are sufficiently diffuse that this is not a significant issue.

\section{Density Fitting}
\label{sec:DensityFitting}

In this section we lay out a forward-modeling approach to describe the spatial distribution of RRab stars within and near overdensities such as dwarf galaxies and globular clusters. We are using a local halo model describing the background of field stars from the halo as in \cite{Hernitschek2018} and a smooth spheroidal distribution describing the overdensity itself.
Regarding the dwarf galaxies, this approach is similar to \cite{Martin2008}, who derived structural parameters of Milky Way satellites from SDSS data.

We presume that the local stellar halo distribution can be sensibly approximated by a spheroidal distribution with a parameterized radial profile. 
(See also \cite{Hernitschek2018}.) The overdensities can be described by multivariate Gaussians in $(l,b,D)$.
Following a number of previous studies \citep[e.g.][]{Sesar2013b, Xue2015, Cohen2015, Iorio2017, Hernitschek2018}, we presume that the overall radial density profile
of the halo outside of overdensities -- the distance-dependent ``background" of field stars in a given direction -- can be described by a power law profile.
We attempt to carry out the fit on small $5 \times 5 \, \deg^2$ patches on the sky around the assumed position prior $(l_{\mathrm{prior}}, b_{\mathrm{prior}})$. We thus
neglect the selection effects and halo flattening described in \cite{Hernitschek2018} and instead fit for a local stellar halo distribution as a function of a power-law index $n$ only.

We apply a forward-modeling process to fit stellar-density models to the data
by generating the expected observed distribution of stars in the RRab sample, based on our model for the halo background and overdensity. The predicted distribution is then automatically compared to the observed star counts to calculate the likelihood.

\subsection{Stellar Density Model}
\label{sub:StellarDensityModel}

For the overall radial density profile of the halo stars, we assume a power-law model
\begin{equation}
\rho_{\mathrm{halo}}(X,Y,Z)=\rho_{\odot \mathrm{RRL}}\left(R_{\odot}/r_q \right)^{n}
\label{eq:halomodel}
\end{equation}
where $\rho_{\odot \mathrm{RRL}}$ gives the number density of RR Lyrae at the position of the Sun, $n$ is the power-law index where larger values of $n$ indicate a steeper profile, $R_{\odot}$ the distance of the Sun from the Galactic center, and $ r_q = \sqrt{(X^2+Y^2+(Z/q)^2)}$ is the flattening-corrected radius.

In \cite{Hernitschek2018}, we derived the halo flattening on concentric ellipsoids.
However, as we deal here with very small patches on the sky, and don't want to derive a reliable fit for the halo but want to remove any background distribution, we decided to set $q=0.75$, leaving $n$ the only free parameter for the halo component. Also, we are not fitting for $\rho_{\odot \mathrm{RRL}}$.

The spatial density of an overdensity, i.e. a dwarf galaxy or globular cluster, is described by a multivariate Gaussian $\mathcal{G}$:
\begin{align}
\rho_*(l,b,D)  =& \mathcal{G}(l,b,D) \label{eq:3dgauss}\\
=& \frac{  \exp \left(-0.5 \left[ \left(\frac{l-\bar{l}}{\sigma_l}\right)^2 + \left(\frac{b-\bar{b}}{\sigma_b}\right)^2 + \left(\frac{D-\bar{D}}{\sigma_D}\right)^2 \right] \right) }{\sigma_l \sigma_b \sigma_D \left(2\pi\right)^{(3/2)} }. \nonumber
\end{align}

The data set $\mathcal{D}$ is given as $\mathcal{D} = (D, \delta D, l,b)$. The parameters are $\boldsymbol{\theta}=(\bar{l},\bar{b},\bar{D}, \sigma_l, \sigma_b, \sigma_D, f_*,n)$, composed of the spatial position of the overdensity $(\bar{l},\bar{b},\bar{D})$, its extent $(\sigma_l,\sigma_b,\sigma_D)$, the fraction of the stars $f_*$ being in the overdensity, and the power-law index $n$ of the halo model. The heliocentric distance distribution of stars is then characterized by 
\begin{align}
p_{\mathrm{RRL}}(\mathcal{D}|\boldsymbol{\theta}) &=& p_{\mathrm{halo}}(\mathcal{D}|\boldsymbol{\theta})+p_*(\mathcal{D}|\boldsymbol{\theta})\label{eq:stream_halo_model}\\
&=&(1-f_*)\times \hat{\rho}_{\mathrm{halo}}(l,b,D,n) \nonumber \\
& &+ f_* \times \hat{\rho}_*(l,b,D,\bar{l},\bar{b},\bar{D},\sigma_{l},\sigma_{b},\sigma_D),\nonumber
\end{align}
where 
\begin{equation}
\hat{\rho}_{\mathrm{halo}}(l,b,D,q,n)\equiv \frac{\rho_{\mathrm{halo}}(l,b,D,q,n)}{\int\rho_{\mathrm{halo}}(l,b,D,q,n)\mathrm{d}D},
\end{equation}
with an analogous definition of $\hat{\rho}_*$.

Although we were aware of a distance uncertainty of 3\% for RRab stars from \cite{Sesar2017}, we have not included it in our calculations yet. The reason for this is that the distance uncertainty can easily be taken into account later on, as the distance uncertainty $\epsilon_D \sim 0.03D$ adds in quadrature to the (true) width in distance:
\begin{equation}
\sigma_D = \sqrt{(\sigma'_D)^2 + \epsilon_D^2} \; .
\label{eq:sigma_D_with_uncertainty}
\end{equation}
We will deal with the distance precision vs. true line of sight depth later when we evaluate our fitting results.

\subsection{Constraining Model Parameters}
\label{sec:ConstrainingModelParameters}

With the model $\rho_{\mathrm{RRL}} (\mathcal{D} \vert \boldsymbol{\theta})$ at hand, we can directly calculate the likelihood of the data $\mathcal{D}$ given the model $\rho_{\mathrm{RRL}}$ and the fitting parameters $\boldsymbol{\theta}$.

The normalized un-marginalized logarithmic likelihood for the $i$-th star with the observables $\mathcal{D}_i$ is then
\begin{equation}
\ln p(\mathcal{D}_i \vert \boldsymbol{\theta}) = \frac{\rho_{\mathrm{RRL}}(\mathcal{D}_i \vert \boldsymbol{\theta}) \vert \mathbf{J} \vert }{\int \int \int  \rho_{\mathrm{RRL}}(l,b,D \vert \boldsymbol{\theta})  \vert \mathbf{J}\vert   \mathrm{d}l \mathrm{d}b \mathrm{d}D  }      
\label{eq:likelihood}
\end{equation}
where the normalization integral is over the observed volume.
The Jacobian term $\vert \mathbf{J}\vert=D^2 \cos b$ reflects the transformation from $(X,Y,Z)$ to $(l,b,D)$ coordinates.

We evaluate the logarithmic posterior probability of the parameters $\boldsymbol{\theta}$ of the halo model, given the full data $\mathcal{D}$ and a prior $p (\boldsymbol{\theta})$, $\ln p(\boldsymbol{\theta} \vert \mathcal{D}) = \ln p (\mathcal{D} \vert \boldsymbol{\theta}) + \ln p (\boldsymbol{\theta})$ with 
\begin{equation}
\ln p (\mathcal{D} \vert \boldsymbol{\theta}) = \sum_i \ln p(\mathcal{D}_i \vert \boldsymbol{\theta})
\label{eq:posterior}
\end{equation}
being the marginal log likelihood for the full data set.

To determine the best-fit parameters and their uncertainties, we sample the posterior probability over the parameter space with Goodman \& Weare's Affine
Invariant Markov Chain Monte Carlo \citep{Goodman2010}, making use of the Python module \texttt{emcee} \citep{Foreman2013}. 

The final best-fit values of the model parameters have been estimated using the median of the posterior distributions; the uncertainties have been estimated using the 15.87th and 84.13th percentiles. For a parameter whose \textit{pdf} can be well-described by a Gaussian distribution, the difference between the 15.87th and 84.13th percentile is equal to 1$\sigma$.

Our model has 8 free parameters, $\boldsymbol{\theta}=(\bar{l},\bar{b},\bar{D}, \sigma_l, \sigma_b, \sigma_D, f_*,n)$, and each star has three-dimensional coordinates $(l,b,D)$. As we attempt to carry out the fit on small $5 \times 5 \, \deg^2$ patches on the sky around the assumed position prior $(l_{\mathrm{prior}}, b_{\mathrm{prior}})$, we always have a sufficient number of overdensity and background RRab stars for the fit as shown in Fig. \ref{fig:Sextans_dSph_fits_Draco_dSph_fits} to \ref{fig:Pal_3_fits_Pal_5_fits}. The overdensity itself is fit by a three-dimensional Gaussian describing the spatial position of the overdensity $(\bar{l},\bar{b},\bar{D})$ and its extent $(\sigma_l,\sigma_b,\sigma_D)$. Thus a minimum number of two RRab stars within the overdensity, in addition to the background, is needed for a successful fit, albeit with large uncertainties when the number of RRab is small. All overdensities for which we later claim fitted positions and spatial extents meet these criteria, i.e. the minimum number of RRab stars is 3.

\subsubsection{Model Priors}

We now lay out the ``pertinent range'', across which the priors for the model parameters $\boldsymbol{\theta}=(\bar{l},\bar{b},\bar{D}, \sigma_l, \sigma_b, \sigma_D, f_*,n)$ are given. 
Based on Table \ref{tab:dwarfs_planned} and \ref{tab:gc_planned}, we can set priors on $\bar{l}$, $\bar{b}$, $\bar{D}$.
In general, we allow a rather wide prior, i.e. allow the on-sky positions to vary around the object's center by $\pm 5\arcdeg$, and the heliocentric mean distance $\bar{D}$ to vary by $\pm 20 \, \mathrm{kpc}$.
In cases with a second overdensity nearby (i.e. NGC 5024, NGC 5053), we set more rigid priors on $\bar{l}$, $\bar{b}$, $\bar{D}$.
As the halo profile tends to vary a lot on such small patches on the sky as are evaluated here, we allow the power-law index $n$ to vary between 1.7 and 5.

Our complete prior function is then:

\begin{align}
\ln p(\boldsymbol{\theta}) =& \mathrm{Uniform}( 0.05 \leq f_* < 1) \nonumber \\&+ \mathrm{Uniform}( 1.7  \leq n < 5.0) \nonumber \\&+ 
\mathrm{Uniform}(\log|\bar{l}-l_{\mathrm{tab}}|<\log\texttt{offset\_deg}) \nonumber \\ &+
\mathrm{Uniform}(\log|\bar{b}-b_{\mathrm{tab}}|<\log\texttt{offset\_deg}) \nonumber \\&+
\mathrm{Uniform}(\log|\bar{D}-D_{\mathrm{tab}}|<\log\texttt{offset\_kpc}) \nonumber \\&+
\mathrm{Uniform}(\log(0.1) < \log(\sigma_l) <\log(10)) \nonumber \\&+
\mathrm{Uniform}(\log(0.1) < \log(\sigma_b) <\log(10)) \nonumber \\&+
\mathrm{Uniform}(\log(0.1) < \log(\sigma_D) <\log(20)), \label{eqn:prior}
\end{align}    
where the index "tab" denotes the values from Tab. \ref{tab:dwarfs_planned} and \ref{tab:gc_planned}, respectively, and $\texttt{offset\_deg} = 5\arcdeg$, $\texttt{offset\_kpc} = 20 \, \mathrm{kpc}$ for all overdensities except for NGC 5024, NGC 5053. For NGC 5024 and NGC 5053, we set the offset to $\texttt{offset\_deg} = 1\arcdeg$.

\subsection{Fitting Tests on Mock Data}
\label{sec:TestsOnMockData}

In order to test the methodology for fitting the density as discussed in Section \ref{sec:DensityFitting}, we created mock data samples of RRab stars in the Galactic halo,
superimposed Gaussian mock overdensities with typical distance, extent and star count to mimic dwarf galaxies and globular clusters, and finally added noise in distance to mimic the distance uncertainty of 3\%. We then applied our fitting method.

\begin{figure*}
\begin{center}  
\subfigure[]
    {
\includegraphics[]{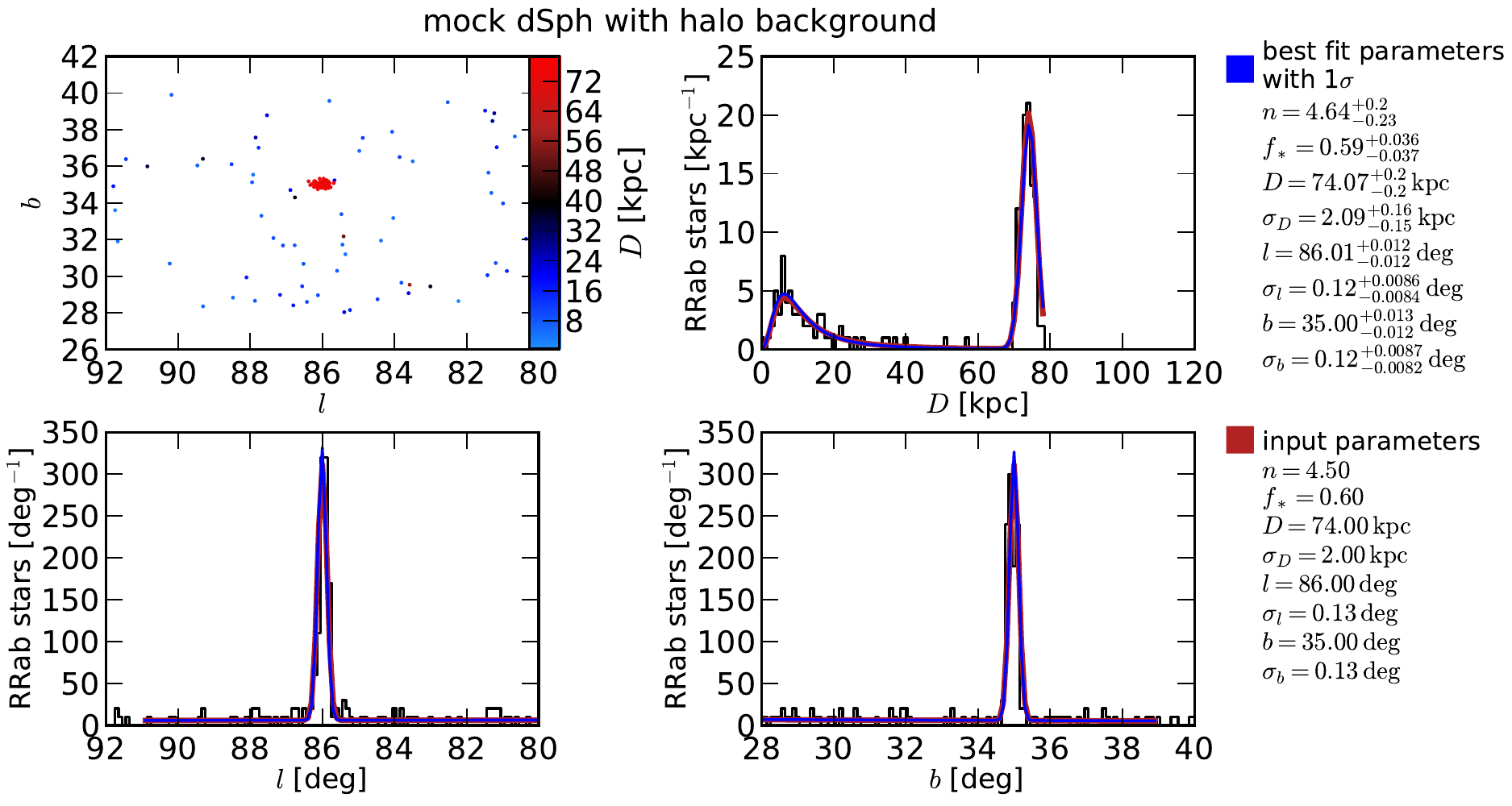}
}
\subfigure[]
    {
\includegraphics[]{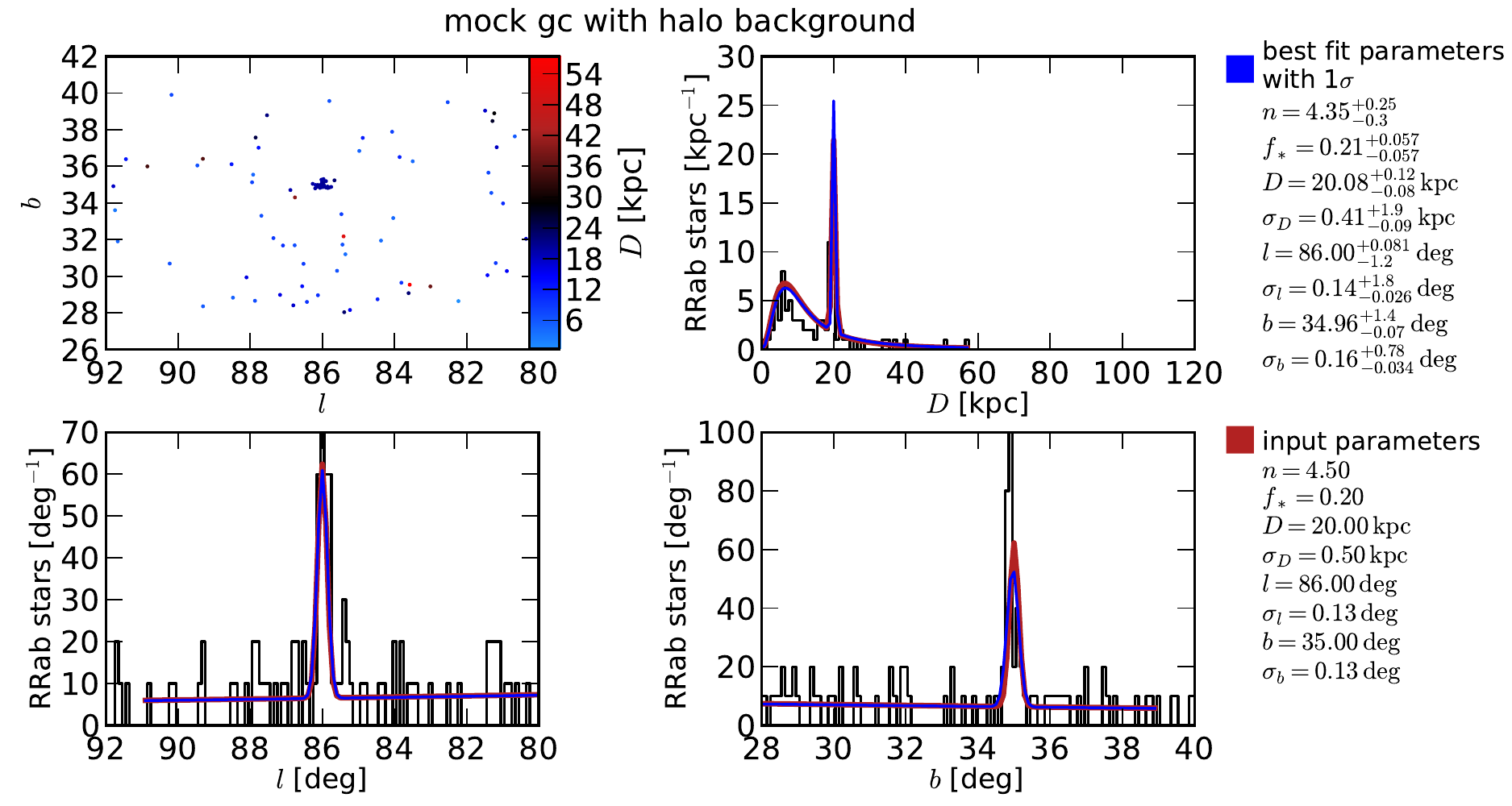}
}
\caption{{
Fit to two mock overdensities, where the one in subfigure (a) resembles a typical dSph with 166 sources, and the one in subfigure (b) resembles a typical GC with 91 sources. We used mock overdensities to test the methodology for fitting overdensities, as well as estimate typical error ranges. The best-fit set of parameters along with their $1\sigma$ intervals, as well as the input parameters used to generate the mock overdensities and background distribution of halo stars, are given in the right part of each panel.
In each case, the first panel shows a map of the star distribution near the overdensity, where the stars are color-coded according to their heliocentric distance.
The other three panels show the histograms in $l$, $b$, $D$ for the stars from the first panel. Overplotted are both the distribution the mock stars were drawn from (red) and the best-fit model from Sec. \ref{sec:DensityFitting} (blue). We find results that are consistent with the input model within reasonable uncertainties, which means that we are able to recover the input parameters for all models in their assumed parameter range.}
\label{fig:fit_mockoverdensities}}
\end{center}  
\end{figure*}

Fig. \ref{fig:fit_mockoverdensities} shows two examples of a simulated distribution of halo and overdensity RRab along with a fit, analogous to the plots of the GC and dSph shown in Fig. \ref{fig:Sextans_dSph_fits_Draco_dSph_fits} to Fig. \ref{fig:Pal_3_fits_Pal_5_fits}. One of them is an overdensity with a high contrast against the background of (mock) field stars that resembles a dSph, whereas the other is a sparse overdensity, comparable to typical GC. We give their distribution in $(l,b,D)$ space as well as their marginalized distributions, the distribution these mock stars were drawn from and the best-fit distribution after applying our fitting methodology.

We find results that are consistent with the input model within reasonable uncertainties, which means that we are able to recover the input parameters for the models within their assumed parameter range.

\section{Results from Fitting}
\label{sec:ResultsfromFitting}

When trying to fit all dSph from Tab. \ref{tab:dwarfs_planned} and globular clusters from Tab. \ref{tab:gc_planned}, we found that the data available for some of them do not allow for a reasonable fit. For the Crater II dSph at a distance of ${\sim}$120 kpc, the PS1 RRab catalog does not contain enough sources for a good fit, not surprising given the large distance. As Sagittarius dSph lies at the edge of the PS1 3$\pi$ footprint, we cannot successfully fit its on-sky position $(l,b)$, but we can fit its heliocentric distance $D$.
Among the globular clusters, we have to exclude NGC 4147, NGC 5634, NGC 5694, NGC 5897, IC 1257, NGC 6093 (M80), NGC 6171 (M107), NGC 6356, NGC 6402 (M14), NGC 6426, NGC 7099 (M30), Pal 1, Pal 13, as in those cases, we find too few, if any, RRab stars associated with these overdensities picked up by the PS1 3$\pi$ survey.

In Tables \ref{tab:dwarfs} and \ref{tab:gc} as well as in Figures \ref{fig:Sextans_dSph_fits_Draco_dSph_fits} to \ref{fig:Pal_3_fits_Pal_5_fits}
we give the dSph and GC best-fit parameters we get from successfully carrying out the fitting as described in Section \ref{sec:DensityFitting}.
Tables \ref{tab:dwarfs} and \ref{tab:gc} list the name, fitted position $(l,b,D)$, fitted angular extent $\sigma_l, \sigma_b$ and depth $\sigma_D$ assuming a multivariate Gaussian, as well as the derived linear extents $(\Delta l, \Delta b, \Delta D)$ as

\begin{align}
\Delta l &=  2\sqrt{2((D\cos(b))^2)(1-\cos(\sigma_l))}\label{eq:delta_l}\\
\Delta b &=  2D \tan(\sigma_b) \label{eq:delta_b}\\
\Delta D &= 2 \sigma_D. \label{eq:delta_D}
\end{align}

We give also the axis ratios $\Delta D / \Delta l$ and $\Delta D / \Delta b$ describing the morphology, as well as the number of sources found in the dSph or globular cluster in each case.

\subsection{Comments on Individual Dwarf Galaxies and Globular Clusters}
\label{sec:CommentsonIndividualGlobularClustersandDwarfGalaxies}

We now comment on the fitting results for individual dwarf galaxies and globular clusters. 

We fitted the five dwarf galaxies Sagittarius dSph, Sextans dSph, Draco dSph, Ursa Minor Dwarf dE, Ursa Major I dSph. 

As the Sagittarius dSph lies at the edge of the PS1 3$\pi$ footprint, we cannot successfully fit its on-sky position $(l,b)$, but we can fit its heliocentric distance $D$. This can be clearly seen from Fig. \ref{fig:Sagittarius_dSph_fits}. For this reason, we don't give any values dependent on the fitted $(l,b)$ position in the tables.

For the other dwarf galaxies, we found that the fit is well defined as these
overdensities have typically ${\gtrsim}$100 sources, except for Ursa Major I. In the case of the latter, the marginalized best-fit model for $l$ as given in the lower left panel of Fig. \ref{fig:Ursa_Minor_Dwarf_dE4_fits_Ursa_Major_I_dSph_fits} does not seem to match the histogram well. However, this is
only an effect due to the marginalization in the histogram, as it shows all sources independent of their distance, whereas the Gaussian is centered
on the fitted $(l, b, D)$. Comparing the best-fit model to the map of stars as given in the upper left panel reveals the accuracy of the fit.

Regarding globular clusters, we have to deal with overdensities of fewer sources, given that the central regions are too crowded for the PS1 RRab catalog \citep{Sesar2017}. 
Comparing for example the panel of NGC 5024 in Fig. \ref{fig:gc_compare} to the histograms in Fig. \ref{fig:NGC_5024__M53_fits_NGC_5053_fits}, we clearly see that stars are missing near the assumed center of the globular cluster. However, the fit is stable enough to identify a reasonable center position.

As for the dwarf galaxy Ursa Minor {Dwarf dE}, in the case of the globular clusters NGC 5024, NGC 5904, NGC 4590, NGC 6864, NGC 7089, the marginalized best-fit model in $l$ and $b$ doesn't seem to match the histogram as well as might be expected, as the histogram shows all sources independent of their distance, while the Gaussian is centered on the
fitted $(l, b, D)$. Again, comparing the best-fit model to the map of stars, we see the fit is reliable.

In some dSphs and GCs, only 3 - 4 RRab were detected. This is the case for Ursa Major I dSph, Ngc 5053, NGC 6864 (M75), NGC 8089 (M2), Pal 3 and Pal 5.
In each of those cases, the overdensity is clearly visible in the map of RRab stars. The histograms might look unconvincing, as they show all sources independent of 
their distance and thus are heavily influenced by field stars. The corresponding best-fit model assigns a distance and position $(l,b,D)$ corresponding well with the 
map of RRab stars. However, we expect that the center, $(l, b, D)$, and the extent of the overdensity, $(\sigma_l, \sigma_b, \sigma_D)$ can not be determined accurately from only 3 - 4 stars, as there is a high chance that the stars are not representative, i.e. are at the bright or shallow end or are not located symmetrically with respect to the center.

\subsection{Uncertainty-Corrected Depth of Dwarf Galaxies and Globular Clusters}
\label{sec:Uncertainty-CorrectedDepthofDwarfGalaxiesandGlobularClusters}

In Table \ref{tab:dwarfs} and \ref{tab:gc}, we give the axis ratios $\Delta D / \Delta l$ and $\Delta D / \Delta b$ describing the morphology of each fitted overdensity.
Starting with the dwarf galaxies, we find that all five of them are found to be quite elongated in the direction of $D$. In contrast, their values for $\Delta D / \Delta l$ and $\Delta D / \Delta b$ are comparable (except for Sagittarius dSph, where we cannot calculate them), which is not surprising from Fig. \ref{fig:Sextans_dSph_fits_Draco_dSph_fits} and \ref{fig:Ursa_Minor_Dwarf_dE4_fits_Ursa_Major_I_dSph_fits} showing a  very round shape in $(l,b)$.
We now check if the elongation in $D$ direction can be explained by the distance uncertainty.

For Sextans dSph, Draco dSph, Ursa Minor Dwarf dE4 and Ursa Major I dSph, we find a $\Delta D / \Delta l$ of 9.83, 14.48, 19.77, 4.84, respectively.
Using Equ. \eqref{eq:sigma_D_with_uncertainty} with $\epsilon_D = 0.03D$, we calculate the uncertainty-corrected (true) line-of-sight depth $(\Delta D)' = 2 \sigma'_D = 2 \sqrt{(\sigma_D ^2- \epsilon_D^2)} = 2 \sqrt{(\sigma_D ^2- (0.03D)^2)}$.
We find thus $(\Delta D)'=$ 4.28 for the Sextans dSph, $(\Delta D)' =$ 1.79  for the Draco dSph, $(\Delta D)' = $ 1.41 for the Ursa Minor Dwarf dE4, $(\Delta D)' = $ 0.94 for the Ursa Major I dSph.

For the Ursa Major I dSph, $\epsilon_D$ must be ${\lesssim} 0.027D$ instead of $0.03D$ to make the result sensible.
With the values for $\Delta l$ given in Tab. \ref{tab:dwarfs} and $\epsilon_D=0.03D$, we calculate $\Delta D / \Delta l$ = 7.78 for Sextans dSph, 6.36 for Draco dSph, 6.40 for Ursa Minor Dwarf dE4, 1.44 for Ursa Major I dSph.

This means that introducing a distance uncertainty, which is appropriate as shown by \cite{Sesar2017}, reduces the elongation in $D$ direction, but is still far away from an axis ratio of 1.
As the distance uncertainty $\epsilon_D$ and the true line-of-sight extent $\sigma'_D$ add in quadrature to make the fitted line-of-sight extent $\sigma_D$, a slightly variation in $\epsilon_D$ from the nominal 3\% derived by \cite{Sesar2017} would be able to explain all of the asymmetry. E.g. for the Sextans dSph, $\epsilon_D=0.0396D$ instead of $0.030D$ would result into a $(\Delta D)'=0.55$ and thus 
$(\Delta D)' / \Delta l \sim 1$. Similarly, an axis ratio of 1 could be achieved for the Draco dSph with $\epsilon_D=0.0322D$, for the Ursa Minor Dwarf dE4 with $\epsilon_D=0.0316D$, and for the Ursa Major I dSph with $\epsilon_D=0.027D$.

In addition to variations in the distance uncertainties, for globular clusters we suggest that fitting uncertainties and especially shot noise are responsible for the non-spherical axis ratios we find. This means, because of the expected small line-of-sight extent, even with a distance uncertainty of 3\% we cannot say much about the line-of-sight extent for GC and dwarf galaxies at the distances considered here.

\subsection{The Radii of Globular Clusters and Dwarf Galaxies}
\label{sec:TheRadiiofGlobularClustersandDwarfGalaxies}

We also compared our fitted extent $\sigma_l$, $\sigma_b$ to the tidal radius and core radius. 

For globular clusters, we use the tidal radius $r_t$ and core radius $r_c$ from the 2010 update of \cite{Harris1996_2010}.
We use the core radii from \cite{Stoehr2002} for the dwarf galaxies Draco dSph, Sextans dSph and Ursa Minor Dwarf dE4. A core radius for Ursa Major I dSph is provided by \cite{Simon2007}. We use tidal radii from \cite{Odenkirchen2001} (Draco dSph), \cite{Roderick2016} (Sextans dSph) and \cite{Kleyna1998} (Ursa Minor Dwarf dE4).
A tidal radius for Ursa Major I dSph is not available.

\begin{figure*}
\begin{center}  
\includegraphics{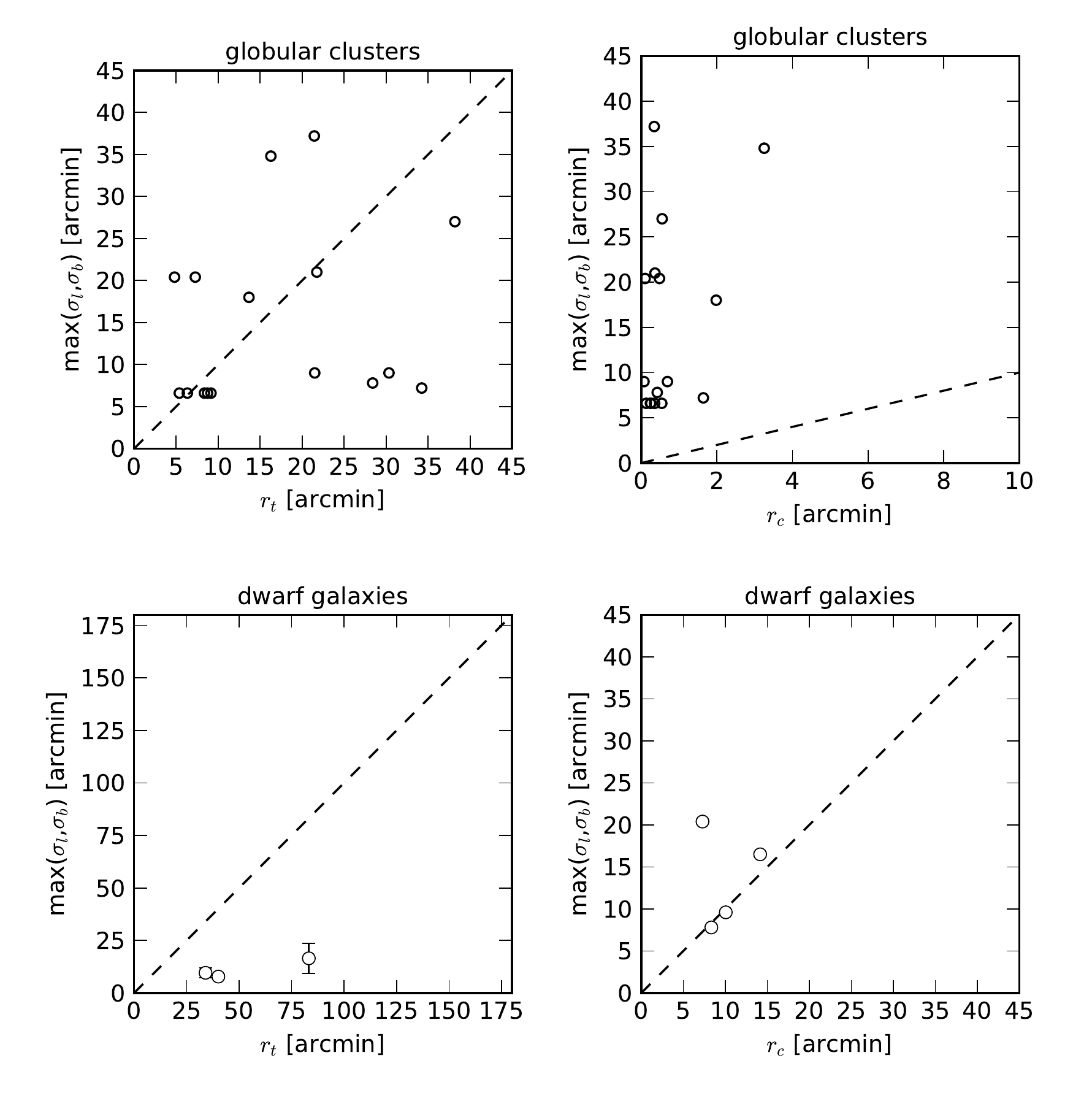}
\caption{{
Comparison between the estimated extent of the overdensities (here we take the maximum of $\sigma_l$ and $\sigma_b$, max($\sigma_l$,$\sigma_b$)) in the present study and the tidal radii $r_t$ and core radii $r_c$ for globular clusters \citep[no uncertainties are available]{Harris1996_2010} and for dwarf galaxies (from different sources, see Table \ref{tab:dwarf_radii}; uncertainties are partially available).
For all of our remote globular clusters and for three of our dwarf galaxies, we were able to look up $r_t$, $r_c$. The dwarf galaxy Ursa Major I lacks a published tidal and core radius.
The diagonal line represents the one-to-one relation.\newline
We find that for globular clusters, whereas the distribution shows a lot of scatter, our estimated extent from max($\sigma_l$,$\sigma_b$) represents significant fractions of the tidal radius. We find sources ways out of the core radius. For dwarf galaxies, our estimated extent from max($\sigma_l$,$\sigma_b$) matches quite good the core radius. We don't pick up sources being as distant as the tidal radius.\newline
Tables \ref{tab:dwarf_radii} and \ref{tab:gc_radii} list the data used for this Figure.}
\label{fig:dwarf_gc_all_radii}}
\end{center}  
\end{figure*}

Tables \ref{tab:dwarf_radii} and \ref{tab:gc_radii} and Fig. \ref{fig:dwarf_gc_all_radii} summarize these comparisons.

We find that for globular clusters, while the distribution shows a lot of scatter, our estimated extent from max($\sigma_l$,$\sigma_b$) represents significant fractions of the tidal radius. We find sources far beyond the core radius. For the dwarf galaxies in our sample, except Ursa Major I dSph, our estimated extent from max($\sigma_l$,$\sigma_b$) matches quite well the core radius. In all cases, our estimated extent is below the tidal radius. We don't pick up sources within or near the core radius for GC.
We attribute this to the fact that in the case of GC, we are losing the RRab stars in the cores, as mentioned in Sec. \ref{sec:SampleOfGlobularClustersAndDwarfGalaxies}.

\section{Mean Distances to Other Globular Clusters and Dwarf Galaxies}
\label{sec:MeanDistancestoOtherGlobularClustersandDwarfGalaxies}

For those dwarf galaxies and globular clusters from Table \ref{tab:dwarfs_planned} and \ref{tab:gc_planned} which don't have enough sources to carry out the fitting process described in Section \ref{sec:DensityFitting}, we derived their mean distances (in the case of finding multiple RRab stars likely associated with the overdensitiy in case) or give the distance to the single RRab star found within this overdensity.

Tab. \ref{tab:dwarfs_mean} and \ref{tab:gc_mean} list the distances, along with the number of RRab stars found in each case.

In total, for each of 13 GC and 6 dSph we found at least one RRab star associated with the overdensity. For those with at least two RRab stars, we give the mean distance,
and otherwise, the distance from the only RRab star found. In most cases, our distance estimate lies well within the 3\% uncertainty we claim for the PS1 3$\pi$ sample of RR Lyrae stars.

For some overdensities, we find a lot of stars in the field at about the same distance. This is the case for NGC 5897, where one RRab star is very close to the assumed
coordinates of this GC, but there are in total up to 4 stars that could be associated with that GC.
For NGC 6402 (M14), the field is even more crowded; there are many sources in the field at that distance without revealing an overdensity.

In the case of NGC 6356, we find one source close to the coordinates given for this GC, but at a different distance: The RRab star from our catalog is at a heliocentric distance of 11.02 kpc, whereas \cite{Harris1996_2010} gives a heliocentric distance of 15.1 kpc for NGC 6356.
In the case of Pal 1, we find no RRab stars clearly associated with this GC.
For Pal 13, we find three RRab stars for which we calculate a mean distance of 23.59 kpc, which is about 2.5 kpc lower than the distance given by the recent compilation of \cite{Harris1996_2010}. \cite{Siegel2010} claim a distance of 24.8 kpc, which is within our distance precision.
The same is the case for the dwarf galaxy Bootes I dSph. We find a mean heliocentric distance of 60.61 kpc, which is about 5 kpc off from the distance given by  \cite{Okamoto2012}. However, our distance estimate matches very well the distance of 60.4 kpc by \cite{Hammer2018}.
For Segue 1 dSph, it is difficult to identify which sources are likely associated with this dwarf galaxy, as there are many sources in the field at about that distance.
For the farthest dwarf galaxy in our sample, Crater II dSph, our heliocentric distance estimate of 105.48 kpc is somewhat lower than distance estimates in the literature, i.e. the 112 kpc claimed by \cite{Joo2018}.

\section{Period-Luminosity Relations}
\label{sec:PeriodLuminosityRelations}
In one of our previous papers \citep{Sesar2017}, we used the period--absolute magnitude--metallicity (PLZ) relation known for RR Lyrae stars to calculate the RRab star's distances we use in the paper at hand. 
The PLZ relation is given as \cite[e.g.][]{Catelan2004, Sollima2006}:
\begin{equation}
M_{\lambda} = \alpha_{\lambda} \log_{10}(P/P_{\mathrm{ref}}) + \beta_{\lambda} (\mathrm{[Fe/H]} - \mathrm{[Fe/H]}_{\mathrm{ref}}) + M_{\mathrm{ref},{\lambda}} + \epsilon
\label{eq:PLZ}
\end{equation}
where $\lambda$ denotes the bandpass, $P$ is the period of pulsation, $M_{\mathrm{ref}}$ is the absolute magnitude at some reference period and metallicity (here chosen to be $P_{\mathrm{ref}}$ = 0.6 days, $\mathrm{[Fe/H]}_{\mathrm{ref}}$ = -1.5 dex), and $\alpha$, $\beta$ describe the dependence of the absolute magnitude on period and metallicity.
The $\epsilon$  is a standard normal random variable centered on 0 and with a standard deviation of the uncertainty in $M_{\lambda}$ in order to model the intrinsic scatter in the absolute magnitude convolved with unaccounted measurement uncertainties.

As PS1 3$\pi$ itself has no metallicity information available, we constrained the PLZ relation Equ. \eqref{eq:PLZ} in PS1 bandpasses using metallicities and distance moduli of PS1 RRab stars in the five Galactic globular clusters NGC 6171, NGC 5904, NGC 4590, NGC 6341, NGC 7078 \citep{Sesar2017}.
Table 1 in \cite{Sesar2017} gives the fitted PLZ relations in all bandpasses. The resulting relation in the $i_{\mathrm{P1}}$ band is then used to constrain distances also for stars where no metallicity is available.

In the case of no available metallicity, the expression for the absolute magnitude in the $i_{\mathrm{P1}}$ band, $M_{i_{\mathrm{P1}}}$, simplifies to \citep[][Equ. (5)]{Sesar2017}: $M_{i_{\mathrm{P1}}}=-1.77\log_{10}(P/0.6)+0.46$, 
and in general, the expression for the absolute magnitude in the $\lambda_{\mathrm{P1}}$ band simplifies to
\begin{equation}
M_{\lambda_{\mathrm{P1}}}=\alpha_{\lambda}\log_{10}(P/0.6)+M_{\mathrm{ref},\lambda}.
\label{eq:PLZ_simple}
\end{equation}

Distances are then constrained using the dereddened flux-averaged $i_{{\mathrm{P1}}}$-band magnitude (this is $i_{\mathrm{F}}$ in \cite{Sesar2017}) and Equ. \eqref{eq:PLZ_simple}. This equation was used to calculate the PS1 3$\pi$ RRab distances we use in this paper.

Now, after having fitted the spatial extent and distance of 16 globular clusters, we are interested in comparing the fitted PLZ relation from \cite{Sesar2017} to fits carried out for each cluster using the [Fe/H] from spectra of red giants of \cite{Carretta2009} (Table A1).
For each RRab star in each GC, we have the period from the RRab catalog, the [Fe/H] -- assumed to be constant for a given GC -- from \cite{Carretta2009} and the fitted distance modulus (DM) and distance from the RRab catalog.

\begin{figure*}
\begin{center}  
\includegraphics{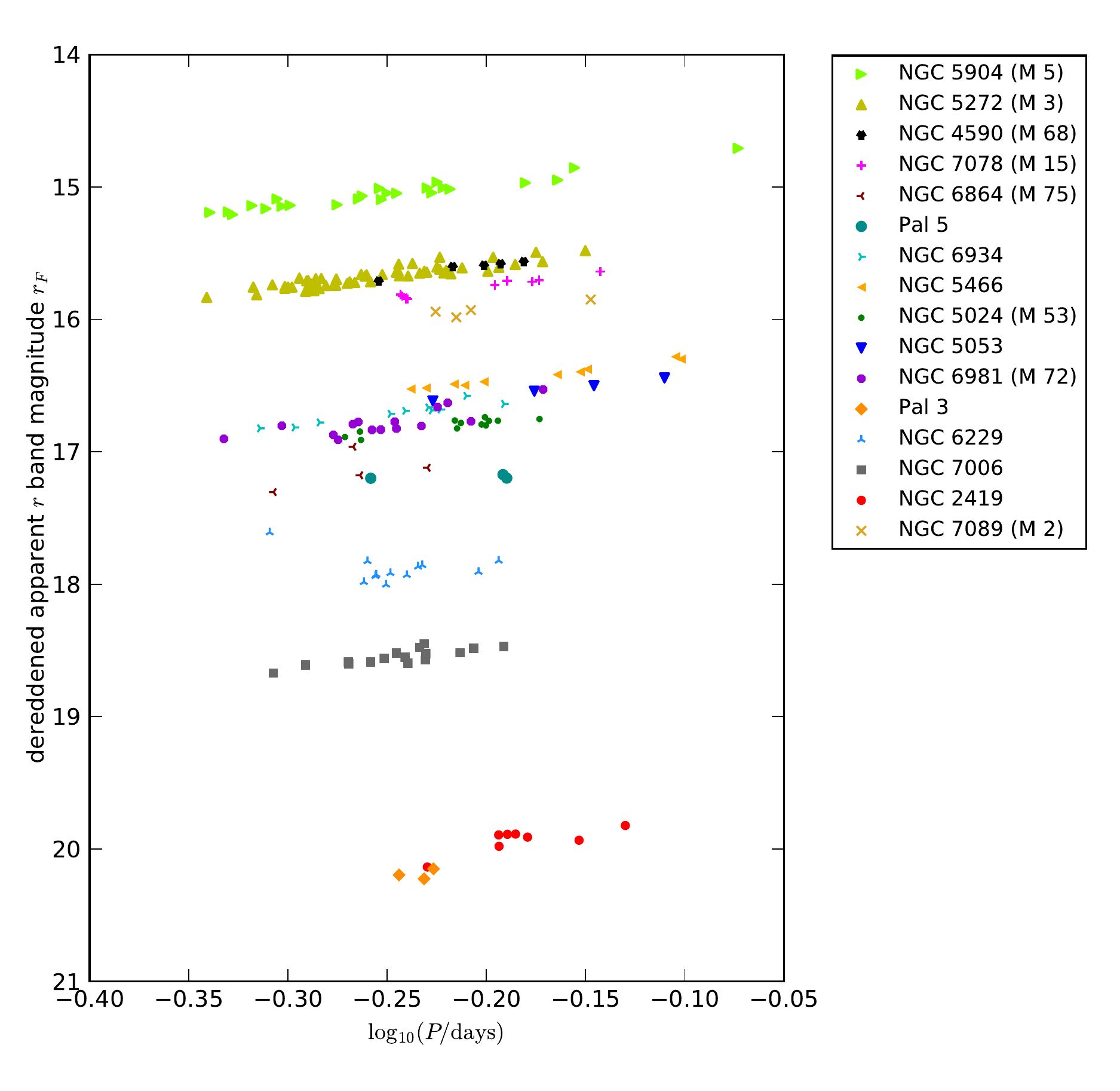}
\caption{{
We plot the dereddened apparent $r$ band magnitude ($r_F$ in the PS1 RRab catalog) for each of the RRab in our sample for each globular cluster vs. their period. The typical trend of a PL relation is clearly visible.}
\label{fig:all_period_rF}}
\end{center}  
\end{figure*}

For Fig. \ref{fig:all_period_rF}, we selected the RRab stars for each globular cluster and plot their dereddended apparent $r$ band magnitude ($r_F$ in the PS1 RRab catalog) vs. their period. The typical trend of a PL relation is clearly visible.

We then took a closer look at the individual globular clusters. For each of them, in each bandpass $\lambda \in \{g_{\mathrm{P1}},...,z_{\mathrm{P1}}\}$ we plot each star's dereddened apparent magnitude $m_{\lambda}$ vs. $P$. We also plot the apparent (unreddened) magnitude as expected from the PLZ relation
$\alpha_{\lambda} \log_{10} (P/P_{\mathrm{ref}}) + \beta_{\lambda}(\mathrm{[Fe/H]} - \mathrm{[Fe/H]}_{\mathrm{ref}}) + M_{\mathrm{ref}} + \mathrm{DM}$ vs. $P$, as well as the apparent magnitude as expected from the equation without metallicity $\alpha_{\lambda} \log_{10} (P/P_{\mathrm{ref}}) + M_{\mathrm{ref}} + \mathrm{DM}$ vs. $P$.

\begin{figure*}
\begin{center}  
\includegraphics{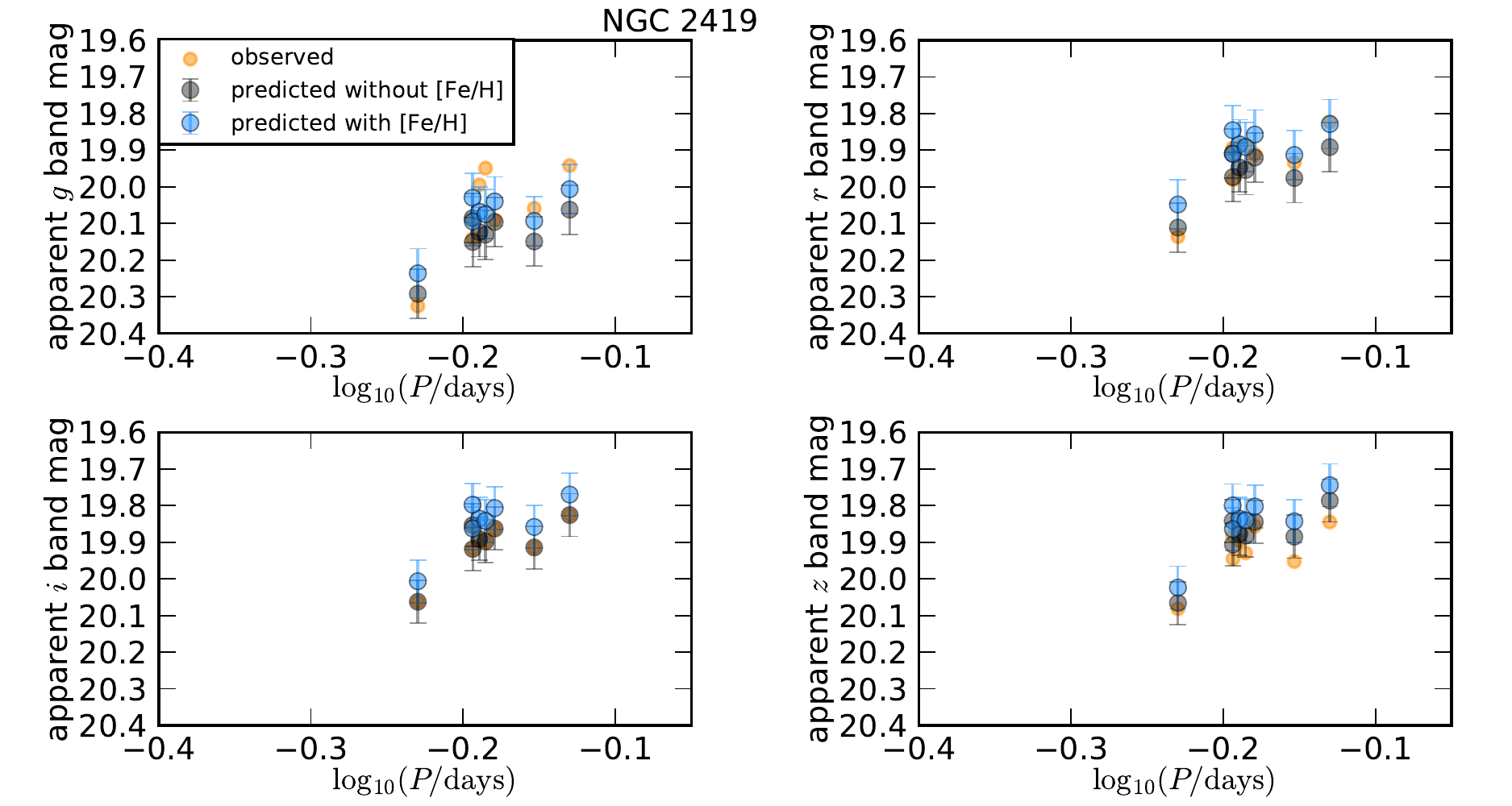}
\caption{{
In this and the following plots, we select the RRab stars for each globular cluster and plot their dereddended apparent $g$, $r$, $i$, $z$ magnitudes ($g_F$,...,$z_F$ in the PS1 RRab catalog) vs. their periods. These are the orange points in each panel.
Along with that, we plot the apparent magnitude one would get from the PLZ or PL relation for each of the periods. The grey points describe the predicted apparent magnitude based on period without any assumption on metallicity (Equ. \eqref{eq:PLZ_simple}) whereas the blue points describe the predicted apparent magnitude based on period when taking the metallicity [Fe/H] from Equ. \eqref{eq:PLZ} into account.\newline
We find that for most of the 16 globular clusters we have evaluated, the predicted apparent magnitude with [Fe/H] (blue markers in the Figures) is a bit brighter than the predicted apparent magnitude without [Fe/H] (black markers), and this is again a bit brighter than the observed dereddened apparent magnitude (orange markers).\newline
This Figure was made for the globular cluster NGC 2419. In the following figures, we show similar plots for all the globular clusters discussed in this paper.
}
\label{fig:NGC_2419_period}}
\end{center}  
\end{figure*}

\begin{figure*}
\begin{center}  
\includegraphics{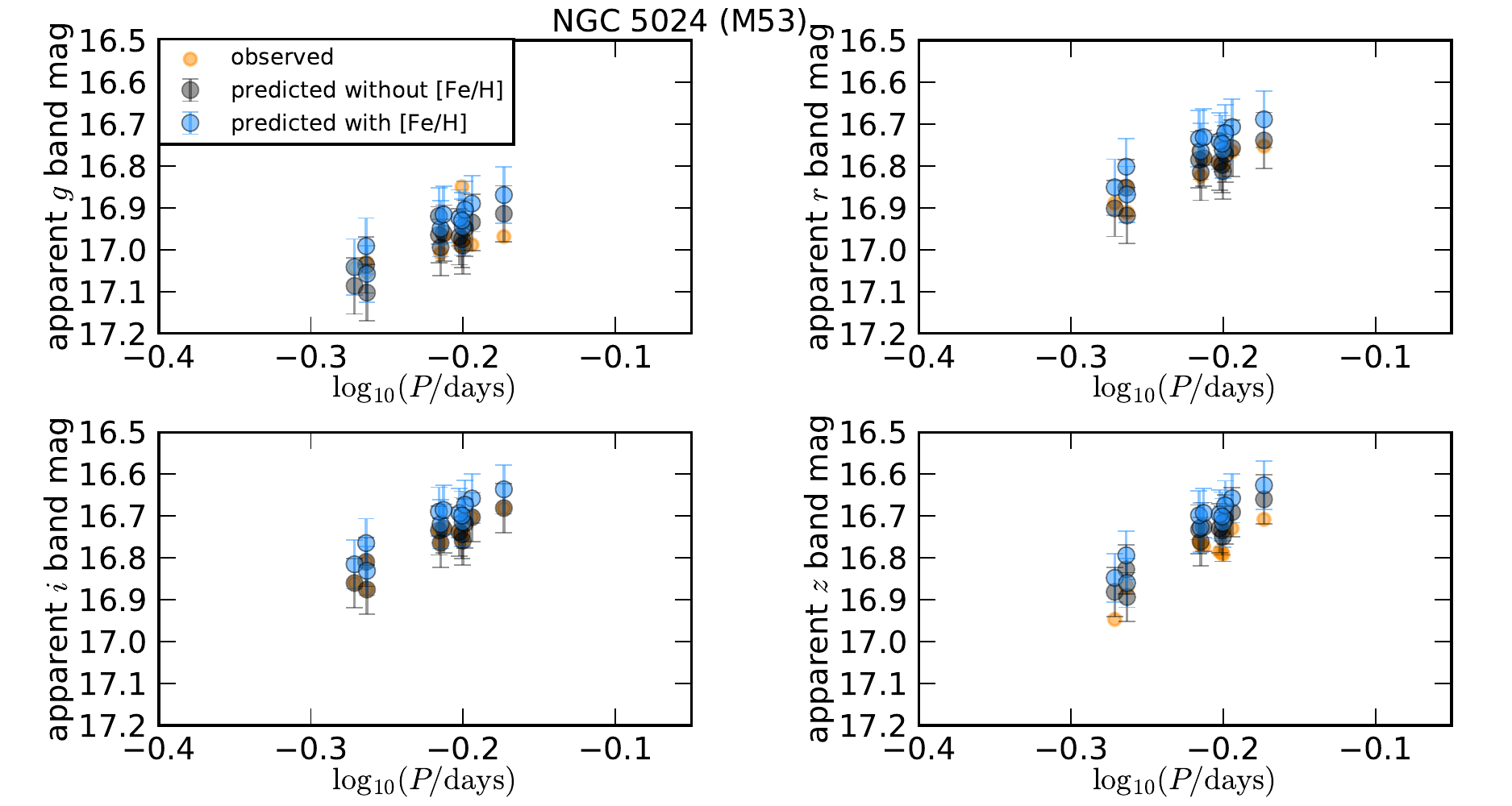}
\caption{{
Observed and predicted apparent magnitudes for the globular cluster NGC 5024 (M53). See Fig. \ref{fig:NGC_2419_period} for a detailed description.}
\label{fig:NGC_5024__M_53_period}}
\end{center}  
\end{figure*}

\begin{figure*}
\begin{center}  
\includegraphics{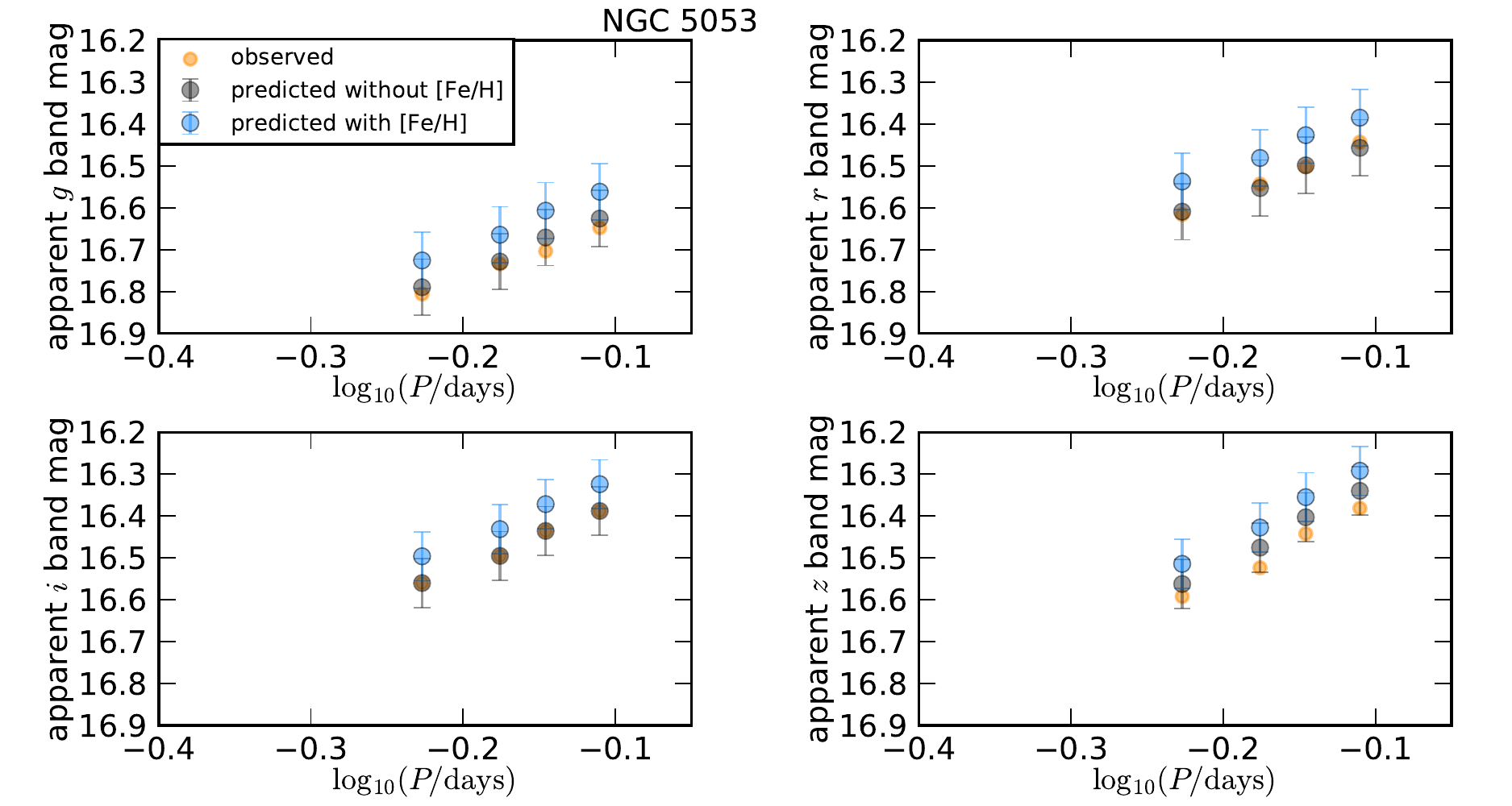}
\caption{{
Observed and predicted apparent magnitudes for the globular cluster NGC 5053. See Fig. \ref{fig:NGC_2419_period} for a detailed description.}
\label{fig:NGC_5053_period}}
\end{center}  
\end{figure*}

\begin{figure*}
\begin{center}  
\includegraphics{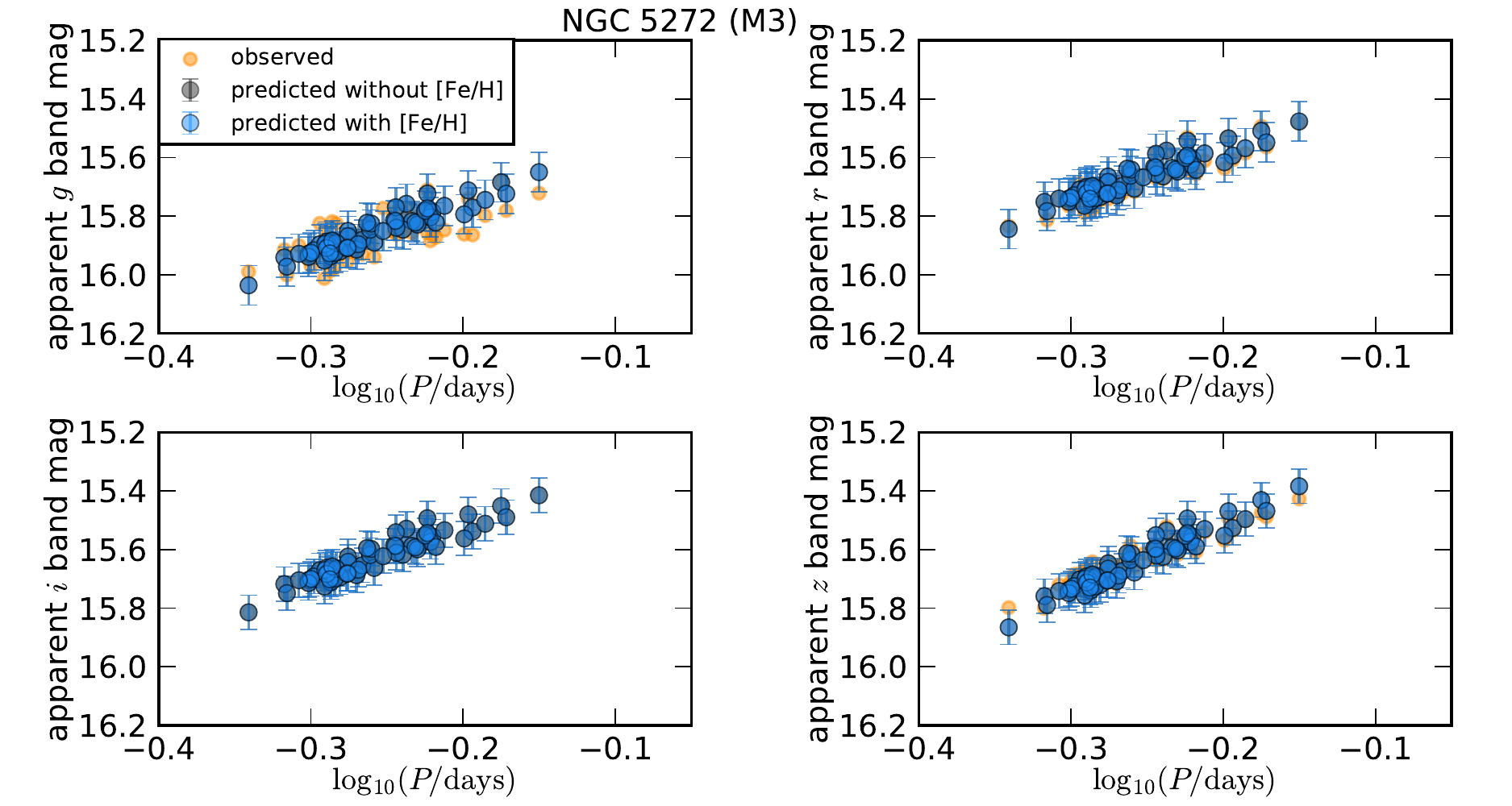}
\caption{{
Observed and predicted apparent magnitudes for the globular cluster NGC 5272 (M3). See Fig. \ref{fig:NGC_2419_period} for a detailed description.}
\label{fig:NGC_5272__M_3_fits }}
\end{center}  
\end{figure*}

\begin{figure*}
\begin{center}  
\includegraphics{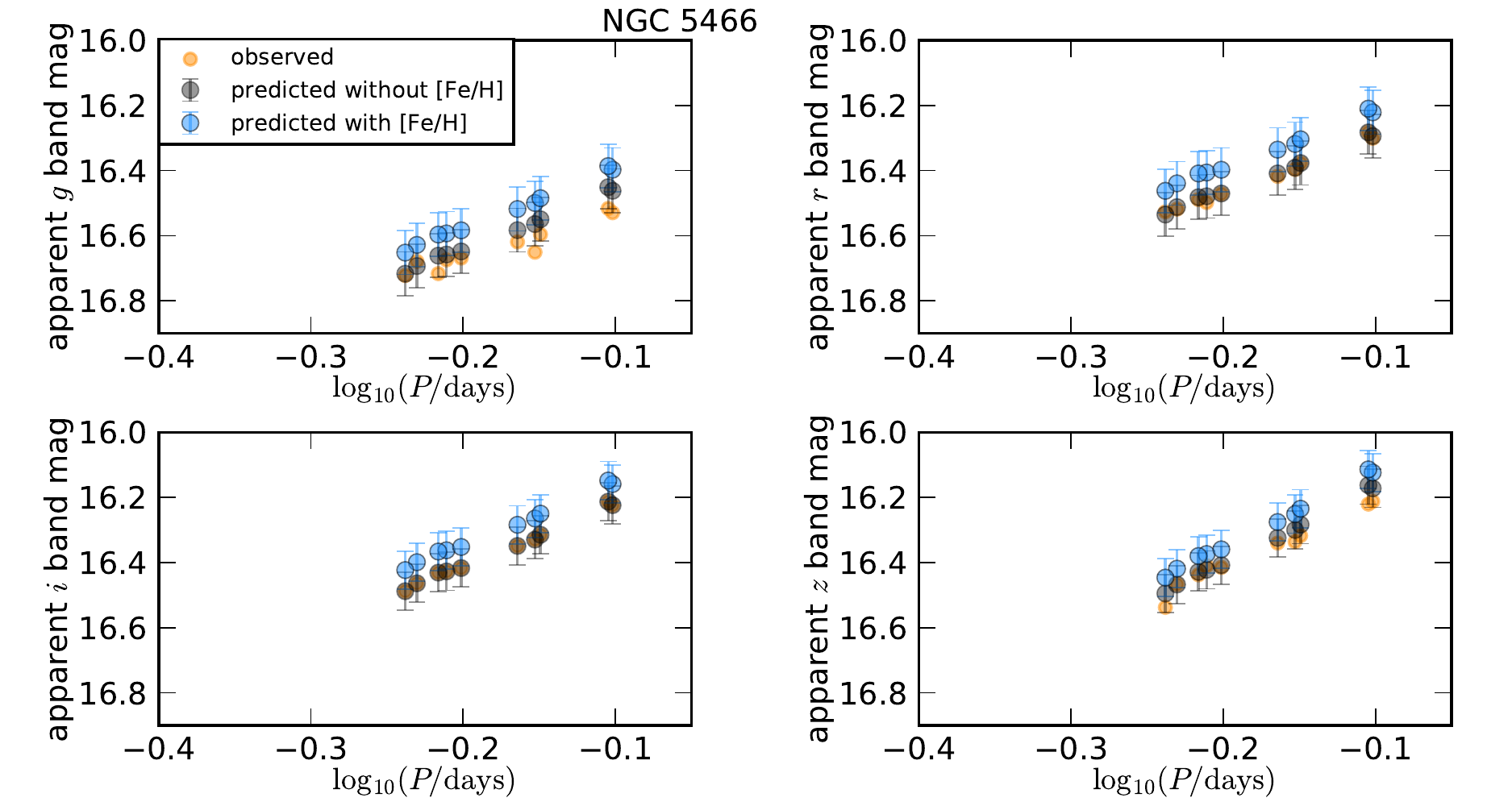}
\caption{{
Observed and predicted apparent magnitudes for the globular cluster NGC 5466. See Fig. \ref{fig:NGC_2419_period} for a detailed description.}
\label{fig:NGC_5466_period}}
\end{center}  
\end{figure*}

\begin{figure*}
\begin{center}  
\includegraphics{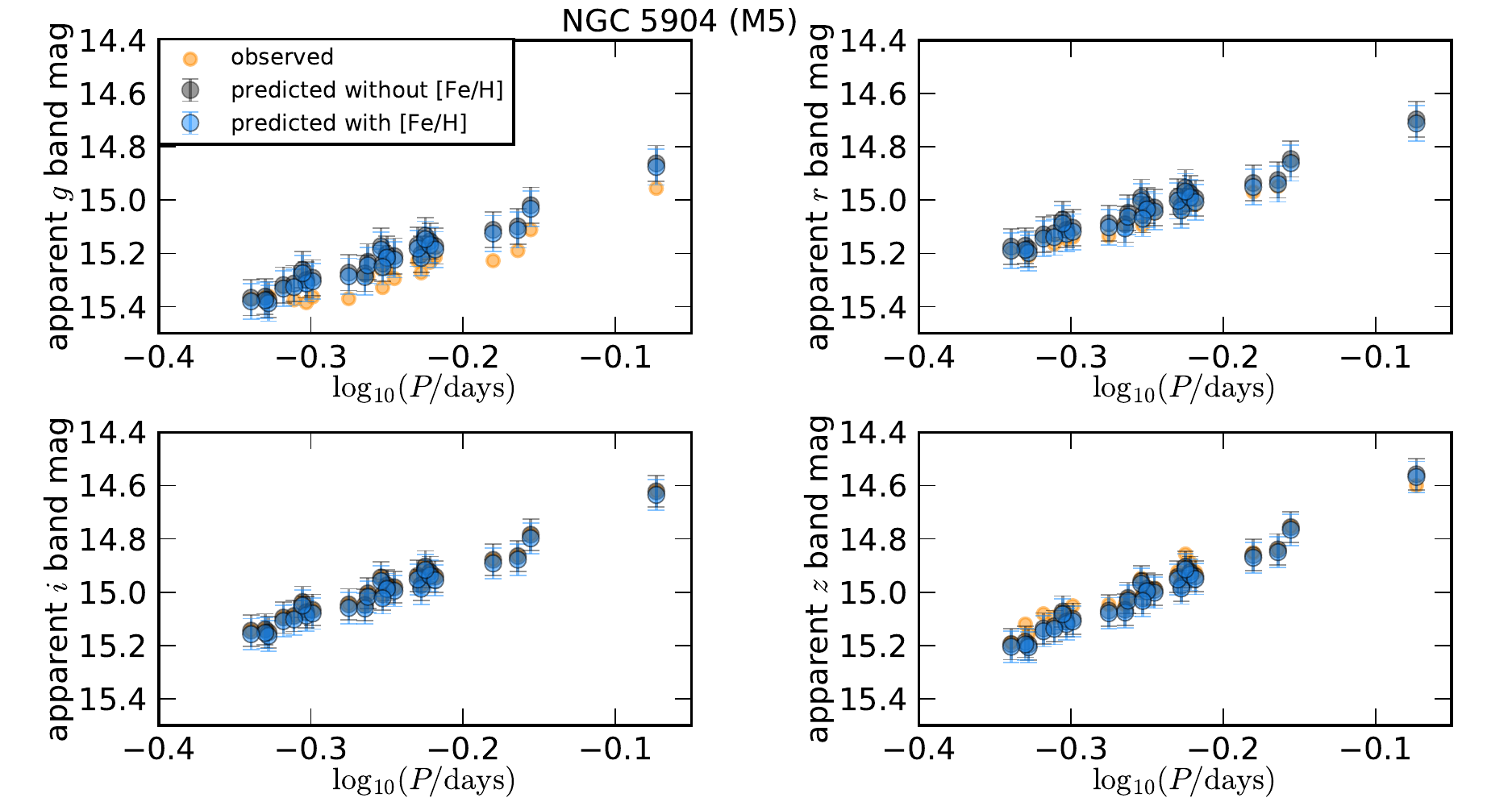}
\caption{{
Observed and predicted apparent magnitudes for the globular cluster NGC 5904 (M5). See Fig. \ref{fig:NGC_2419_period} for a detailed description.}
\label{fig:NGC_5904__M_5_period}}
\end{center}  
\end{figure*}

\begin{figure*}
\begin{center}  
\includegraphics{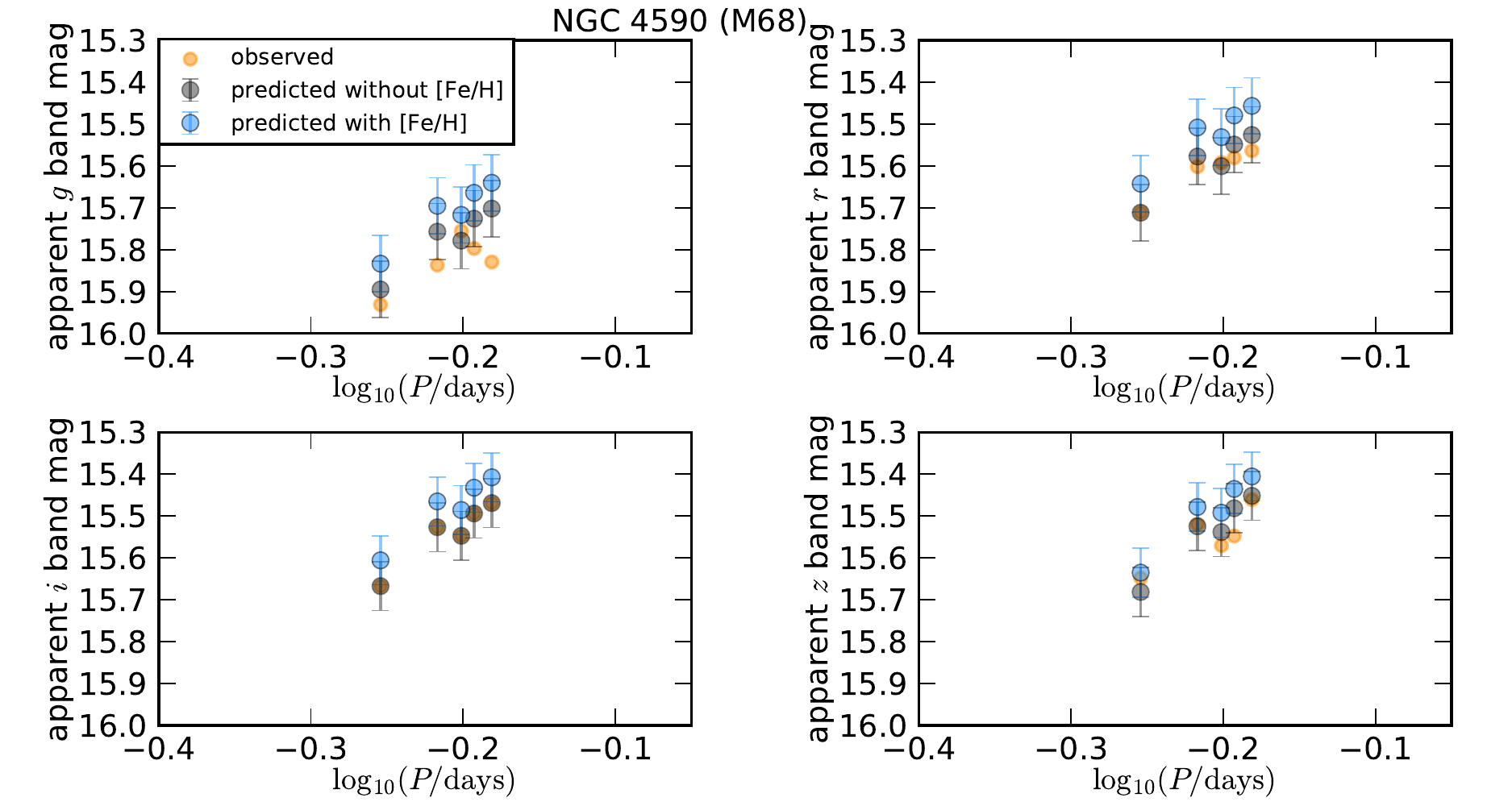}
\caption{{
Observed and predicted apparent magnitudes for the globular cluster NGC 4590 (M68). See Fig. \ref{fig:NGC_2419_period} for a detailed description.}
\label{fig:NGC_4590__M_68_period}}
\end{center}  
\end{figure*}

\begin{figure*}
\begin{center}  
\includegraphics{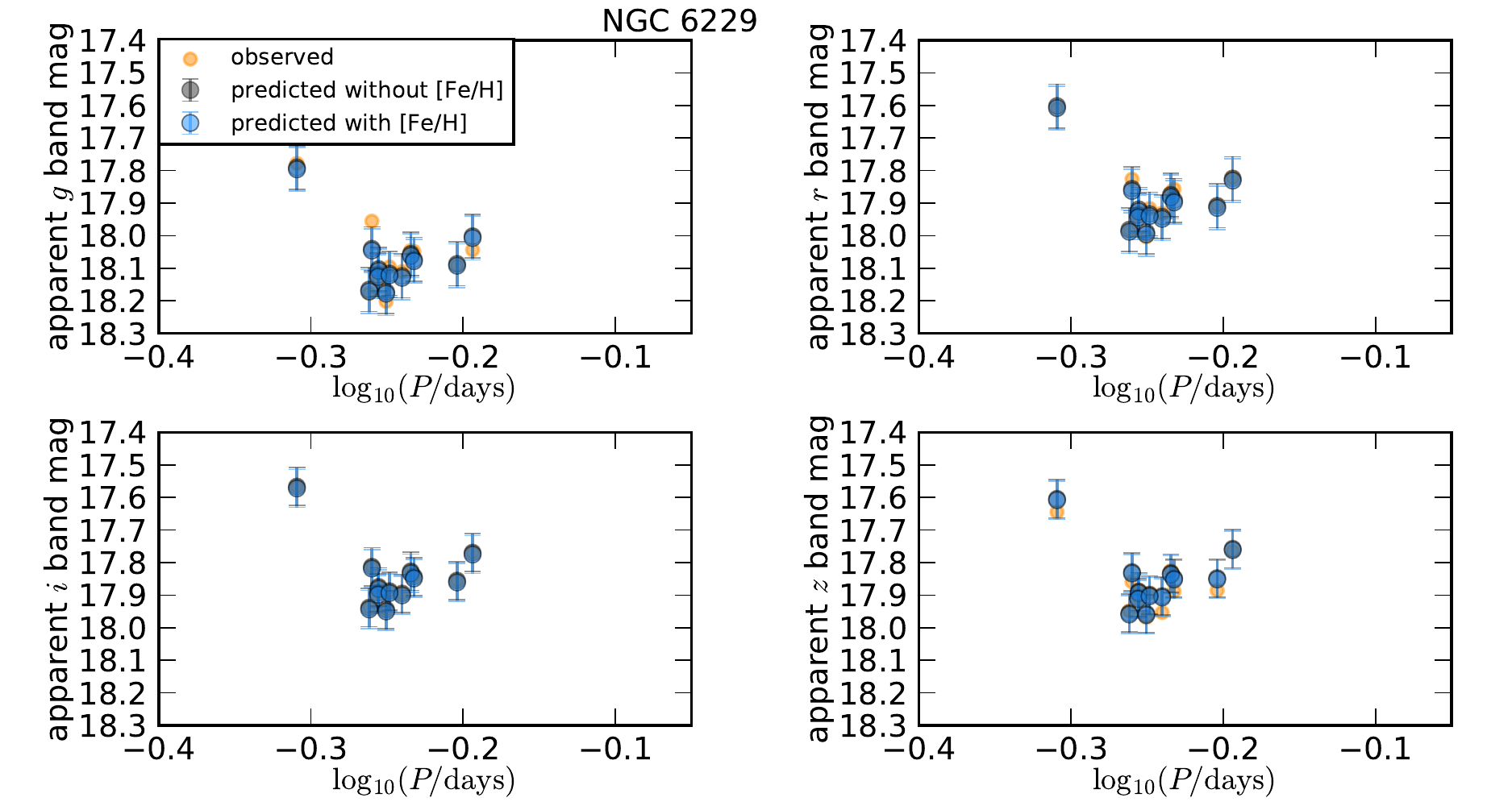}
\caption{{
Observed and predicted apparent magnitudes for the globular cluster NGC 6229. See Fig. \ref{fig:NGC_2419_period} for a detailed description.}
\label{fig:NGC_6229_period}}
\end{center}  
\end{figure*}

\begin{figure*}
\begin{center}  
\includegraphics{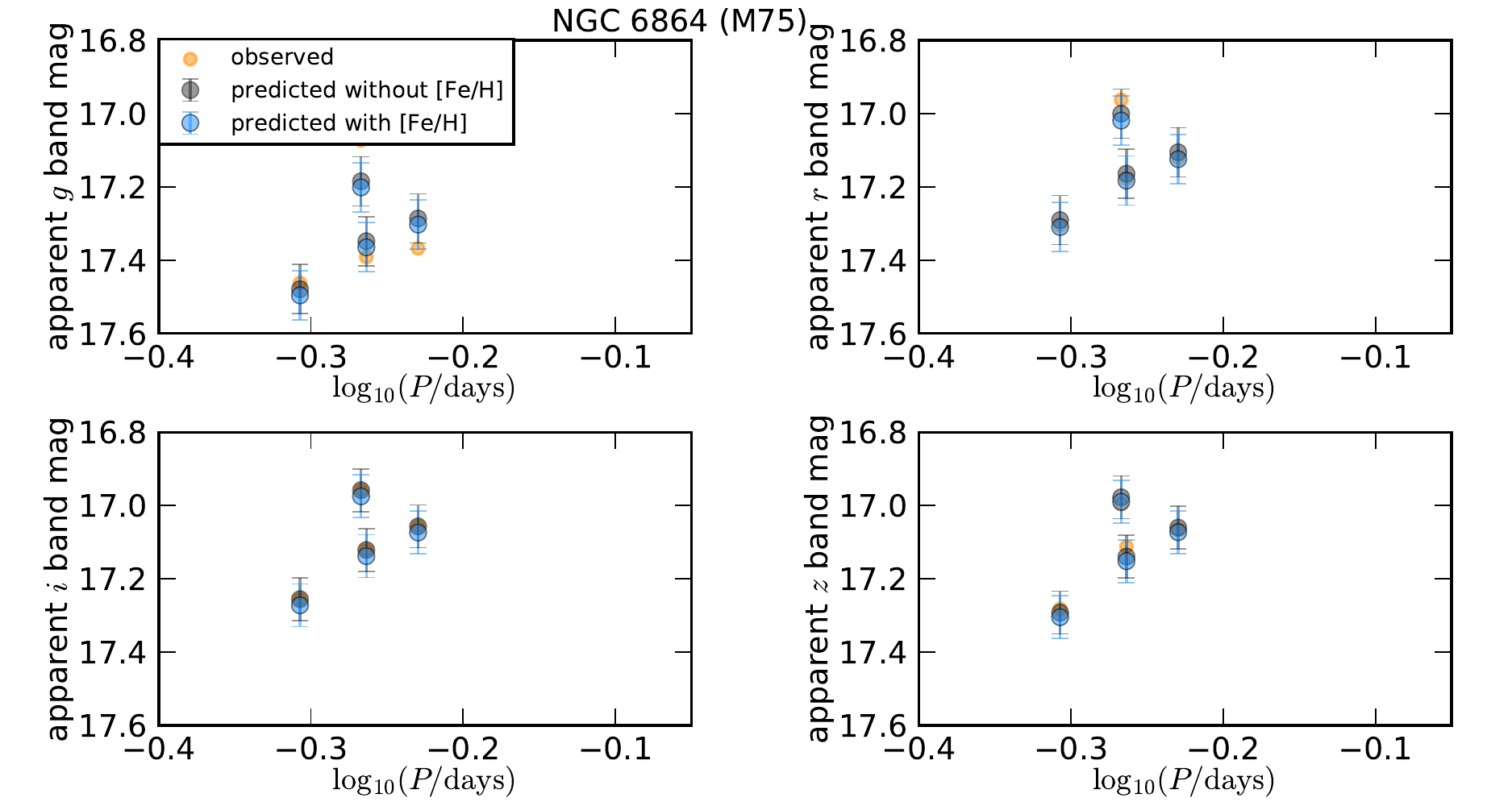}
\caption{{
Observed and predicted apparent magnitudes for the globular cluster NGC 6864 (M75). See Fig. \ref{fig:NGC_2419_period} for a detailed description.}
\label{fig:NGC_6864__M_75_period}}
\end{center}  
\end{figure*}

\begin{figure*}
\begin{center}  
\includegraphics{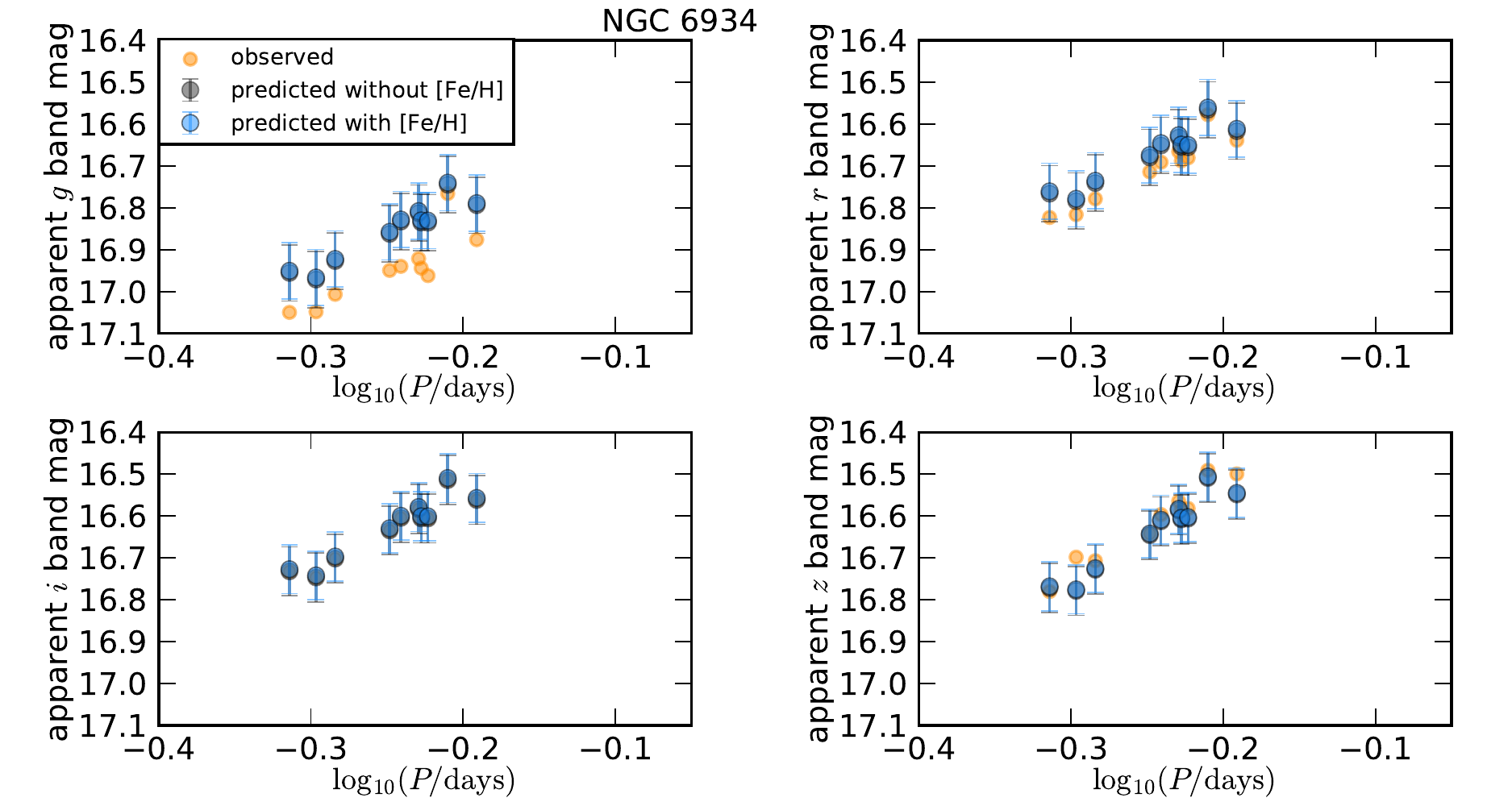}
\caption{{
Observed and predicted apparent magnitudes for the globular cluster NGC 6934. See Fig. \ref{fig:NGC_2419_period} for a detailed description.}
\label{fig:NGC_6934_period}}
\end{center}  
\end{figure*}

\begin{figure*}
\begin{center}  
\includegraphics{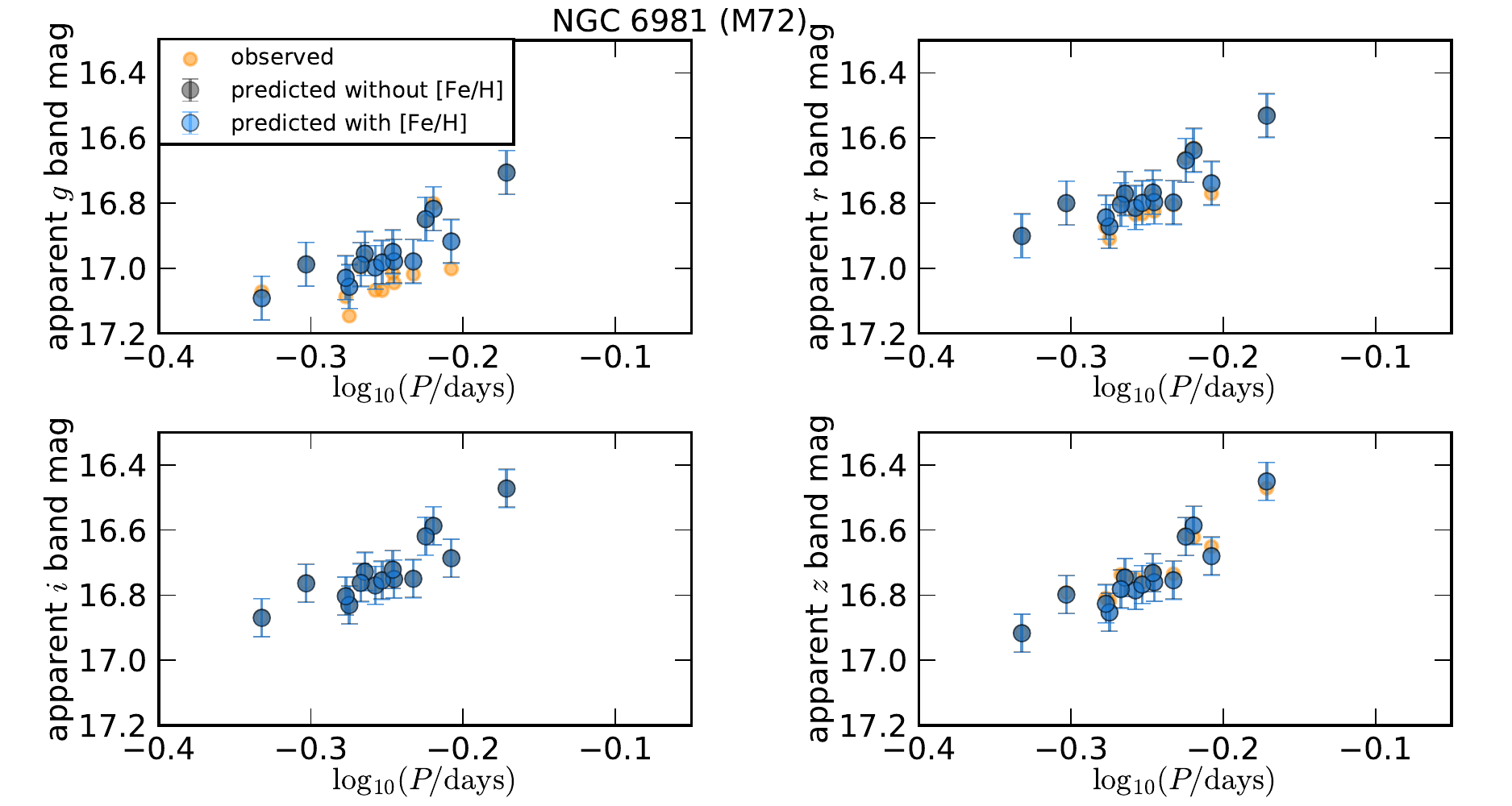}
\caption{{
Observed and predicted apparent magnitudes for the globular cluster NGC 6981 (M72). See Fig. \ref{fig:NGC_2419_period} for a detailed description.}
\label{fig:NGC_6981__M_72_period}}
\end{center}  
\end{figure*}

\begin{figure*}
\begin{center}  
\includegraphics{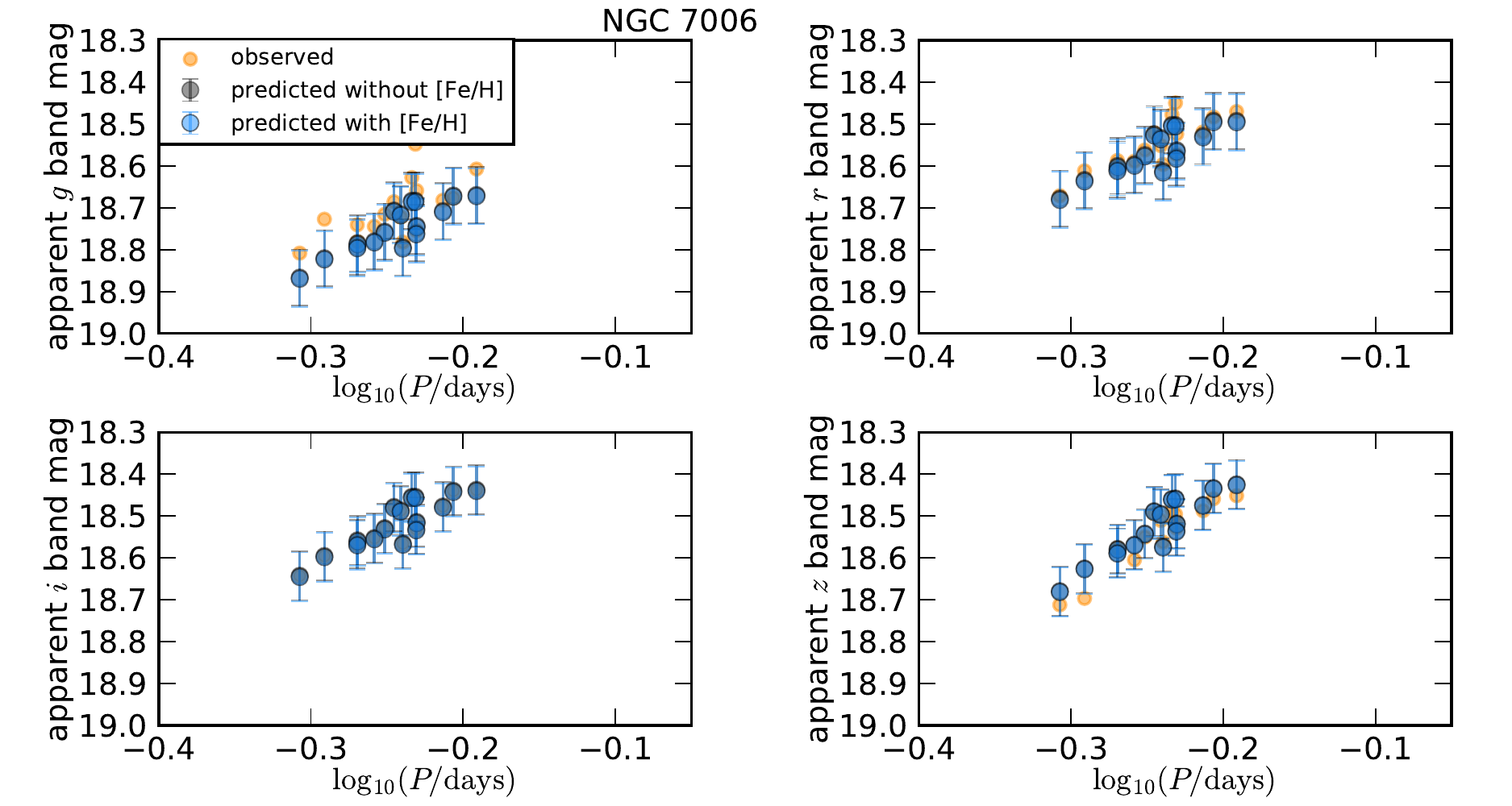}
\caption{{
Observed and predicted apparent magnitudes for the globular cluster NGC 7006. See Fig. \ref{fig:NGC_2419_period} for a detailed description.}
\label{fig:NGC_7006_period}}
\end{center}  
\end{figure*}

\begin{figure*}
\begin{center}  
\includegraphics{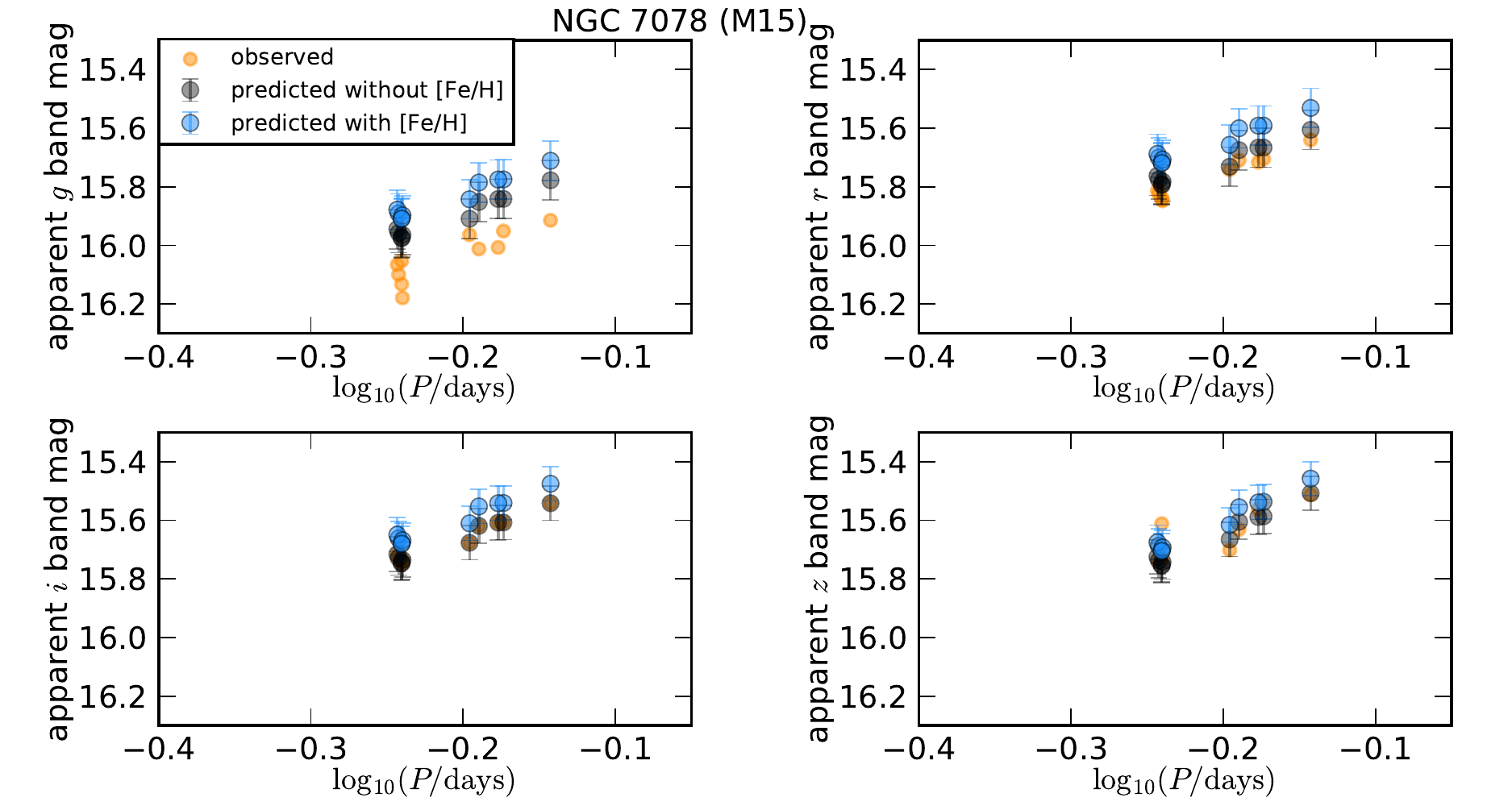}
\caption{{
Observed and predicted apparent magnitudes for the globular cluster NGC 7078 (M15). See Fig. \ref{fig:NGC_2419_period} for a detailed description.}
\label{fig:NGC_7078__M_15_period}}
\end{center}  
\end{figure*}

\begin{figure*}
\begin{center}  
\includegraphics{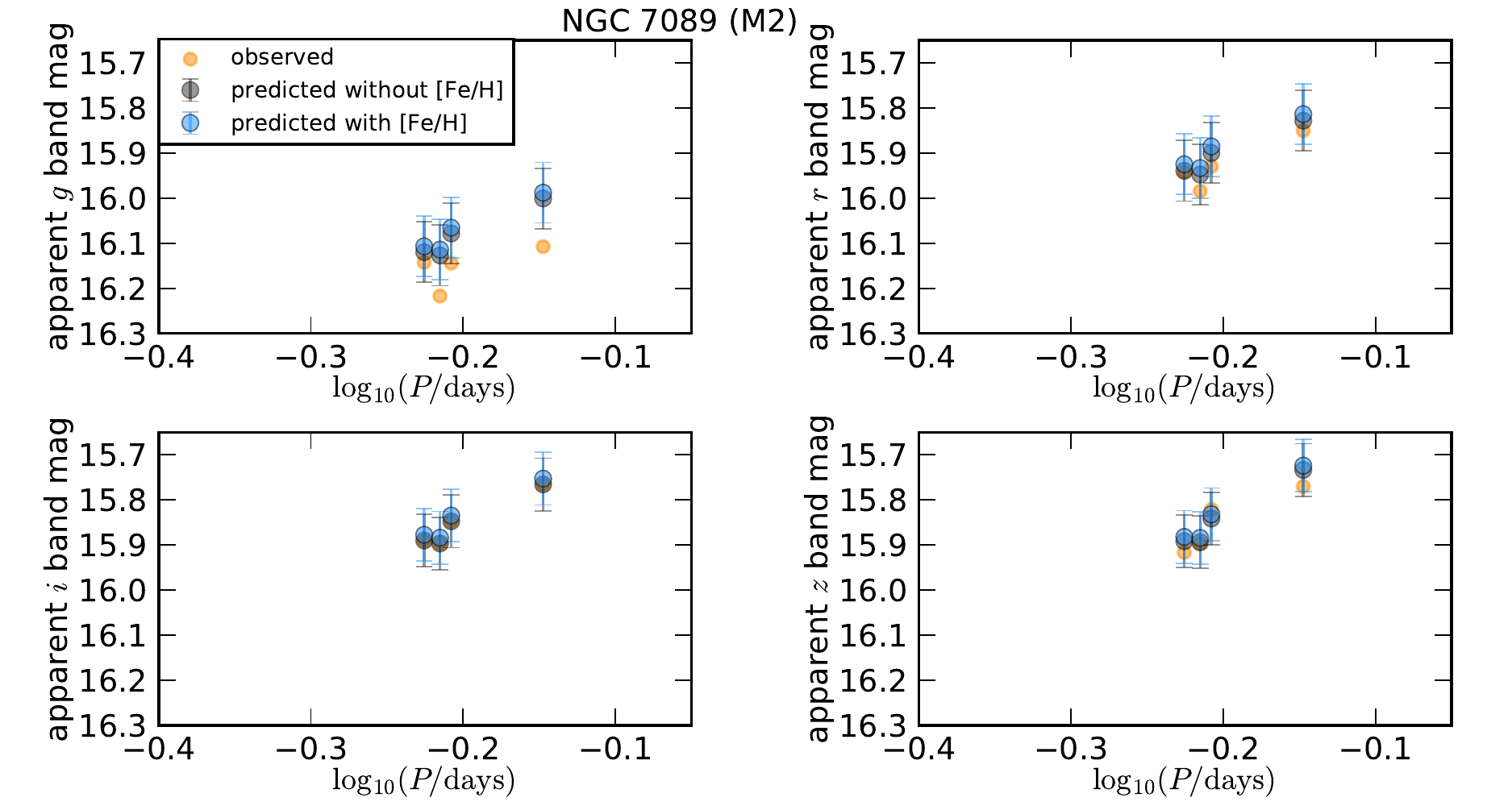}
\caption{{
Observed and predicted apparent magnitudes for the globular cluster NGC 7089 (M2). See Fig. \ref{fig:NGC_2419_period} for a detailed description.}
\label{fig:NGC_7089__M_2_period}}
\end{center}  
\end{figure*}

\begin{figure*}
\begin{center}  
\includegraphics{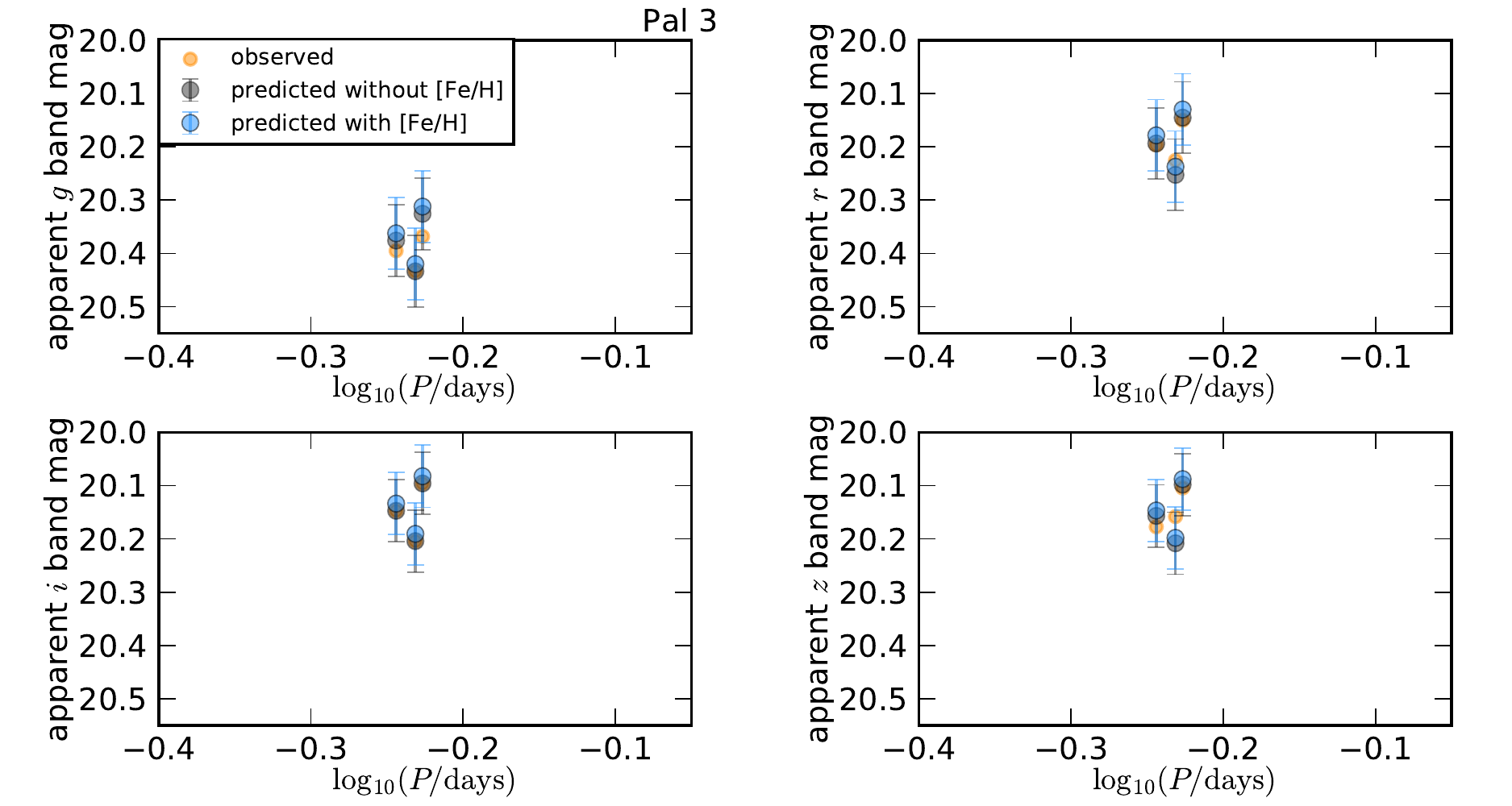}
\caption{{
Observed and predicted apparent magnitudes for the globular cluster Pal 3. See Fig. \ref{fig:NGC_2419_period} for a detailed description.}
\label{fig:Pal_3_period}}
\end{center}  
\end{figure*}

\begin{figure*}
\begin{center}  
\includegraphics{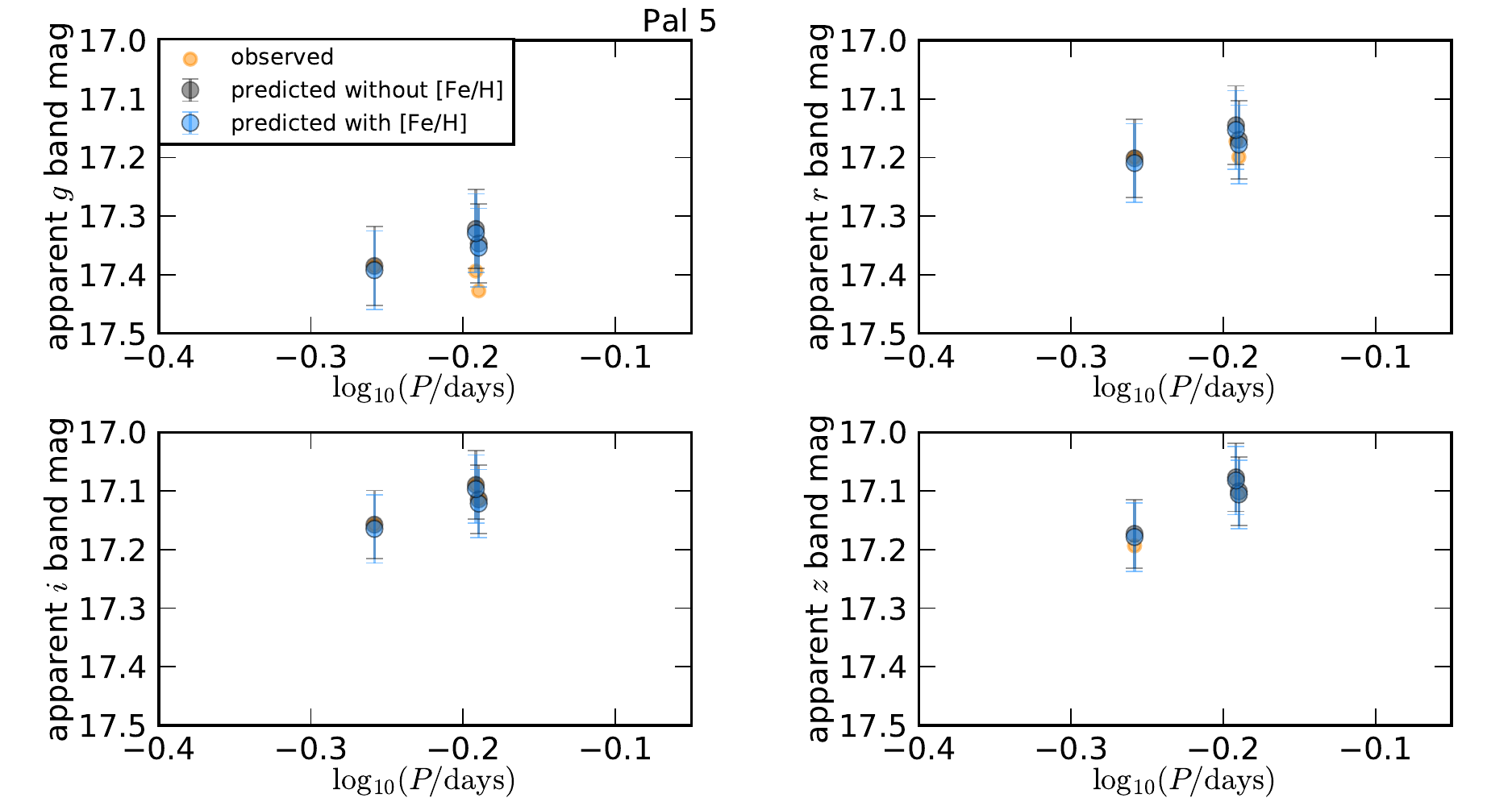}
\caption{{
Observed and predicted apparent magnitudes for the globular cluster Pal 5. See Fig. \ref{fig:NGC_2419_period} for a detailed description.}
\label{fig:Pal_5_period}}
\end{center}  
\end{figure*}

This results in Figures \ref{fig:NGC_2419_period} to \ref{fig:Pal_5_period}.
The error bars in the predicted apparent magnitudes correspond to an uncertainty of ${\sim}0.06$ mag in the absolute magnitude and thus distance modulus \citep{Sesar2017}.
As the distance modulus was derived from the $i_{\mathrm{P1}}$ band, in this band, per definition, the observed magnitude and the apparent magnitude as predicted without [Fe/H] are the same.
We find that for most of the 16 globular clusters we have evaluated, the predicted apparent magnitude with [Fe/H] (blue markers in the Figures) is a bit brighter than the predicted apparent magnitude without taking into account [Fe/H] (black markers), and this is again a bit brighter than the observed dereddened apparent magnitude (orange markers).
NGC 6864 (Fig. \ref{fig:NGC_6864__M_75_period}) is the only GC where the predicted apparent magnitude with [Fe/H] is higher than the one predicted without [Fe/H].
It has the highest metallicity in the sample of GC studied here.

The offset in the predicted apparent magnitude is depending on the metallicity of the GC in case. The predicted apparent magnitudes with [Fe/H] (blue markers in the Figures) are calculated using Equ. \eqref{eq:PLZ} and the distance modulus for each RRab star given in the PS1 RRab catalog \citep{Sesar2017}.
For the black markers, we neglect the term $\beta_{\lambda} (\mathrm{[Fe/H]} - \mathrm{[Fe/H]}_{\mathrm{ref}})$ from Equ. \eqref{eq:PLZ}, and again use the distance modulus. 
As $\beta_{\lambda}$ ranges from 0.06 to 0.09, $\mathrm{[Fe/H]}_{\mathrm{ref}}=-1.5$, and [Fe/H], which ranges from -2.33 to -1.29, is $<-1.5$ for most GC, this term is negative.
Thus, for most of the 16 globular clusters we have evaluated, the predicted apparent magnitude with [Fe/H] (blue markers in the Figures) is a bit brighter 
than the predicted apparent magnitude without taking into account [Fe/H] (black markers).

The offset between observed and predicted apparent magnitudes can be explained by the way the distance to the RRab stars was derived. The distance modulus we use here to calculate the predicted magnitudes was derived in \cite{Sesar2017} using the flux-averaged $i_{{\mathrm{P1}}}$-band magnitude.
Thus, for the $i$ band, the observed magnitude (orange markers) and the magnitude predicted without [Fe/H]
agree, while for the other bands, we find the predicted magnitudes to deviate slightly from the observed ones (black vs. orange markers).
This deviation is in the order of the uncertainty claimed for the distance modulus, 0.06(rnd) $\pm$ 0.03 (sys) mag \citep{Sesar2017}.

\section{Gaia DR2 Distances}
\label{sec:GaiaDR2Distances}

In this section, we compare the PS1 3$\pi$ distances to those published for the second \textit{Gaia} data release (hereafter \textit{Gaia} DR2; \cite{GaiaCollaboration2018a}).

According to \cite{BailerJones2018}, for the vast majority of stars in \textit{Gaia} DR2, reliable distances cannot be obtained by inverting the parallax, but should be derived
by an algorithm that accounts for the nonlinearity of the transformation. Carrying out a probabilistic approach, \cite{BailerJones2018} published distances to 1.33 billion stars in \textit{Gaia} DR2.

\begin{figure*}
\begin{center}  
\includegraphics[trim=0cm 0.6cm 0cm 0.8cm, clip=true]{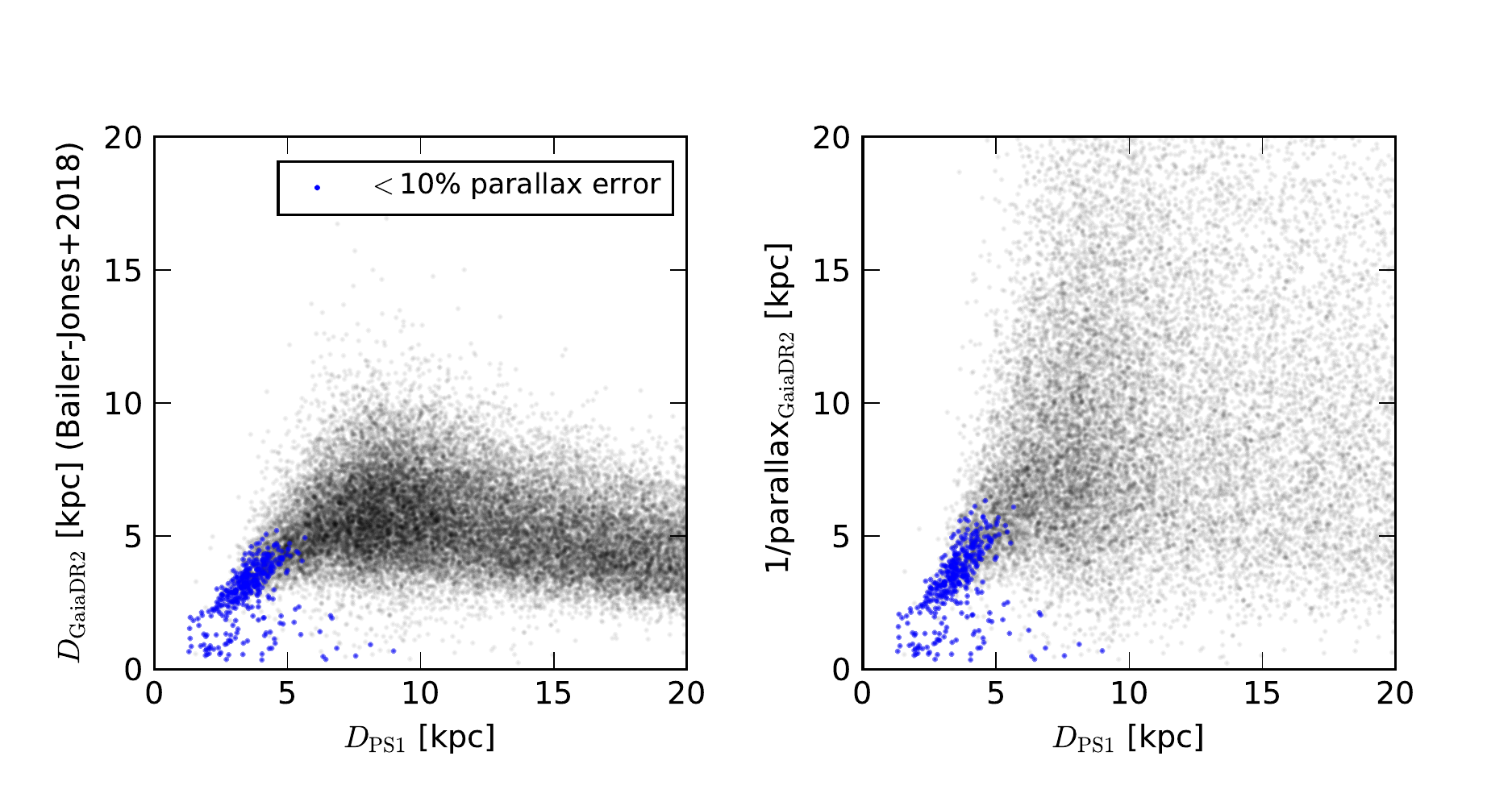}
\caption{{
To compare our distance estimates $D_{\mathrm{PS1}}$ from \cite{Sesar2017} to those of Gaia DR2 \citep{BailerJones2018}, we cross-matched our PS1 RRab sample with Gaia DR2, and found a cross-match for 43,791 sources.
The plot shows the $D_{\mathrm{GaiaDR2}}$ and a rough distance estimate 1/parallax vs. $D_{\mathrm{PS1}}$ for the RRab stars within $D_{\mathrm{PS1}}<20\;\mathrm{kpc}$. Blue points indicate RRab stars with parallax errors of $<$10\%.
We find a rather large scatter in the distances, ways larger than the PS1 distance uncertainty of 3\% or an estimate of the distance uncertainty from the parallax error. Beyond about 5 kpc, where also the parallax error exceeds 10\%, the distances from Gaia DR2 become very unreliable.}
\label{fig:gaia_ps1_rrab_dist}}
\end{center}  
\end{figure*}

To compare our heliocentric distance estimates $D_{\mathrm{PS1}}$ from \cite{Sesar2017} to those of Gaia DR2 \cite{BailerJones2018}, we cross-matched our PS1 RRab sample of 44,403 stars with \textit{Gaia} DR2, and found a cross-match for 43,791 sources.
In Fig. \ref{fig:gaia_ps1_rrab_dist}, we show the $D_{\mathrm{GaiaDR2}}$ and a rough distance estimate 1/parallax vs. $D_{\mathrm{PS1}}$ for the RRab stars within $D_{\mathrm{PS1}}<20\;\mathrm{kpc}$. 

\begin{figure*}
\begin{center}  
\includegraphics[]{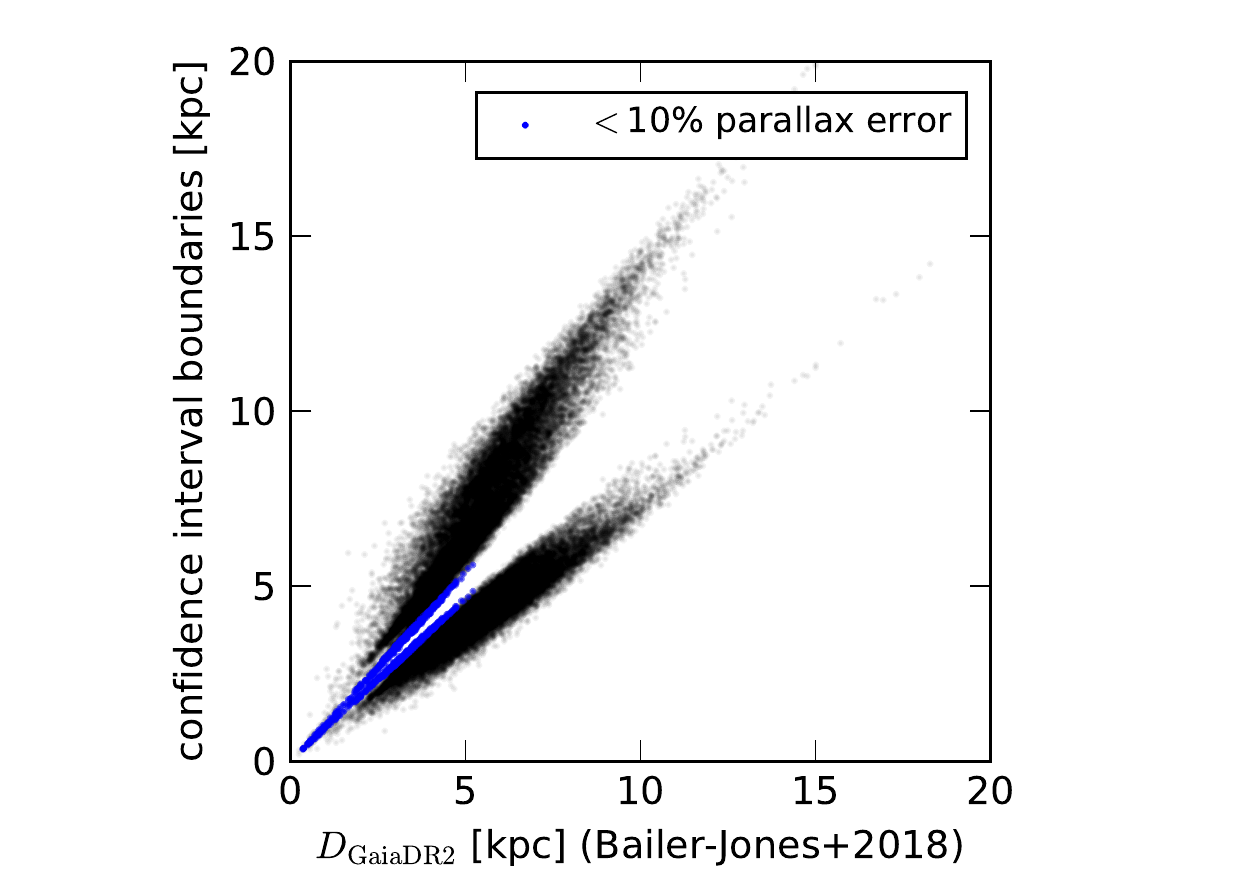}
\caption{{\cite{BailerJones2018} claim quite large uncertainties of their distance estimates. The plot shows the upper and lower 68\% confidence interval boundaries of their distance estimates for our cross-matched sample of RRab stars. Again, blue points indicate RRab stars with $<$10\% parallax error (blue points).
For those RRab stars with a parallax error of ${\sim}$10\%, we find that the uncertainty is about 10\%, and otherwise, can increase to as much as 
50\% for many RRab stars.}
\label{fig:gaia_dist_confidenceinterval}}
\end{center}  
\end{figure*}

At the present time, we find a rather large scatter in the distances, much larger than the PS1 distance uncertainty of 3\% or an estimate of the distance uncertainty from the parallax error. When the parallax error exceeds 10\%, which happens at a heliocentric distance of about 5 kpc, the distances from Gaia DR2 become very unreliable. \cite{BailerJones2018} already claim quite large uncertainties of their distance estimates. In 
Fig. \ref{fig:gaia_dist_confidenceinterval}, we plot the upper and lower 68\% confidence interval boundaries of their distance estimates for our cross-matched sample of RRab stars. According to \cite{BailerJones2018}, the confidence intervals are asymmetric with respect to the distance estimate $D_{\mathrm{GaiaDR2}}$ as a consequence of the nonlinear transformation from parallax to distance.

Up to now, the precision of those distances -- which are calculated only from geometric principles without any assumptions on astrophysics such as pulsation or PLZ relationships -- is not competitive with the precision of our distances from \cite{Sesar2017}, and especially cannot be used for the distant GC and dwarfs. 
However, this might change at least partially for the end-of-mission data. Until then, we strongly recommend using the RRab distances from \cite{Sesar2017} and subsequent analysis instead of those from \textit{Gaia} DR2.

\section{Discussion And Summary}
\label{sec:Discussion}

We started our analysis based on a list of dwarf galaxies within 3 Mpc by \cite{McConnachie2012} 
and a list of currently known globular clusters from a current online version of the \cite{Harris1996_2010} database. We excluded those outside the PS1 3$\pi$ footprint and attempted to fit the $(l,b,D)$ distribution of the remaining ones. For the ones for which a fit was not possible due to only a small number of RRab stars available, we instead selected those stars and give their mean distance.

In Tab. \ref{tab:dsph_comparison} and \ref{tab:gc_comparison}, we compare our distances for dwarf galaxies and GC to literature distances. 
For almost all dwarf galaxies and globular clusters, the estimated distances compared to those in the recent literature are well within our distance precision of 3\%.
However, there are a few dwarf galaxies and globular clusters for which our distances deviate significantly from those given in the literature.

In the case of NGC 6356, we find only one source close to the coordinates given for this GC, but at a different distance: While the RRab star from our catalog is at a heliocentric distance of 11.02 kpc, the updated catalog of \cite{Harris1996_2010} gives a heliocentric distance of 15.1 kpc for NGC 6356.
For Pal 3, we find three stars within a small distance range, resulting in a heliocentric distance estimate of 85.05 kpc, whereas \cite{Harris1996_2010} gives a distance of 92.5 kpc. For Pal 13, we find three RRab stars resulting in a mean distance of 23.59 kpc, which is about 2.5 kpc off from the distance given by \cite{Harris1996_2010}. However, this star might be a member of Pal 13, as \cite{Siegel2010} claim a distance of 24.8 kpc, which is within our distance precision. For IC 1257, we have found only 1 RRab star associated with this GC and estimated its distance to 27.24 kpc, in comparison to the 25 kpc given by \cite{Harris1996_2010}.

For the dwarf galaxy Bootes I dSph, we find a mean heliocentric distance of 60.61 kpc, which is about 5 kpc off from the distance given by \cite{Okamoto2012} and about 1.5 kpc off from the distance given by \cite{Siegel2006}. However, our distance estimate matches very well the distance of 60.4 kpc given by \cite{Hammer2018}. 
The most distant dwarf galaxy in our sample is Crater II dSph. Our heliocentric distance estimate of 105.48 kpc for this dwarf galaxy is somewhat off from the 117.5 kpc distance estimate reported by \cite{McConnachie2012}, but there are closer distance estimates like the $112 \pm 5$ kpc claimed by \cite{Joo2018}, which agrees within the uncertainties with our result. For Sagittarius dSph, our distance estimate of 28.18 kpc is slightly larger than, but within the uncertainties of the $26.3 \pm 1.8$ kpc result obtained by \citep{Monaco2004}. For Ursa Minor Dwarf dE4, we find a distance of 68.41 kpc, whereas the distance is given as 76 kpc from \cite{Bellazzini2002} as well as \cite{Carrera2002}.

In some cases, it is hard to find stars associated with the overdensities, especially if there are many field stars within the assumed distance range of the dwarf galaxy or GC, and there is no apparent overdensity.
This is the case for NGC 5897, where one RRab star is very close to the assumed coordinates of this GC, but there are in total up to 4 stars that could be associated with that GC. For the GC NGC 6402 (M14), the field is even more crowded, so many sources in the field are at its assumed distance without revealing an overdensity.
For NGC 6356, we find one source close to the coordinates given for this GC, but at a different distance: The RRab star from our catalog is at a heliocentric distance of 11.02 kpc, whereas \cite{Harris1996_2010} give a heliocentric distance of 15.1 kpc for NGC 6356.
In the case of Pal 1, we find no stars clearly associated with this GC.
For Segue 1 dSph, it is difficult to identify the sources that might be associated with this dwarf galaxy, as there are many sources in the field at about that distance.

For the GC, even though these sources are closer, in most cases our distance uncertainties are larger than for dwarf galaxies. This probably occurs because we are missing sources near the centers of the GCs as the spatial density of sources is higher that which can be handled by the PS1 analysis codes.

In this paper we have computed the extents $\Delta l, \Delta b, \Delta D$ and the axis ratios
$\Delta D/ \Delta l$ and $\Delta D/ \Delta l$  for all dwarf galaxies and GCs in which a sufficient number of
RRab were picked up in the PS1 RR Lyr database of \cite{Sesar2017}.
Our distances to the stellar overdensities we study here, including both GCs and dwarf galaxy satellites of the Milky Way,
are accurate to better than 3\% when more than 8 RRab occur within a system (i.e. a GC or a dSph).

In the past 5 years many groups have attempted to determine distances to some of the dwarf galaxies using RRab, but they
have each used their own procedures and calibrations to convert a mean RRab magnitude into a distance (see Tab.  \ref{tab:dsph_comparison} and \ref{tab:gc_comparison} for references). Our work is unique in that every object, within the large sample we study,
is treated identically and comes from the same survey, including the metallicity dependence of the RR luminosity, so that the distances for our large
sample of objects are on the same scale across the entire part of the sky covered by the PS1 3$\pi$ survey, for all halo stellar overdensities
within which a sufficient number of RRab could be detected. Thus overall we believe that our distances for the sample
of stellar overdensities in the Milky Way halo, i.e. globular clusters and dwarf galaxies, that we study here are more precise and more
homogeneous than those in the published literature.

\acknowledgments

H.-W.R. acknowledges funding from the
European Research Council under the European Unions
Seventh Framework Programme (FP 7) ERC Grant Agreement n. [321035].

The Pan-STARRS1 Surveys (PS1) have been made possible through contributions of the Institute for Astronomy, the University of Hawaii, the Pan-STARRS Project Office, the Max-Planck Society and its participating institutes, the Max Planck Institute for Astronomy, Heidelberg and the Max Planck Institute for Extraterrestrial Physics, Garching, The Johns Hopkins University, Durham University, the University of Edinburgh, Queen's University Belfast, the Harvard-Smithsonian Center for Astrophysics, the Las Cumbres Observatory Global Telescope Network Incorporated, the National Central University of Taiwan, the Space Telescope Science Institute, the National Aeronautics and Space Administration under Grant No. NNX08AR22G issued through the Planetary Science Division of the NASA Science Mission Directorate, the National Science Foundation under Grant No. AST-1238877, the University of Maryland, and Eotvos Lorand University (ELTE) and the Los Alamos National Laboratory. 

\clearpage

\appendix

\section{Figures}
\label{sec:Figures}

% left bottom right top
\clearpage 
\begin{figure*}
\begin{center}  
\subfigure[]
    {
\includegraphics[]{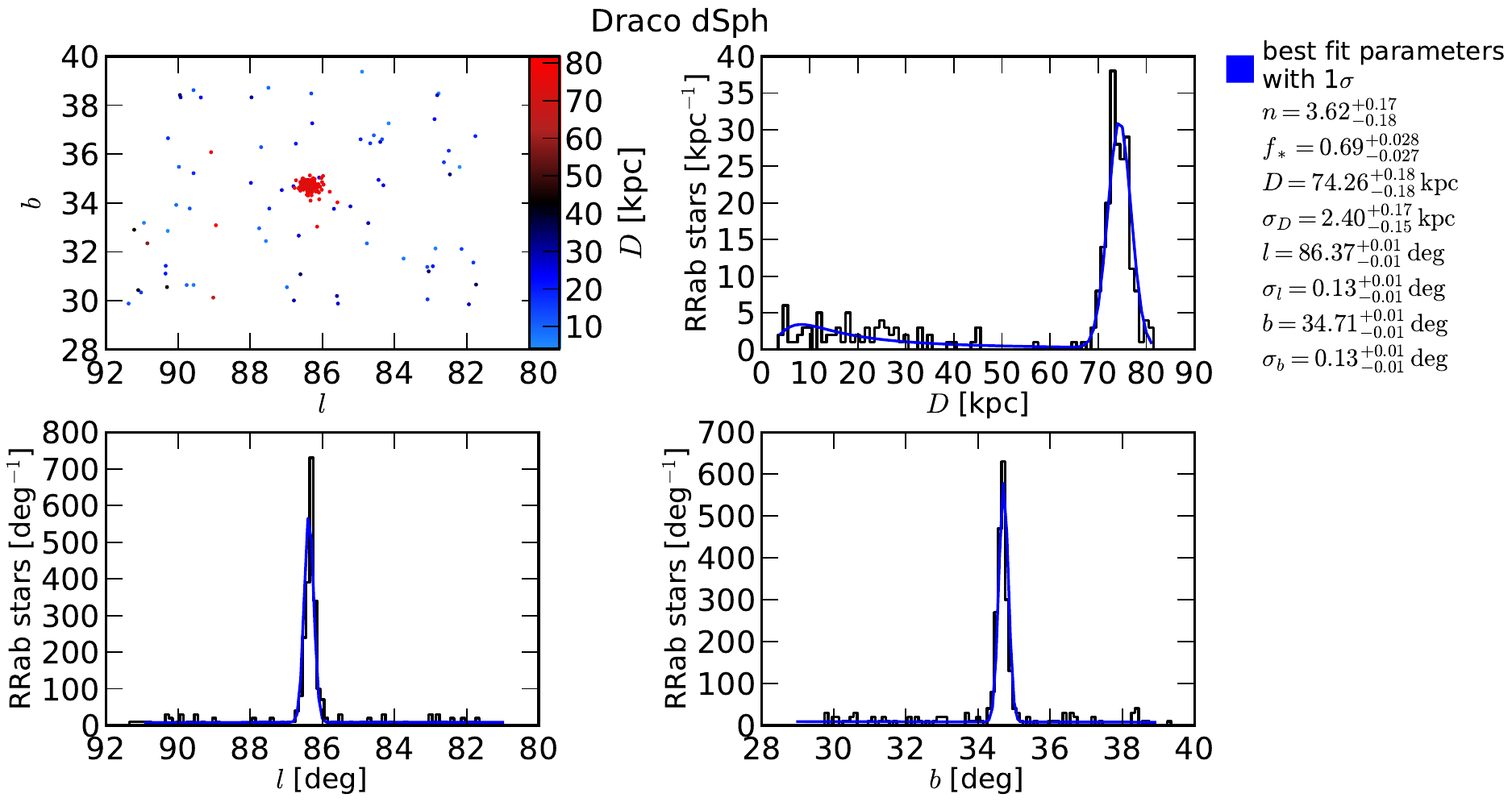}
}
\subfigure[]
    {
\includegraphics[]{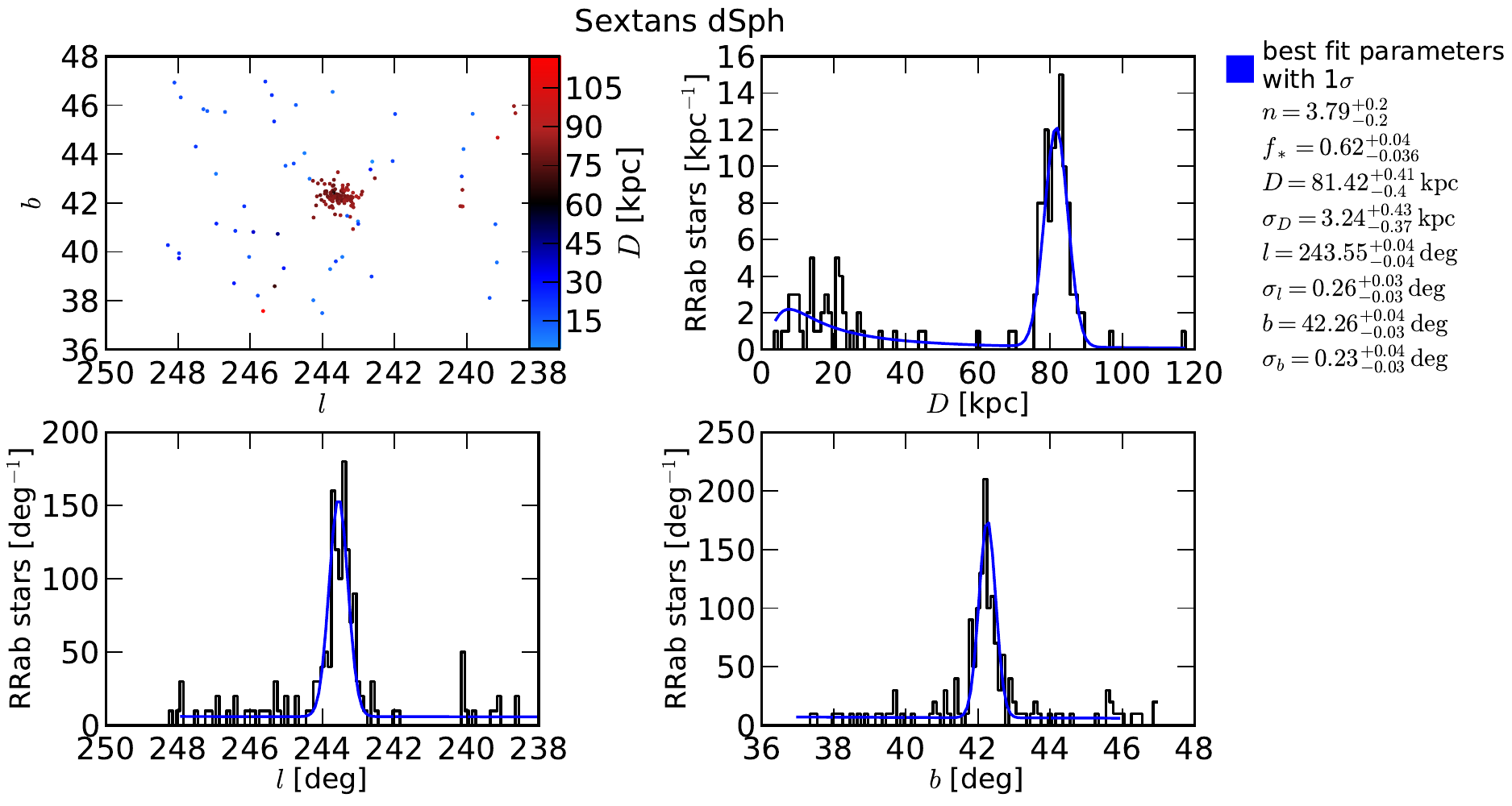}
}
\caption{{
The dwarf spheroidals Draco dSph (a) and Sextans dSph (b). For both subfigures, the first panel shows a map of RRab stars near the dwarf galaxy, 270 stars in the figure for Draco dSph, and 157 for Sextans dSph, respectively.
The stars are color-coded according to their heliocentric distance.
The other three panels show the histograms in $l$, $b$, $D$ for the stars from the first panel. Overplotted is the best-fit model from \ref{sec:DensityFitting} with the parameters given on the right.
}
\label{fig:Sextans_dSph_fits_Draco_dSph_fits}}
\end{center}  
\end{figure*}

% left bottom right top
\clearpage 

\clearpage

%%left bottom right top

\clearpage

%left bottom right top

\clearpage

\section{Tables}
\label{sec:Tables}
\newpage

\capstartfalse
\begin{deluxetable*}{lrrr}
\tablecolumns{2}
\tablecaption{Dwarf Galaxies, attempted\label{tab:dwarfs_planned}}
\tablewidth{0pt}
\tablehead{
\colhead{name} &  \colhead{ ($l_{\mathrm{prior}}$ [$\arcdeg$],} & \colhead{ $b_{\mathrm{prior}}$ [$\arcdeg$],} & \colhead{ $D_{\mathrm{prior}}$ [kpc])}
}
\startdata
Aquarius II dSph & 55.0 & -53.0 & 108 \\
Bootes I dSph & 358.036 & 69.642 & 60 \\ 
Crater II dSph & 282.9084 & 42.0276 & 117.5\\
Draco dSph       &  	86.3747 & 34.7171 & 79 	 	\\
Sagittarius dSph & 5.6081 & -14.0858 & 26.3\\
Segue 1 dSph & 220.488 & 50.420 & 23 	\\
Segue 2 dSph & 149.433 & -38.1352 & 35 	 \\
Sextans dSph &  243.4981 & 42.2721 & 86	   \\
Ursa Minor Dwarf dE4	&	104.9527 & 44.8028 & 63 			\\
Ursa Major I dSph	&	159.4311 & 54.4143 & 100 		\\
Ursa Major II Dwarf & 152.464 & 37.443 & 30 \\
\enddata
\tablecomments{The list of dwarf galaxies within the PS1 3$\pi$ footprint for which we want to determine their distance. Their $(l,b,D)$ coordinates are from \cite{McConnachie2012}.}
\end{deluxetable*}   
\capstarttrue

\capstartfalse
\begin{deluxetable*}{lrrr}
\tablecolumns{2}
\tablecaption{Globular clusters, attempted\label{tab:gc_planned}}
\tablewidth{0pt}
\tablehead{
\colhead{name} &  \colhead{ ($l_{\mathrm{prior}}$ [$\arcdeg$],} & \colhead{ $b_{\mathrm{prior}}$ [$\arcdeg$],} & \colhead{ $D_{\mathrm{prior}}$ [kpc])}
}
\startdata
IC 1257 & 16.54 & 15.15 & 25			\\
NGC 2419 & 180.37 & 25.24 & 82.6		 \\
NGC 4147 & 252.85 & 77.19 & 19.3 		\\
NGC 4590 (M68) & 299.63 & 36.05 & 10.3	\\
NGC 5024 (M53) & 332.96 & 79.76 & 17.9 \\	
NGC 5053 & 335.7 & 78.95 & 17.4		\\
NGC 5272 (M3) & 42.22 & 78.71 & 10.2	\\
NGC 5466 & 42.15 & 73.59 & 16 		\\
NGC 5634 & 342.21 & 49.26 & 25.2 		\\
NGC 5694 & 331.06 & 30.36 & 35 		\\
NGC 5897 & 342.95 & 30.29 & 12.5 		\\
NGC 5904 (M5) & 3.86 & 46.8 & 7.5 		\\
NGC 6093 (M80) & 352.67 & 19.46 & 10 	\\
NGC 6171 (M107) & 3.37 & 23.01 & 5.4 	\\
NGC 6229 & 73.64 & 40.31 & 30.5 		\\
NGC 6356 & 6.72 & 10.22 & 15.1 		\\
NGC 6402 (M14) & 21.32 & 14.81 & 9.3 	\\
NGC 6426 & 28.09 & 16.23 & 20.6 		\\
NGC 6864 (M75) & 20.3 & -25.75 & 20.9	\\
NGC 6934 & 52.1 & -18.89 & 15.6		\\
NGC 6981  (M72) & 35.16 & -32.68 & 17	\\
NGC 7006 & 63.77 & -19.41 & 41.2		\\
NGC 7078 (M15) & 65.01 & -27.31 & 10.4	\\
NGC 7089 (M2) & 53.37 & -35.77 & 11.5 	\\
NGC 7099 (M30) & 27.18 & -46.84 & 8.1 	\\
Pal 1 & 130.06 & 19.03 & 11.1			\\
Pal 3 & 240.15 & 41.86 & 92.5			\\
Pal 5 & 0.85 & 45.86 & 23.2 			\\
Pal 13 & 87.1 & -42.7 & 26			\\        
\enddata
\tablecomments{The list of globular clusters within the PS1 3$\pi$ footprint for which we want to determine their distance. Their $(l,b,D)$ coordinates are from a current update of \cite{Harris1996_2010}.}
\end{deluxetable*}   
\capstarttrue

\capstartfalse
\begin{deluxetable*}{lrrrrrrrr}
\tabletypesize{\scriptsize}
\tablecolumns{9}
\tablecaption{Fitted Dwarf Galaxies\label{tab:dwarfs}}
\tablewidth{0pt}
\tablehead{
\colhead{name} &  \colhead{fitted}& \colhead{fitted} &
\colhead{$\Delta l$ [kpc]} & \colhead{$\Delta b$ [kpc]}& \colhead{$\Delta D$ [kpc]} & \colhead{$\Delta D / \Delta l$} &\colhead{$\Delta D / \Delta b$}& \colhead{sources}  \\
\colhead{ } &  \colhead{($l$ [$\arcdeg$], $b$ [$\arcdeg$], $D$ [kpc])}& \colhead{($\sigma_l$ [$\arcdeg$],$\sigma_b$ [$\arcdeg$],$\sigma_D$ [kpc])} &
\colhead{ } & \colhead{ }& \colhead{ } & \colhead{ } &\colhead{ }& \colhead{ }
}
\startdata
Draco dSph        &   $86.37_{-0.01}^{+0.01}$, $34.71_{-0.01}^{+0.01}$, $74.26_{-0.18}^{+0.18}$ & $0.13_{-0.01}^{+0.01}$, $0.13_{-0.01}^{+0.01}$, $2.40_{-0.15}^{+0.17}$ & $0.28$ & $0.33$ & $4.81$ & $14.48$ & $14.66$ &    191		\\
Sagittarius dSph        &   -,-, $28.18_{-0.10}^{+0.10}$ & -, -, $1.01_{-0.01}^{+0.02}$ & $2.02$ & - & - & - & - &    538		\\
Sextans dSph            &    $243.55_{-0.04}^{+0.04}$, $42.26_{-0.03}^{+0.04}$, $81.42_{-0.40}^{+0.41}$ & $0.26_{-0.03}^{+0.03}$, $0.23_{-0.03}^{+0.04}$, $3.24_{-0.37}^{+0.43}$ & $0.55$ & $0.66$ & $6.49$ & $9.83$ & $9.87$ &   99 \\
Ursa Major I dSph	&	$159.38_{-1.48}^{+2.07}$, $54.43_{-0.39}^{+2.79}$, $94.33_{-4.94}^{+10.80}$ & $0.34_{-0.20}^{+2.53}$, $0.33_{-0.19}^{+2.05}$, $ 2.59_{-1.11}^{+5.90}$ & $0.65$  & $1.07$ & $5.18$ & $4.84$ & $4.84$ &    4		\\
Ursa Minor Dwarf dE4	&	$105.00_{-0.06}^{+0.62}$, $44.74_{-0.06}^{+0.36}$, $68.41_{-0.51}^{+0.51}$ & $0.16_{-0.03}^{+0.86}$, $0.14_{-0.02}^{+1.45}$, $2.17_{-0.59}^{+3.10}$ & $0.22$ & $0.33$ & $4.35$ & $19.77$ & $13.11$ &  53		\\
\enddata
\tablecomments{The best-fit positions $(l,b,D)$ and extent $(\sigma_l, \sigma_b, \sigma_D)$ of dwarf galaxies from Table \ref{tab:dwarfs_planned},
along with their 1$\sigma$ uncertainties. Assuming an ellipsoidal shape for each dwarf galaxy, we calculate their axis ratios by first translating their angular extent $(\sigma_l, \sigma_b)$ into a linear extent $(\Delta l, \Delta b)$ as given by Equ. \eqref{eq:delta_l} and Equ. \eqref{eq:delta_b}, and then use the line-of-sight depth $\Delta D = 2\sigma_D$ \eqref{eq:delta_D} to calculate the axis ratios $\Delta D / \Delta l$, $\Delta D / \Delta b$.\newline
In this table, Sagittarius Dwarf dSph and Crater II dSph from Table \ref{tab:dwarfs_planned} are missing, as we were not able to determine reasonable fits for them. For details on this, see Sec. \ref{sec:DensityFitting}.}
\end{deluxetable*}   
\capstarttrue

\capstartfalse
\begin{deluxetable*}{lrrrrrrrr}
\tablecolumns{9}
\tablecaption{Fitted Globular Clusters\label{tab:gc}}
\tablewidth{0pt}
\tablehead{
\colhead{name} &  \colhead{fitted}& \colhead{fitted} &
\colhead{$\Delta l$ [kpc]} & \colhead{$\Delta b$ [kpc]}& \colhead{$\Delta D$ [kpc]} & \colhead{$\Delta D / \Delta l$} &\colhead{$\Delta D / \Delta b$}& \colhead{sources}  \\
\colhead{ } &  \colhead{($l$ [$\arcdeg$], $b$ [$\arcdeg$], $D$ [kpc])}& \colhead{($\sigma_l$ [$\arcdeg$],$\sigma_b$ [$\arcdeg$],$\sigma_D$ [kpc])} &
\colhead{ } & \colhead{ }& \colhead{ } & \colhead{ } &\colhead{ }& \colhead{ }
}
\startdata
NGC 2419            &    $180.36_{-0.03}^{+0.03}$, $25.24_{-0.03}^{+0.04}, 79.70_{-0.37}^{+0.32}$ & $0.11_{-0.01}^{+0.01}$, $0.11_{-0.01}^{+0.02}$, $4.24_{-1.14}^{+1.33}$ & $0.28$    & $0.30$ & $8.47$ & $30.25$ & $27.81$  &    8 \\
NGC 4590 (M68)	    &    $299.62_{-0.08}^{+0.12}$, $36.07_{-0.11}^{+0.10}$, $10.48_{-0.28}^{+0.26}$ & $0.15_{-0.04}^{+0.68}$, $0.15_{-0.04}^{+0.65}$, $0.66_{-0.13}^{+1.29}$ & $0.04$ & $0.05$ & $1.32$ & $33.0$ & $24.15$	 &    5\\
NGC 5024 (M53) 	&    $332.93_{-0.12}^{+0.12}$, $79.74_{-0.04}^{+0.04}$, $18.25_{-0.14}^{+0.13}$ & $0.35_{-0.08}^{+0.12}$, $0.11_{-0.01}^{+0.02}$, $0.55_{-0.03}^{+0.08}$ & $0.04$ & $0.07$ & $1.09$ & $15.57$ & $15.31$  &    12\\
NGC 5053            &    $335.78_{-0.18}^{+0.19}$, $78.93_{-0.14}^{+0.16}$, $16.66_{-0.26}^{+0.28}$ & $0.30_{-0.16}^{+1.08}$, $0.18_{-0.06}^{+1.10}$, $0.75_{-0.20}^{+3.69}$ & $0.03$ & $0.10$ & $1.50$ &  $50.0$   & $14.54$  &    4\\
NGC 5272 (M3)      &    $42.20_{-0.07}^{+0.08}$, $78.71_{-0.02}^{+0.02}$, $10.48_{-0.07}^{+0.07}$ & $0.45_{-0.05}^{+0.05}$, $0.10_{-0.002}^{+0.01}$, $0.51_{-0.01}^{+0.02}$ & $0.03$ & $0.04$ & $1.02$ & $34.0$  & $27.05$  &    56\\
NGC 5466            &    $42.13_{-0.04}^{+0.04}$, $73.59_{-0.03}^{+0.04}$, $15.76_{-0.14}^{+0.14}$ & $0.12_{-0.02}^{+0.03}$, $0.11_{-0.01}^{+0.02}$, $0.54_{-0.03}^{+0.07}$ & $0.02$ & $0.06$ & $1.08$ & $54.0$  & $17.76$  &    10\\
NGC 5904 (M5)      &    $3.88_{-0.04}^{+0.03}$, $46.77_{-0.02}^{+0.03}$, $7.87_{-0.19}^{+0.19}$ & $0.13_{-0.02}^{+0.04}$, $0.11_{-0.005}^{+0.02}$, $0.53_{-0.02}^{+0.08}$ & $0.02$  & $0.03$ & $1.05$ & $52.5$  & $36.01$  &    24\\
NGC 6229            &    $73.64_{-0.04}^{+0.04}$, $40.31_{-0.04}^{+0.04}$, $29.94_{-0.19}^{+0.17}$ & $0.11_{-0.01}^{+0.02}$, $0.11_{-0.01}^{+0.03}$, $0.61_{-0.08}^{+0.19}$ & $0.09$ & $0.12$ & $1.22$ & $13.5$ & $10.31$  &    12\\
NGC 6864 (M75)	    &    $20.28_{-0.23}^{+0.25}$, $-25.76_{-0.28}^{+0.20}$, $20.79_{-0.35}^{+0.32}$ & $0.25_{-0.13}^{+1.46}$, $0.34_{-0.21}^{+2.70}$, $1.59_{-0.93}^{+4.27}$ & $0.16$ & $0.24$ & $3.17$ & $19.8$ & $13.01$  &    4\\
NGC 6934            &    $52.12_{-0.04}^{+0.04}$, $-18.87_{0.04}^{+0.03}$, $16.77_{-0.21}^{+0.21}$ & $0.11_{-0.01}^{+0.03}$, $0.11_{-0.01}^{+0.02}$, $0.57_{-0.05}^{+0.16}$ & $0.06$ & $0.07$ & $1.14$ & $19.00$ & $17.14$  &    10\\
NGC 6981 (M72)     &    3$35.16_{-0.03}^{+0.03}$, $-32.70_{-0.03}^{+0.03}$, $17.51_{-0.17}^{+0.15}$ & $0.11_{-0.01}^{+0.01}$, $0.11_{-0.01}^{+0.01}$, $0.56_{-0.05}^{+0.11}$ & $0.06$ & $0.07$ & $1.12$ & $18.70$ & $16.90$  &    15\\
NGC 7006            &    $63.78_{-0.03}^{+0.03}$, $-19.39_{-0.03}^{+0.03}$, $40.12_{-0.15}^{+0.16}$ & $0.11_{-0.004}^{+0.01}$, $0.11_{-0.01}^{+0.01}$, $0.59_{-0.06}^{+0.11}$ & $0.15$ & $0.15$ & $1.17$ & $7.80$ & $7.87$   &    16\\
NGC 7078 (M15)     &    $65.03_{-0.08}^{+0.11}$, $-27.32_{-0.08}^{+0.07}$, $11.07_{-0.22}^{+0.24}$ & $0.15_{-0.04}^{+1.07}$, $0.13_{-0.03}^{+0.95}$, $0.60_{-0.08}^{+2.69}$ & $0.05$ & $0.05$ & $1.21$ & $24.20$  & $23.42$  &    10\\
NGC 7089 (M2)      &    $53.46_{0.25}^{+0.22}$, $-35.81_{-0.29}^{+0.29}$, $12.11_{-0.28}^{+0.39}$ & $0.35_{-0.21}^{+2.38}$, $0.62_{-0.47}^{+3.60}$, $1.47_{-0.84}^{+6.28}$ & $0.12$ & $0.26$ & $2.95$ & $24.58$   & $11.24$ &     4\\			
Pal 3               &    $240.14_{-0.09}^{+0.10}$, $41.95_{-0.18}^{+0.21}$, $85.05_{-0.34}^{+0.32}$ & $0.14_{-0.03}^{+0.17}$, $0.34_{-0.17}^{+0.31}$, $1.76_{-1.08}^{+1.77}$ & $0.31$ & $1.01$ & $3.53$ & $11.39$ & $3.50$  &     3\\
Pal 5               &    $0.87_{-0.27}^{+0.29}$, $45.80_{-0.28}^{+0.26}$, $21.66_{-0.30}^{+0.33}$ & $0.58_{-0.43}^{+2.82}$, $0.31_{-0.19}^{+2.37}$, $1.61_{-0.98}^{+7.52}$ & $0.31$ & $0.24$ & $3.23$ & $10.41$  & $13.56$  &     3\\
\enddata
\tablecomments{The best-fit positions $(l,b,D)$ and extent $(\sigma_l, \sigma_b, \sigma_D)$ of globular clusters from Table \ref{tab:gc_planned},
along with their 1$\sigma$ uncertainties. Assuming an ellipsoidal shape for each globular cluster, we calculate their axis ratios by first translating their angular extent $(\sigma_l, \sigma_b)$ into a linear extent $(\Delta l, \Delta b)$ as given by Equ. \eqref{eq:delta_l} and Equ. \eqref{eq:delta_b}, and then use the line-of-sight depth $\Delta D = 2\sigma_D$ \eqref{eq:delta_D} to calculate the axis ratios $\Delta D / \Delta l$, $\Delta D / \Delta b$.\newline
In this table, NGC 4147, NGC 5634, NGC 5694, NGC 5897, IC 1257, NGC 6093 (M80), NGC 6171 (M107), NGC 6284, NGC 6356, NGC 6402 (M14), NGC 6426, NGC 7099 (M30), Pal 1 and 
Pal 13 from Table \ref{tab:gc_planned} are missing, as we were not able to determine reasonable fits for them. For details on this, see Sec. \ref{sec:DensityFitting}.}
\end{deluxetable*}   
\capstarttrue

    \capstartfalse
\begin{deluxetable*}{lrrrrrrrr}
\tabletypesize{\scriptsize}
\tablecolumns{4}
\tablecaption{Dwarf Galaxies with too few RRab stars\label{tab:dwarfs_mean}}
\tablewidth{0pt}
\tablehead{
\colhead{name} &  \colhead{mean}& \colhead{sources} &
\colhead{comment}  \\
\colhead{ } &  \colhead{ $\bar{D}$ [kpc]} &\colhead{ }& \colhead{ }
}
\startdata
Aquarius II dSph     &   $107.87$ & 1 &  	\\
Bootes I dSph & $60.61$ & 2 & \\
Crater II dSph & $105.48$ &  2 & \\			
Segue 1 dSph & $23.24$ & 1 & up to 2 additional sources nearby \\
Segue 2 dSph & $33.31$ & 1 & \\
Ursa Major II Dwarf & $33.02$ & 1 & \\
\enddata
\tablecomments{The mean positions $(l,b,D)$ of those dwarf galaxies from Table \ref{tab:dwarfs_planned} which don't have enough sources in the PS1 catalog of RR Lyrae stars for carrying out the fitting process described in Sec. \ref{sec:DensityFitting}.}
\end{deluxetable*}   
\capstarttrue

%\capstartfalse
\begin{deluxetable*}{lrrrrrrrr}
\tabletypesize{\scriptsize}
\tablecolumns{4}
\tablecaption{Globular Clusters with too few RRab stars\label{tab:gc_mean}}
\tablewidth{0pt}
\tablehead{
\colhead{name} &  \colhead{mean}& \colhead{sources} &
\colhead{comment}  \\
\colhead{ } &  \colhead{ $\bar{D}$ [kpc]} &\colhead{ }& \colhead{ }
}
\startdata
IC 1257 & $27.24$ & 1 & \\
NGC 4147  &  $18.54$ & 1 & \\
NGC 5634 & $25.81$ & 1 & \\
NGC 5694 & $33.96$ & 1 & \\
NGC 5897 &  $12.91$ & 1 &   up to 3 additional sources nearby \\
NGC 6093 & $10.74$ & 2 & \\
NGC 6171 (M107) & $6.01$ & 7 & \\
NGC 6356 & $11.02$ & 1 &   \\
NGC 6402 (M14) & & & many field stars at this distance\\
NGC 6426 & $19.83$ & 5 & \\
NGC 7099 (M30) & $8.41$ &  2 & \\
Pal 1 & & &		nothing detected at this position: mean D = 10.08 kpc \\                               
Pal 13 & $23.59$ & 3 & \\
\enddata
\tablecomments{The mean positions $(l,b,D)$ of those globular clusters from Table \ref{tab:gc_planned} which don't have enough sources in the PS1 catalog of RR Lyrae stars for carrying out the fitting process described in Sec. \ref{sec:DensityFitting}.}
\end{deluxetable*}   
\capstarttrue

\capstartfalse
\begin{deluxetable*}{llll}
\tablecolumns{4}
\tablecaption{Radii of Dwarf Galaxies\label{tab:dwarf_radii}}
\tablewidth{0pt}
\tablehead{
\colhead{name} &  \colhead{$r_{t}$ [arcmin]}&  \colhead{$r_{c}$ [arcmin]}& \colhead{max($\sigma_l$,$\sigma_b$) [arcmin]}  }
\startdata
Draco dSph            & 40.1 $\pm$ 0.9 \citep{Odenkirchen2001} & 8.33 \citep{Stoehr2002} & 7.8  \\
Sextans dSph           & 83.2 $\pm$ 7.1\citep{Roderick2016} & 14.14 \citep{Stoehr2002} &  16.5  \\
Ursa Minor Dwarf dE4           &  34 $\pm$ 2.4 \citep{Kleyna1998} & 10.5 \citep{Stoehr2002} & 9.6  \\
Ursa Major I dSph           &   & 7.28 \citep{Simon2007} & 20.4 \\                           
\enddata
\tablecomments{
Comparison of tidal radii $r_t$ and core radii $r_c$ to the angular extent max($\sigma_l$,$\sigma_b$) from our analysis for the four fitted dwarf galaxies. Ursa Major I dSph lacks a published tidal radius. We give uncertainties as far as provided. See also Fig. \ref{fig:dwarf_gc_all_radii}.}
\end{deluxetable*}   
\capstarttrue

\capstartfalse
\begin{deluxetable*}{lrrr}
\tablecolumns{4}
\tablecaption{Radii of Globular Clusters\label{tab:gc_radii}}
\tablewidth{0pt}
\tablehead{
\colhead{name} &  \colhead{$r_{t}$ [arcmin]}&  \colhead{$r_{c}$ [arcmin]}& \colhead{max($\sigma_l$,$\sigma_b$) [arcmin]}  }
\startdata
NGC 2419            & 8.74 & 0.35 & 6.6  \\
NGC 4590 (M68)	    &  30.34 & 0.69 & 9.0  \\
NGC 5024 (M53) 	& 21.75 & 0.36 & 21  \\
NGC 5053            & 13.67 & 1.98 & 18  \\
NGC 5272 (M3)      & 38.19 & 0.55 & 27   \\
NGC 5466            & 34.24 & 1.64 & 7.2 \\
NGC 5904 (M5)      &  28.40 & 0.42 & 7.8  \\
NGC 6229            & 5.38 & 0.13 & 6.6 \\
NGC 6864 (M75)	    &  7.28 & 0.10 & 20.4   \\
NGC 6934            & 8.37 & 0.25 & 6.6   \\
NGC 6981 (M72)     & 9.15 & 0.54 & 6.6 \\
NGC 7006            & 6.34 & 0.24 & 6.6  \\
NGC 7078 (M15)     & 21.50 & 0.07 & 9.0  \\
NGC 7089 (M2)      & 21.45 & 0.34  & 37.2  \\			
Pal 3               & 4.81 & 0.48 & 20.4  \\
Pal 5               & 16.28 & 3.25 & 34.8  \\
\enddata
\tablecomments{Tidal radius $r_t$, core radius $r_c$ from the 2010 version of \cite{Harris1996_2010}, as well as the angular extent max($\sigma_l$,$\sigma_b$) from our analysis for all the fitted globular clusters. See also Fig. \ref{fig:dwarf_gc_all_radii}.}
\end{deluxetable*}   
\capstarttrue

\begin{deluxetable*}{lllll}
\tabletypesize{\scriptsize}
\tablecolumns{5}
\tablecaption{Distance Comparison for Dwarf Galaxies\label{tab:dsph_comparison}}
\tablewidth{0pt}
\tablehead{
\colhead{name} &\colhead{$D$ [kpc] }& \colhead{method} & \colhead{literature $D$ [kpc]} & \colhead{method and reference}
}
\startdata
Aquarius II dSph  &       107.87 & 1 RRab &  $107.900 \pm 3.300$ & BHB stars \citep{Torrealba2016}\\
Bootes I dSph  & 60.61  & mean of 2 RRab &  65.3  & (BHB + RR Lyr) HSC/Subaru imaging \citep{Okamoto2012}\\
   &  &  &   $62 \pm 4$ &  15 RR Lyr \citep{Siegel2006}\\
   &  &  &   60.4 &  \citep{Hammer2018}\\
Crater II dSph & 105.48 & mean of 2 RRab & ${\sim} 120$ & CMD \citep{Torrealba2016a}\\
   &  &  & $112 \pm 5$ &  RR Lyr \citep{Joo2018}\\
   &  &  & $117.5 \pm 5$ & \citep{McConnachie2012}\\   
Draco dSph     &  $74.26_{-0.18}^{+0.18}$  & fit & $71 \pm 7$ & SDSS photometry \citep{Odenkirchen2001}\\
   &  &  &   $75.8 \pm 5.4$  &  94 RR Lyr \citep{Bonanos2004}\\
Sagittarius dSph &  $28.18_{-0.10}^{+0.10}$ & fit & $26.3 \pm 1.8$ & RGB tip \citep{Monaco2004} \\
Segue 1 dSph & 23.24 & 1 RRab & $23 \pm 2$ & CMD fitting, esp. horizontal branch \citep{Belokurov2007}\\
Segue 2 dSph & 33.31 & 1 RRab & 35 &  from 4 BHB stars \citep{Belokurov2009}\\
Sextans dSph & $81.42_{-0.40}^{+0.41}$ & fit &  $84.2 \pm 3.3$ & RR Lyr from Dark Energy Camera imaging \citep{Medina2018}\\
Ursa Minor Dwarf dE4 & $68.41_{-0.51}^{+0.51}$ & fit & 76 & RR Lyr and HB \citep{Bellazzini2002}  \\
  &   &   & $76 \pm 4$  & distance from HB, considering metallicity of UMi \citep{Carrera2002}  \\
Ursa Major I dSph &   $94.33_{-4.94}^{+10.80}$ & fit & $96.8 \pm 4$ & V mag of HB, HSC/Subaru imaging study \citep{Okamoto2008}\\
  &   &   &   $97.3_{-5.7}^{+6.0}$  &  variable stars in UMa 1 \citep{Garofalo2013}\\
Ursa Major II Dwarf & 33.02 & 1 RRab & $34.7_{-1.9}^{+2.0}$ &  1 RR Lyr \citep{DallOra2012}\\
\enddata
\tablecomments{Comparison of our distance estimates (column $D$ [kpc]) to reference values from literature. For our distance estimates, we give whether they are a result from the fitting process or are determined as a mean distance from a small number of RRab stars. For literature values, we give the method used to obtain the distance and the reference. }
\end{deluxetable*}   
\capstarttrue

\begin{deluxetable*}{llll}
\tabletypesize{\scriptsize}
\tablecolumns{4}
\tablecaption{Distance Comparison for Globular Cluster\label{tab:gc_comparison}}
\tablewidth{0pt}
\tablehead{
\colhead{name} &\colhead{$D$ [kpc] }& \colhead{method} & \colhead{literature $D$ [kpc]} 
}
\startdata
IC 1257 &   $27.24$ & 1 RRab &      25 \\
NGC 2419 &   $79.70_{-0.37}^{+0.32}$ &  fit& 82.6 \\
NGC 4147 &  $18.54$ & 1 RRab &  19.3    \\
NGC 4590 (M68) &  $10.48_{-0.28}^{+0.26}$ & fit & 10.3  \\
NGC 5024 (M53) &   $18.25_{-0.14}^{+0.13}$ & fit & 17.9  \\
NGC 5053 &  $16.66_{-0.26}^{+0.28}$  & fit & 17.4  \\
NGC 5272 (M3) &  $10.48_{-0.07}^{+0.07}$  & fit & 10.2  \\
NGC 5466 &     $15.76_{-0.14}^{+0.14}$  & fit& 16  \\
NGC 5634 &   $25.81$ & 1 RRab &  25.2  \\
NGC 5694 & $33.96$ & 1 RRab  &  35   \\
NGC 5897 &   $12.91$ & 1 RRab  &   12.5 \\
NGC 5904 (M5) &  $7.87_{-0.19}^{+0.19}$  & fit & 7.5 	\\
NGC 6093 (M80) &  $10.74$ & 2 RRab  &  10  \\
NGC 6171 (M107) & $6.01$ & 7 RRab  &   5.4 \\
NGC 6229 &   $29.94_{-0.19}^{+0.17}$   &  fit & 30.5  \\
NGC 6356 & $11.02$ & 1 RRab  &  15.1   \\
NGC 6426 & $19.83$ & 5 RRab  & 20.6   \\
NGC 6864 (M75) &    $20.79_{-0.35}^{+0.32}$  & fit& 20.9   \\
NGC 6934 &  $16.77_{-0.21}^{+0.21}$  & fit & 15.6    \\
NGC 6981  (M72) &  $17.51_{-0.17}^{+0.15}$  & fit& 17  \\
NGC 7006 &  $40.12_{-0.15}^{+0.16}$   & fit& 41.2  \\
NGC 7078 (M15) &    $11.07_{-0.22}^{+0.24}$   & fit& 10.4  \\
NGC 7089 (M2) &      $12.11_{-0.28}^{+0.39}$  & fit& 11.5   \\
NGC 7099 (M30) & $8.41$ &  2 RRab  & 8.1   \\
Pal 3 &  $85.05_{-0.34}^{+0.32}$   &fit &  92.5   \\
Pal 5 &  $21.66_{-0.30}^{+0.33}$  & fit & 23.2   \\
Pal 13 &  $23.59$ & 3 RRab  &  26	  \\
\enddata
\tablecomments{Comparison of our distance estimates (column $D$ [kpc]) to reference values from literature. For our distance estimates, we give whether they are a result from the fitting process or are determined as a mean distance from a small number of RRab stars. For literature values, we use the 2010 online version of the \citep{Harris1996_2010} catalog.}
\end{deluxetable*}   
\capstarttrue

\end{document}